\documentclass[journal]{IEEEtran}
\ifCLASSINFOpdf
  \usepackage[pdftex]{graphicx}
\else
   \usepackage[dvips]{graphicx}
\fi
\usepackage[caption=false,font=footnotesize]{subfig}

\usepackage{amsmath}
\usepackage{amssymb}
\usepackage{bm}

\usepackage{mathrsfs}
\usepackage{amsfonts}
\usepackage{euscript}

\newtheorem{theorem}{Theorem}
\newtheorem{definition}{Definition}
\newtheorem{propos}{Proposition}
\newtheorem{lemma}{Lemma}

\def\be{\begin{eqnarray}}
\def\bea{\begin{eqnarray}}
\def\bma{\begin{mathletters}}
\def\ee{\end{eqnarray}}
\def\eea{\end{eqnarray}}
\def\ema{\end{mathletters}}
\def\Tr{{\rm Tr}}

\newcommand{\arcosh}{\mathop{\mathrm{arcosh}}\nolimits}
\newcommand{\diag}{\mathop{\mathrm{diag}}\nolimits}

\def\cn{\mathfrak{cn}}
\def\sn{\mathfrak{sn}}

\begin{document}
%
\title{Methods for Estimating Capacities and Rates of Gaussian Quantum Channels}

\author{Oleg V. Pilyavets, Cosmo Lupo
        and~Stefano~Mancini
\thanks{O. Pilyavets is with the
School of Science and Technology, Physics Division, University of Camerino, I-62032 Camerino, Italy and with the
P. N. Lebedev Physical Institute, Leninskii Prospect 53, Moscow 119991, Russia (email: pilyavets@gmail.com).}
\thanks{C. Lupo is with the School of Science and Technology, Physics Division, University of Camerino,
62032 Camerino, Italy (email: cosmo.lupo@unicam.it).}
\thanks{S. Mancini is with the School of Science and Technology, Physics Division, University of Camerino,
62032 Camerino, Italy (email: stefano.mancini@unicam.it).}
\thanks{}}

\markboth{Methods for Estimating Capacities and Rates of Gaussian Quantum
Channels}%
{Pilyavets \MakeLowercase{\emph{et al.}}:
Methods for Evaluating Capacities of Gaussian Quantum Channels}

\maketitle

\begin{abstract}
Optimization methods aimed at estimating the capacities of a general Gaussian channel are
developed. Specifically evaluation of classical capacity as maximum of the Holevo
information is pursued over all possible Gaussian encodings for the lossy bosonic channel,
but extension to other capacities and other Gaussian channels seems feasible.

Solutions for both memoryless and memory channels are presented.  It is first dealt with
single use (single-mode) channel where the capacity dependence from channel's parameters
is analyzed providing a full classification of the possible cases.  Then it is dealt with
multiple uses (multi-mode) channel where the capacity dependence from the (multi-mode)
environment state is analyzed when both total environment energy and environment purity
are fixed.  This allows a fair comparison among different environments, thus understanding
the role of memory (inter-mode correlations) and phenomenon like superadditivity of the
capacity.

The developed methods are also used for deriving transmission rates with
heterodyne and homodyne measurements at the channel output.  Classical capacity
and transmission rates are presented within a unique framework where the rates
can be treated as logarithmic approximations of the capacity.
\end{abstract}


\IEEEpeerreviewmaketitle

\section{Introduction}
\IEEEPARstart{Q}{uantum} channels
are every means that convey quantum systems on whose states information is
encoded. Formally they are quantum maps from input to output
states~\cite{Hol72}. The maximum rate at which information can be reliably
transmitted through a quantum channel defines its capacity. Actually one can
define several capacities depending on the kind of information transmitted
(classical or quantum) and on the additional resources used in
transmission~\cite{BS98}.

Evaluation of quantum channel capacities is one of the most important and
difficult problems of quantum information theory.  Gaussian channels, which maps
input Gaussian states into output Gaussian states, are among the simplest models
allowing capacities investigation~\cite{HW}. They are also relevant for
experimental implementations in quantum optics~\cite{Bra} and for security
analysis in continuous variables quantum key distribution \cite{optgaus}.

A paradigmatic example of Gaussian quantum channel is the lossy bosonic channel
\cite{HW,EisertWolf} where states lose energy `en route' from the sender to the
receiver. The term bosonic arises because each input (respectively output) is
represented by an optical bosonic field mode. In turn, the effect of losses is
usually modeled by letting each input mode interact with an environment mode
through a rotation (beam splitter) transform whose angle (transmissivity)
determines the loss rate~\cite{Bra}.

The classical capacity and the classical assisted capacity for such a channel
were evaluated in Refs.~\cite{GG2,GG1} by assuming each environment mode in the
vacuum state.  Subsequently, also the quantum capacity has been
derived~\cite{WGG}. However, when more general states of the environment are 
taken into account, e.g.\ non-separable ones giving rise to memory
effect~\cite{GM}, the evaluation of capacities becomes much more demanding.
Attempts have been carried out in \cite{PRA,NJP} by resorting to specific
parameters' ranges and numerics.

There are different ways to introduce memory effects in such channels (see
e.g.~\cite{PRA} and~\cite{CosmoLBCPRL}).  Here, we shall refer to the method
first presented in~\cite{GM}.  Moreover, we will solely consider classical
capacity and classical information transmission rates.  

Finding classical capacity results in the constrained maximization of Holevo
information \cite{Haus,Sch97,Hol98} over input states, where constraints appear
due to the restriction on input energy.  We shall confine our attention to
Gaussian inputs which in practice are the most important set of states and are
also conjectured to be optimal \cite{minoutent}. However, this gives rise to a
maximization problem which in general might be not spectral, therefore we shall
consider only that class of memory models which result in a spectral problem.
The latter will allow us to split the maximization for memory channel in two
steps: the maximization inside each channel's mode (use) respecting its own
energy restriction (it gives the capacity for the single channel use
(single-mode)), and a further optimization of the distribution of total input
energy over different channel modes (uses).  This essential simplification is
possible thanks 
to the obtained proof of concavity for one-shot capacity over input energy
and to the additivity of the Holevo function in the memoryless case (see also~\cite{HiroshimaPRA}).  

As far as the first maximization step involves the optimization inside each
channel mode separately, we shall first discuss the single channel use
(single-mode).  It can be shown that its environment is characterized by two
parameters: the amount of squeezing and the average amount of thermal photons.
To completely specify the channel usage we also have to consider the transmissivity
value and the input energy restriction.  Thus, the classical capacity is found  
to be a monotonic function of all these parameters except of the environment squeezing.
This makes the latter a specific parameter indicating different channel's
regimes. In particular, it turns out that the capacity does not depend on any
parameters except of the input energy if the environment squeezing tends to infinity (see
also~\cite{PRA,NJP}).  Then we shall deeply study this behavior putting forward
the existence of \emph{critical parameters} that characterize the general
behavior of the channel.  We will also find out \emph{supercritical parameters},
which in turn characterize the behavior of the critical parameters, and can be somehow
regarded as fundamental constants.

We shall then move to the multi-mode channel setting to address the second
maximization step.  This will be done by resorting to convex separable
programming techniques~\cite{StefStef,StefBook} and will allow us to draw
conclusions about the memory channel.  This has became a palatable subject
because of the possibility of enhancing the memoryless capacity~\cite{Cerf2005}.
This fact gives evidence of the superadditive phenomenon for quantum memory
channels. However, in order to establish the superadditivity of the memory channel, one
has to fairly compare different environments, by e.g. using the same energy
constraints and purity. We shall investigate this problem showing optimality of
non-homogenuous distribution of energy over modes for some channel's parameters,
which happens due to non-monotonic dependence of the one-shot capacity from the
environment squeezing discussed above. That can be interpreted as violation of
mode symmetry, because the optimization problem is completely symmetric over
channel modes. In turn, this mode symmetry violation can be related to the
quadrature symmetry violation occurring in the single-mode channel.  Then we can
conclude that capacity is superadditive if mode symmetry is violated and
additive otherwise.

It worth noticing that also the recent study \cite{NewNoiseChannel} about the
effect of noise correlation on the capacity of additive Gaussian noise channel
can be brought back to the above sketched approach. 

Finally, we will make use of the developed methods for deriving transmission
rates which are even more relevant than capacity for practical purposes.
Specifically we will account for  the most common continuous variable
measurements at the channel output, namely heterodyne and homodyne
measurements~\cite{dyne}.  Preliminary studies on such rates for lossy memory
channel have been performed in \cite{CosmoRates}.  Here, throughout the paper,
capacity and transmission rates are presented in the same framework showing an
unexpected parallelism between these quantities.  Actually, within this
framework the rates result as logarithmic approximations to the capacity.
Similarly to the capacity, in general they are also subjected to violation of
quadrature and mode symmetry, which will allow us to pose the optimal memory
problem and calculate the critical parameters for the rates as well.

The paper is organized as follows.  In Sec.~\ref{GaussQuantChannelsSec}
Gaussian channels are introduced.  In Sec.~\ref{ClassicalCapacitySec} the
classical capacity together with the information transmission rates are defined.
In Sec.~\ref{MLChannel} classical capacity and transmission rates for
single-mode lossy bosonic channel are evaluated. 
In Sec.~\ref{CritParsSec} the role of single-mode channel parameters
is discussed by evaluating critical and supercritical parameters for
capacity and rates.  In Sec.~\ref{MemoryChannelsSec} the capacity and rates for
the multi-mode channel are evaluated and a particular memory model is
studied.  Sec.~\ref{ConclusionSec} is for conclusions.

\section{Gaussian quantum channels}\label{GaussQuantChannelsSec}

Quantum mechanics in continuous variables can be introduced independently
from Dirac approach as Weyl star-product (also known as Weyl
calculus~\cite{SymplGeom}) which operates with Weyl symbols defined on
system's phase space $\mathbf{(q,p)}$. Quadratures $\mathbf{q}$ and
$\mathbf{p}$ for the system with $n$ degrees of freedom are $n$-dimensional
vectors of canonical variables. Below it will be useful to consider a vector
\begin{equation}\label{Xquadrs}
\mathbf{x}:=\mathbf{(q,p)}=(q_1,\dots,q_n,p_1,\dots,p_n).
\end{equation}
Any quantum state, usually represented as a density operator $\hat\rho$ in the
Hilbert space $\mathcal{H}\equiv\otimes^nL_2(\mathbb{R})$,  can be specified in
the above framework by its \emph{Wigner function} $W(\mathbf{q},\mathbf{p})$,
which is a Weyl symbol of  $\hat\rho$.  Its relation with the density matrix in
the $\mathbf{q}$ representation reads\footnote{Throughout the paper it is
assumed commutation relations between canonical operators $\hat q_h,\hat p_l$
belonging to $\mathcal{H}$ to be $[\hat q_h,\hat p_l]=i\delta_{hl}$ (with
$\delta$ the Kronecker symbol and $\hbar=1$), and normalization of a $n$-mode
Wigner function to be $\int W(\mathbf{x})\,d\mathbf{x}=(2\pi)^n$.}
\begin{align*}
&W(\mathbf{q},\mathbf{p})=
\int\rho\left(\mathbf{q}+\frac{\mathbf{u}}2,\mathbf{q}-\frac{\mathbf{u}}2\right)
e^{-i\mathbf{pu}}d\mathbf{u},\\
&\rho(\mathbf{q},\mathbf{q'})=
\frac1{(2\pi)^n}\int W\left(\mathbf{p},\frac{\mathbf{q+q'}}2\right)
e^{i\mathbf{p}(\mathbf{q-q'})}d\mathbf{p}.
\end{align*}

In this work we apply Weyl calculus to the system of $n$ one-dimensional
harmonic oscillators, therefore we will call these degrees of freedom as
\emph{modes}. Furthermore, we restrict all possible quantum states of these
oscillators by \emph{Gaussian} ones, which are defined as follows. The quantum
state $\hat\rho$ is called Gaussian if its Wigner function is Gaussian,
\emph{i.e.} such state can be completely specified by quadratures covariance
matrix $V$ and vector $\mathbf{a}$ which are parameters (the second and first
moments) of its Wigner function\footnote{Notice, that Eq.~\eqref{WinFunStart}
completely specifies the ordering of covariances in matrix~$V$ as corresponding
to the vector~\eqref{Xquadrs}.}:
\begin{equation}
{\hat\rho}\:\,\leftrightarrow\:\,\{\mathbf{a},V\}
\leftrightarrow\:\,W(\mathbf{x})=
\frac1{\sqrt{\det {V}}}
e^{-\tfrac12\left( \mathbf{x}-\mathbf{a},V^{-1}(\mathbf{x}-\mathbf{a})\right)},
\label{WinFunStart}
\end{equation}
where $(,)$ stands for the real scalar product and the vector $\mathbf a$ represents
displacement in the phase space.  The quantities to be studied do not depend on
this displacement, therefore each quantum state and each classical\footnote{We
will use convention accepted in quantum information theory, where random
variables and probability densities of standard (classical) information theory
are called \emph{classical} to distinguish them from \emph{quasi-probability}
distributions (and variables associated with them) appearing in quantum
setting.} Gaussian distribution will be solely labeled by their quadratures
covariance matrices, e.g.  $\hat\rho\leftrightarrow V$.

Notice, that any Gaussian distribution of the form~\eqref{WinFunStart} is the
Wigner function of some quantum state if its covariance matrix $V$ satisfies the
Heisenberg uncertainty condition~\cite{SymplGeom,Mukunda}
\begin{equation}\label{unc}
V+\frac{i\Sigma}2\geqslant0,
\end{equation}
where
\begin{eqnarray}
\label{symplecticform}
\Sigma = \left( 
\begin{array}{cc}
\mathbb{O}_n&\mathbb{I}_n\\
-\mathbb{I}_n&\mathbb{O}_n
\end{array}
\right)
\end{eqnarray}
is the symplectic form with $\mathbb{O}_n$ and $\mathbb{I}_n$ the $n\times n$
null and identity matrices, respectively. The eigenvalues of $\Sigma V$ are $2n$
purely imaginary numbers  $\{\pm i\nu_k\}$, ${k=1,\dots,n}$, where $\{\nu_k\}$
are called \emph{symplectic eigenvalues} of $V$.  The condition~\eqref{unc} can
be equivalently written as inequalities $\nu_k\geqslant1/2$, which are saturated
by pure Gaussian states~\cite{Bra,SymplGeom}.

A \emph{Gaussian quantum channel} acting on $n$ modes is by
definition a \emph{completely positive} and \emph{trace preserving} map defined
on the set of quantum states, which maps any $n$-mode Gaussian state into a 
$n$-mode Gaussian state.  
As a consequence, it is any map $\Phi_{n}$ of moments~\cite{EisertWolf}
\begin{equation}
\label{Gmap}
\bigl\{\mathbf{a},V\bigr\}\mapsto
\bigl\{X_{n}^\top\mathbf{a}+\mathbf{d}_{n}^{\phantom{\top}},X_{n}^\top
VX_{n}^{\phantom{\top}}+Y_{n}^{\phantom{\top}}\bigr\}
\end{equation}
characterized by the triad $(\mathbf{d}_{n},X_{n},Y_{n})$,
where $X_{n},Y_{n}$ are two real $2n\times2n$-matrices obeying the inequality
\begin{equation*}
Y_{n}^{\phantom{\top}}+
\frac i2\left(\Sigma-X_{n}^\top\Sigma X_{n}^{\phantom{\top}}\right)\geqslant0
\end{equation*}
with $Y_{n}\geqslant0$ and symmetric   
and $\mathbf{d}_{n}\in\mathbb{R}^{2n}$ a displacement vector.

A very special case is that of the \emph{memoryless channel}, for
which $\Phi_{n}=\Phi_{1}^{\otimes n}$ is the direct product of $n$ identical
maps, \emph{i.e.} a single-mode Gaussian channel used $n$ times.
It is hence characterized by a triad
\begin{equation*}
\bigl({\textstyle\bigoplus^n\mathbf{d}_{1},\bigoplus^nX_{1},\bigoplus^nY_{1}}\bigr),
\end{equation*}
where we have denoted 
\begin{align*}
\mathbf{d}_1&:=(d_q,d_p)^\top,\\
{\textstyle\bigoplus^n}\mathbf{d}_{1}&:=
(d_q,\dots,d_q,d_p,\dots,d_p)^\top\in\mathbb{R}^{2n}.
\end{align*}
Notice, that $X_1,Y_1$ are $2\times2$-matrices, whose entries are scalars, and
the direct sums $\bigoplus^nX_{1}$, $\bigoplus^nY_{1}$ are $2\times2$ matrices,
whose entries are $n\times n$ diagonal matrices\footnote{Such a convention was
chosen to be consistent with the ordering~\eqref{Xquadrs} and the symplectic
form~\eqref{symplecticform}.}.  Loosely speaking, the memoryless channel acts
equally and independently on each of its uses.

More generally we can consider the case of a \emph{quantum channel
with memory} (or simply a \emph{memory channel}). It is any channel which is
\emph{not} memoryless. Making no assumption on additional structures that might
be present (e.g.\ causality, invariance under time translations), we can only
say that $\Phi_{n}\neq\Phi_{1}^{\otimes n}$ or
\begin{equation*}
(\mathbf{d}_{n},X_{n},Y_{n})\neq
\bigl({\textstyle\bigoplus^n\mathbf{d}_{1},\bigoplus^nX_{1},\bigoplus^nY_{1}}\bigr).
\end{equation*}
The memory channel can be interpreted as a framework to describe correlations
between channel actions corresponding to different channel uses.

\section{Classical Capacity}\label{ClassicalCapacitySec}

The Gaussian quantum channel can be used to transmit classical information by
\emph{encoding} a classical stochastic continuous variable
$\bm\alpha\in\mathbb{R}^{2n}$, distributed according to a probability density
$P_{\bm\alpha}$, into a set of quantum states (Wigner functions)
$W_{\bm\alpha}$. The maximum rate at which classical information can be reliably
sent through the channel defines its \emph{classical capacity}.

In the case of a memoryless quantum channel, its classical capacity is given by
\cite{Sch97, Hol98}
\begin{equation}\label{Holevo}
C=\lim_{n\to\infty}\frac{\chi\left(\Phi_n\right)}n,
\end{equation}
where the \emph{Holevo function} $\chi$ evaluated on $n$ channel uses is defined
as\footnote{Contrarily to the original definition~\cite{Hol98}, in this paper we
incorporate the maximum over input states in the Holevo function.}
\begin{multline}
\chi\left(\Phi_n\right)=
\max_{\{W_{\bm\alpha},P_{\bm\alpha}\}}
\biggl\{
S\left[
\int \Phi_n\left(W_{\bm\alpha}\right)P_{\bm\alpha}d\bm\alpha\right]\\
-\int S\bigl[
\Phi_n\left(W_{\bm\alpha}\right)\bigr]P_{\bm\alpha}d\bm\alpha
\biggr\},
\label{HolevoGen}
\end{multline}
with $\Phi_n=\Phi_1^{\otimes n}$ and $S$ the von Neumann entropy. Thus, the
computation of the memoryless capacity\footnote{As far as we consider only
classical channel capacity, it will be often called as simply the capacity.} is
based on the optimization over all input ensembles $W_{\bm\alpha}$, including
those made of states which are entangled among different channel uses. 

If the input states are restricted to an ensemble of product
states, it is reasonable to consider the so-called \emph{one-shot capacity}
\begin{equation*}
C_1=\chi\left(\Phi_1\right)
\end{equation*}
obtained from Eqs.~\eqref{Holevo} and~\eqref{HolevoGen} by assuming $n=1$.
Clearly the one-shot capacity is a lower bound on the memoryless capacity.  If
these two quantities coincide, the Holevo function is said to be additive.  In
turn, additivity of the Holevo function dramatically simplifies the problem of
evaluating the memoryless capacity.  Even though the Holevo function has been
shown to be additive for several relevant channels, this property does not
always hold \cite{Hastings}. 

Moving to the general case, one could be tempted to generalize the
formulae~\eqref{Holevo} and~\eqref{HolevoGen} to the case of memory channels by
applying them for $\Phi_n \neq \Phi_1^{\otimes n}$.  Quite generally we can say
that the relation~\eqref{Holevo} only provides an upper bound for the capacity
of the memory channel \cite{BM}.  Indeed, it has been proven~\cite{KW2} that it
coincides with the memory channel capacity for the class of so-called
\emph{forgetful} channels. 

Thus, on the one hand we can define the upper bound
\begin{equation*}
\overline{C}:=\lim_{n\to\infty}\overline{C}_n,\quad\quad
\overline{C}_n:=\frac{\chi(\Phi_n)}n.
\end{equation*}
On the other hand, for any $n$, one can look at $n$ uses of the channel
described by $\Phi_n$ as a single $n$-mode memoryless channel.  Its one-shot
capacity found as maximum over the set of Gaussian states provides a lower bound
on the capacity of the memory channel~\cite{NJP}
\begin{equation}\label{HGaussian}
\underline{C}_n := 
\frac{\chi_{\scriptscriptstyle G}(\Phi_n)}n.
\end{equation}
Taking the limit over $n$, we can as well define the lower
bound\footnote{Throughout the paper, for the sake of simplicity, we will often
refer to this lower bound as simply the capacity.}
\begin{equation}
\underline{C}:=\lim_{n\to\infty}\underline{C}_n.
\label{underC}
\end{equation}

Since the capacity (in sense of the above definitions) in continuous variables
case turns out to be infinite, some physically motivated constraints must be
specified to avoid meaningless results. 
A typical choice in the framework of Gaussian channels is to impose a restriction on the maximal
average input energy per channel use.  As far as we are considering the system
of $n$ single-mode oscillators (see Sec.~\ref{GaussQuantChannelsSec}) with
channel uses corresponding to oscillators modes, this constraint
reads\footnote{We assume quantum states to have zero mean, \emph{i.e.}
$\langle\mathbf x\rangle=0$.}
\begin{equation}\label{chi_constr}
\frac1{2n(2\pi)^n}\int\left(
\int W_{\bm\alpha}(\mathbf x)P_{\bm\alpha}d\bm\alpha
\right)
{\mathbf x^2}d\mathbf x
\leqslant N+\frac{1}{2},
\end{equation}
where $N$ represents the maximum number of excitations (photons) per mode in
average.  

Finally, let us consider \emph{Gaussian encoding} $W_{\bm\alpha},P_{\bm\alpha}$
used to calculate $\chi_{\scriptscriptstyle G}$.  For $n$ uses of the quantum
channel, we fix a reference $n$-mode Gaussian state, with zero mean, which is
described by the Wigner function $\{0,V_\mathrm{in}\}$ (see the
definition~\eqref{WinFunStart}).  A classical variable $\bm\alpha$ will be
encoded by applying a displacement operation on the reference state, thus
obtaining Wigner function $\{\sqrt{2}\bm\alpha,V_\mathrm{in}\}$.  We assume the
stochastic variable $\bm\alpha$ to be itself distributed according to the
Gaussian probability density distribution with zero mean:
\begin{equation*}
P_{\bm\alpha} = \frac{1}{(2\pi)^n\sqrt{\det V_\mathrm{mod}}}
e^{-\left(\bm\alpha,V_\mathrm{mod}^{-1}\bm\alpha\right)}.
\end{equation*}
Hence, the corresponding ensemble state
\begin{equation*}
\int W_{\bm\alpha}P_{\bm\alpha}d\bm\alpha,
\end{equation*}
is also Gaussian and described by a Wigner function $\{0,\overline
V_\mathrm{in}\}$, where
\begin{equation}
\overline V_\mathrm{in}=V_\mathrm{in}+V_\mathrm{mod}.
\label{VinAv}
\end{equation}
Quadratures covariance matrices of output state and output average state below
will be labeled by $V_\mathrm{out}$ and $\overline V_\mathrm{out}$,
respectively:
\begin{equation}
\begin{split}
V_\mathrm{out}\,&\leftrightarrow\,\Phi_n\left(W_{\bm\alpha}\right),\\
\overline V_\mathrm{out}&\leftrightarrow
\int\Phi_n\left(W_{\bm\alpha}\right)P_{\bm\alpha}\,d\bm\alpha.
\end{split}
\label{VoutAndAvGenForm}
\end{equation}

The restriction to Gaussian states, which are mapped into Gaussian states by
Gaussian channels, dramatically simplifies the problem, since the complexity of
specifying Gaussian states is polynomial in the number $n$ of modes (see
Eq.~\eqref{WinFunStart}).  Moreover, Gaussian states are conjectured to be
optimal inputs for Gaussian channels \cite{minoutent}.

The von Neumann entropy of a $n$-mode Gaussian state $\{\mathbf{a},V\}$ is the
function of symplectic eigenvalues $\nu_k$ of matrix $V$~\cite{Bra}:
\begin{equation}
\label{von}
S(V) = \sum_{k=1}^n g\left(\nu_k-\frac12\right),
\end{equation}
where $g$ is defined as
\begin{equation*}
g(v):=(v+1)\log_2(v+1)-v\log_2v.
\end{equation*}
The Holevo-$\chi$ quantity for the set $G$ of Gaussian states can be derived
from Eqs.~\eqref{HolevoGen} and \eqref{von}.  It equals~\cite{HW}
\begin{equation}
\chi_n=\max_{V_\mathrm{in},V_\mathrm{mod}}
\sum_{k=1}^{n}
\left[g\left( \overline\nu_{k} - \frac{1}{2}\right) 
-g\left( \nu_{k} - \frac{1}{2}\right)\right],
\label{HolDef}
\end{equation}
where $\chi_n$ is a shorthand notation for $\chi(\Phi_n)$.  In turn, the
quantities $\overline\nu_{k}$ and $\nu_k$ are the symplectic eigenvalues of
$\overline V_\mathrm{out}$ and $V_\mathrm{out}$, respectively.  Finally, the
input energy constraint~\eqref{chi_constr} for Gaussian states can be written in
terms of the covariance matrices as
\begin{equation}
\frac{\Tr\overline V_\mathrm{in}}{2n}\leqslant N+\frac12.
\label{constr2}
\end{equation}

\subsection{Estimating the classical capacity}\label{EstimClassCapacity}

As we have seen the evaluation of the classical capacity practically reduces to the
evaluation of the function~\eqref{von}.  Notice, that $g(v)$ is not analytic in
the neighborhood of zero where its asymptotic value is $-v\log_2v$. Also, the
function $g\left(v-\frac12\right)$ is not analytic in the neighborhood of infinity, where its
asymptotic value is $\log_2v$.  By subtracting this logarithm part we get the
analytic function in the region $v\geqslant\frac12$ which has its Laurent series (see
also~\cite{HSH})
\begin{equation}
g\left(v-\frac12\right)=\log_2v+\frac1{\ln2}\left[1-
\frac12\sum_{j=1}^{\infty}\frac{(2v)^{-2j}}{j(2j+1)}
\right]
\label{decom}
\end{equation}
written in the neighborhood of infinity.  In particular, to the zeroth-order
approximation it is
\begin{equation}
\label{zeroth}
g\left(v-\frac12\right)=\log_2v+\frac1{\ln2},
\end{equation}
where we have neglected terms of the order $O\left(v^{-2}\right)$. Allowing perturbation of
logarithm by the first terms in the series~\eqref{decom} we can also construct
next-order approximations.

In what follows, it will be convenient to introduce the function 
\begin{equation}
g_{j}(v):=v^{j}g^{(j)}\left(v-\frac12\right),
\label{gkfun}
\end{equation}
where $j=0,1,2,\dots$. Thus, 
\begin{equation}
\begin{split}
&g_0(v)=g\left(v-\frac12\right),\\
&g_1(v)=vg'\left(v-\frac12\right)=v\log_2\frac{v+\frac12}{v-\frac12},\\
&g_2(v)=v^2g''\left(v-\frac12\right)=-\frac{v^2}{\left(v^2-\frac14\right)\ln2},
\end{split}
\label{gExplicit}
\end{equation}
and so on. It also has simple rules for derivatives, e.g.:
\begin{equation*}
g'_1(v)=\frac{g_1(v)+g_2(v)}{v},\qquad g'_2(v)=\frac{2g_2(v)+g_3(v)}{v}.
\end{equation*}
In particular, we have $g_1(v)\equiv(\ln2)^{-1}$ at zeroth-order approximation and 
\begin{equation}
g_1(v)=\frac1{\ln2}\left[1+\frac1{12v^2}\right],
\label{g1v1stOrdApp}
\end{equation}
at first-order approximation. 

Notice, that by using \eqref{zeroth} we have, at the lowest order,
\begin{equation}
\chi_n^{(\mathrm{log})}=\max_{V_\mathrm{in},V_\mathrm{mod}}
\sum_{k=1}^{n}\log_2\frac{\overline\nu_{k}}{\nu_{k}}.
\label{HolDef0app}
\end{equation}
The value $\underline{C}_n$ calculated through approximation~\eqref{HolDef0app}
below will be denoted as $\underline{C}_n^{(\mathrm{log})}$ and called
\emph{logarithmic approximation to capacity}. In turn, the quantity
$\underline{C}_n$ will be called \emph{zeroth-order approximation to capacity}
and denoted by $\underline{C}_n^{(0)}$, if actual maximum over $V_\mathrm{in}$
and $V_\mathrm{mod}$ is not taken in Eq.~\eqref{HolDef}, but symplectic
eigenvalues $\nu_{k}$ and $\overline\nu_{k}$ are chosen instead to be those at
which the maximum in~\eqref{HolDef0app} is achieved.
Thus, $\underline{C}_n^{(0)}$ is given by substitution of the approximate
symplectic eigenvalues into the exact relation for Holevo-$\chi$ quantity.

\subsection{Examples of Gaussian channels}

There are two types of noises that are mostly relevant for
experimental setups: \emph{attenuation} and \emph{addition of classical noise}.
The so-called \emph{lossy (bosonic) channels} describe the attenuation, while
the \emph{additive (classical) noise channels} take into account only the
addition of classical noise. For a discussion of the capacity of the other 
classes of Gaussian channels, in the single-mode case, see \cite{PST}.

The lossy channels play a prominent role and below we will focus our attention
to them.  They are characterized by the map~\eqref{Gmap} with the matrices
\begin{equation}
\label{XYloss}
X= \sqrt{\eta}\,\mathbb{I}_{2n},\quad Y=(1-\eta)V_{\rm env}.
\end{equation}
Here $V_\mathrm{env}$ denotes the $2n\times 2n$ covariance matrix of the channel
environment that `contaminates' the input signal, which is attenuated by the
the channel's \emph{transmissivity} $\eta\in[0,1]$.  In~particular the lossy
bosonic channel acts as a rotation (beam splitter) on the canonical
quadratures and gives rise to the following relation among the covariance
matrices~\cite{EisertWolf}:
\begin{align}
&V_\mathrm{out}\:=\eta\,V_\mathrm{in}+(1-\eta)\,V_\mathrm{env},
\label{VoutInit}\\
&\overline V_\mathrm{out}=
\eta\,(V_\mathrm{in}+V_\mathrm{mod})+(1-\eta)\,V_\mathrm{env}.
\label{VoutAv}
\end{align} 
In fact, these transformations follow from the definitions~\eqref{Gmap},
\eqref{VoutAndAvGenForm} and~\eqref{XYloss}.  Below (see Eq.~\eqref{dCdNGen}) it will be
shown that the capacity is a monotonically increasing function of the average number of
input photons per mode (channel use) $N$, therefore we shall constrain the input
energy using the equality in \eqref{constr2}, \emph{i.e.}
\begin{equation}
\frac1{2n}\Tr({V_\mathrm{in}+V_\mathrm{mod}})=N+\frac12.
\label{enrestr}
\end{equation}

The additive noise channels are described by similar
transformations~\cite{EisertWolf}
\begin{align}
&V_\mathrm{out}\:=\,V_\mathrm{in}+\,V_\mathrm{env},
\label{VoutInitAN}\\
&\overline V_\mathrm{out}=\,V_\mathrm{in}+V_\mathrm{mod}+\,V_\mathrm{env}
\label{VoutAvAN}
\end{align}
following from Eq.~\eqref{Gmap} if $X=\mathbb I_{2n}$ and $Y=V_\mathrm{env}$,
where $V_\mathrm{mod}$ and $V_\mathrm{env}$ correspond to classical
distribution, while $V_\mathrm{in}$ should satisfy the uncertainty relation.
Notice, that similarity between Eqs.~\eqref{VoutInitAN}, \eqref{VoutAvAN}
and~\eqref{VoutInit}, \eqref{VoutAv} makes the extension of the method we are
going to develop to the additive noise channel straightforward.  In particular,
a similar approach has been recently used in Ref.~\cite{NewNoiseChannel}.

\subsection{Heterodyne and homodyne rates}\label{TransRates}

As far as the general optimization approach to find the Holevo function~\eqref{HolDef}
is also applicable to information transmission rates, we are going to consider
these as well and compare them with the capacity.

Suppose, that the matrices $V_\mathrm{in}$, $V_\mathrm{mod}$ and
$V_\mathrm{env}$ are block diagonal, \emph{i.e.} can be written in the form
\begin{equation}
V_\mathrm{ind}=\left(
\begin{array}{cc}
V_{\mathrm{ind},qq} & \mathbb{O}_n  \\
\mathbb{O}_n & V_{\mathrm{ind},pp}
\end{array}
\right),
\label{ind00}
\end{equation}
where $\mathbb{O}_n$ was defined by Eq.~\eqref{symplecticform} and ``ind'' may
stand for ``in'', ``mod'' or ``env''. Moreover, let us assume that their
diagonal blocks mutually commute (including blocks taken from different
matrices).  In such a case by considering the average information accessible by
performing \emph{heterodyne} measurement on each single channel output (joint
measurement of $q$ and $p$ quadratures) one can get the \emph{heterodyne
rate}~\cite{PRA}
\begin{align}
R^{(\rm het)}&:=\lim_{n\to\infty}R_n^{(\rm het)},\\
R_n^{(\rm het)}&=\frac1{2n}\max_{V_{\rm in},V_{\rm mod}}\log_2\det
\nonumber\\
&\qquad\qquad\Biggl[\Biggl(\overline V_{\rm out}+\frac{\mathbb{I}_{2n}}2\Biggr)
\Biggl(V_{\rm out}+\frac{\mathbb{I}_{2n}}2\Biggr)^{-1}\Biggr].
\label{HetR}
\end{align}
Analogously, by considering \emph{homodyne} measurement on each single channel
output (measurement of $u_\star$ quadrature, where $u_\star$ is a placeholder
for $q$ and $p$) one can find the \emph{homodyne rate}~\cite{PRA}
\begin{align}
&R^{(\rm hom)}:=\lim_{n\to\infty}R_n^{(\rm hom)},\nonumber\\
&R_n^{(\rm hom)}=\frac1{2n}\max_{V_{\rm in},V_{\rm mod}}
\log_2\det\Bigl[\overline V_{{\rm out},u_\star u_\star}
V_{{\rm out},u_\star u_\star}^{-1}\Bigr],
\label{HomR}
\end{align}
where notations of matrix blocks are the same as in Eq.~\eqref{ind00}.

\section{Single channel use}\label{MLChannel}

Let us consider single use (single mode) of the lossy bosonic channel.  Its
description requires the consideration of $2\times2$ covariance matrices of the
general form to solve the optimization problem. However, all the properties can
be found by taking all involved matrices in the diagonal form. This can be done
thanks to the following \emph{input purity theorems}:

\bigskip

\begin{theorem}\label{InpPurTheor}
For the single use of the lossy bosonic channel, the $2\times2$
matrices $V_\mathrm{in}$ and $V_\mathrm{mod}$ at which the maximum of the Holevo
function over Gaussian states is achieved, are simultaneously diagonalizable
together with $V_\mathrm{env}$.  Moreover, the optimal matrix $V_\mathrm{in}$
corresponds to pure state.
\end{theorem}

\bigskip

\begin{IEEEproof}
The proof\footnote{In the generic setting, the optimality of pure input states
has been proven in~\cite{Sch97}. However, in our case the Holevo function
has to be optimized under the constraint of Gaussian input states
and energy restriction. For these reasons, it is worth proving 
this property explicitly for the considered setting.} is reported in Appendix~\ref{InPurTh}.
\end{IEEEproof}

\bigskip

\begin{theorem}\label{InpPurTheor2}
Let us consider the single use of the lossy bosonic channel
characterized by $2\times2$ diagonal covariance matrices $V_{\rm env}$, $V_{\rm
in}$, and $V_{\rm mod}$, then the maxima for both heterodyne and homodyne rates
are provided by pure input states\footnote{Notice, that extension of
theorem~\ref{InpPurTheor2} to the case of Holevo function is straightforward,
being it a particular case of theorem~\ref{InpPurTheor}.}.
\end{theorem}

\bigskip

\begin{IEEEproof}
The proof is reported in Appendix~\ref{InPurTh2}.
\end{IEEEproof}

\bigskip

Let us discuss these theorems in the context of rates.  Remember, that in the
case of $2\times2$-matrices the assumptions used to derive general
relations~\eqref{HetR} and~\eqref{HomR} are equivalent to diagonality of all involved
matrices, therefore optimality of pure input states for rates is
guaranteed by theorem~\ref{InpPurTheor2}.  Moreover, if one conjectures that the
relations~\eqref{HetR} and~\eqref{HomR} hold also for $2\times2$ matrices of
general form (\emph{i.e.} non-diagonal), then commutativity of matrices together
with input purity are guaranteed by theorem~\ref{InpPurTheor}, whose extension
to the case of rates is straightforward.

Thus, below it is always assumed without loss of generality that all the
matrices are already diagonalized and the input state is pure. Furthermore,
unless otherwise stated, in the following it is assumed\footnote{This is done
because the limit case $\eta=1$ (noiseless channel) is considered separately in
Subsec.~\ref{NoiseLessChannel} and the limit case $\eta=0$ (infinitely noisy
channel) is trivial giving zero capacity and rates.} that $0<\eta<1$.

\subsection{System of notations}\label{SystemOfNotations}

Let us introduce the system of notations that will be used hereafter.  Any
single-mode state labeled by index ``ind'' will be referred to by its
quadratures covariance matrix $V_\mathrm{ind}$ parametrized by
$\EuScript{N}_\mathrm{ind}\in\mathbb{R}_+$ and $s_\mathrm{ind}\in\mathbb{R}$ as
\begin{equation}
V_\mathrm{ind}:=
V(\EuScript{N}_\mathrm{ind},s_\mathrm{ind})=
\left[\EuScript{N}_\mathrm{ind}+\frac12\right]
\left(
\begin{array}{cc}
e^{s_\mathrm{ind}}&0\\
0&e^{-s_\mathrm{ind}}
\end{array}
\right).
\label{ind1mode}
\end{equation}
In particular, ``ind'' may stand for ``in'', ``mod'', ``env'' or ``out'' for the
cases of input, modulation, environment or output covariance matrices,
respectively.  The quantity $s_\mathrm{ind}$ will be referred to as
\emph{squeezing} in ``ind''.  The quantity $\EuScript{N}_\mathrm{ind}$ will
always be written in ``EuScript'' font and called \emph{average amount of
thermal photons} in the state ``ind''.  We also define the \emph{average amount
of photons} $N_\mathrm{ind}$ in the state ``ind'' as
\begin{equation}
\frac12\Tr(V_\mathrm{ind})=N_\mathrm{ind}+\frac12,
\label{indNsm}
\end{equation}
which is equivalent to the relation
\begin{equation}
N_\mathrm{ind}=
\left(\EuScript{N}_\mathrm{ind}+\frac12\right)\cosh s_\mathrm{ind}-\frac12.
\label{NindDef}
\end{equation}
Below we will usually omit the word ``average'' refering to the quantities
$\EuScript{N}_\mathrm{ind}$ and $N_\mathrm{ind}$.  All the quantities related
with some overlined matrix will be also overlined, \emph{i.e.}
$\overline{V}_\mathrm{ind}$ equals
$V(\overline{\EuScript{N}}_\mathrm{ind},\overline{s}_\mathrm{ind})$ and has
amount of photons $\overline{N}_\mathrm{ind}$. In order to indicate that some
channel parameters are related with homodyne or heterodyne rates (they are
defined below in Subsec.\ref{HetVars}) the upper indices ``(hom)'' and ``(het)''
will be used.  The only exceptions from the above rules are: the index ``env''
will be omitted for quantities which represent the squeezing (or its particular
values) in channel environment, e.g. $s\equiv s_\mathrm{env}$; the index ``in''
and overlining will be omitted for the quantities which represent the average
amount of photons (or its particular values, e.g. thresholds) in averaged input
state $\overline{V}_\mathrm{in}$ (see Eq.~\eqref{VinAv}) and its ``heterodyne
analog'' $\overline{V}_\mathrm{in}^{\mathrm{(het)}}$, e.g. $N\equiv
\overline{N}_\mathrm{in}\equiv\overline{N}_\mathrm{in}^\mathrm{(het)}$.

Notice, that the state $V(\EuScript{N}_\mathrm{ind},s_\mathrm{ind})$ is
\emph{pure} if $\EuScript{N}_\mathrm{ind}=0$ and \emph{mixed} otherwise, is
\emph{squeezed} if $s_\mathrm{ind}\neq0$, is \emph{thermal} if
$\EuScript{N}_\mathrm{ind}\neq0$ and $s_\mathrm{ind}=0$, is \emph{thermal
squeezed} if both $\EuScript{N}_\mathrm{ind}\neq0$ and $s_\mathrm{ind}\neq0$,
and is \emph{vacuum} if both $\EuScript{N}_\mathrm{ind}=0$ and
$s_\mathrm{ind}=0$.

The eigenvalues of each matrix will be denoted by the first character of matrix
index. Then, the eigenvalue which is the first diagonal element corresponds to
quadrature $q$, therefore it will be labeled by index $q$ (analogously, by $p$
for the second diagonal element).  However, as far as both quadratures enter all
the relations in the same way, instead of specifying the quadrature $q$ or $p$
usually we will use index $u$ as a placeholder for $q$ or $p$. Also, we will use
the rule: if $u=q$, then $u_\star=p$, and vice versa.  In particular, we will
refer to the eigenvalues of matrices $V_\mathrm{in}$,
$V_\mathrm{mod}$, $V_\mathrm{env}$, $V_\mathrm{out}$ and $\overline
V_\mathrm{out}$ as $i_{u}$, $m_{u}$, $e_{u}$, $o_{u}$ and $\overline{o}_{u}$,
respectively. For instance, we have $V_\mathrm{env}=\diag(e_q,e_p)$ for the
environment matrix.  Also, without loss of generality, below it is always
assumed that if environment eigenvalues are non-equal, then $e_u>e_{u_\star}$.

As far as only the single-mode case is discussed in this section, index $k$ will
be omitted for symplectic eigenvalues $\nu_k$ and $\overline\nu_k$ (they were
introduced in Eq.~\eqref{HolDef}). Also, index $n$ will be omitted for $\chi$-
and $\underline C$- and $R$-quantities (e.g., see Eqs.~\eqref{HGaussian},
\eqref{HolDef}, \eqref{HolDef0app}, \eqref{HetR} and~\eqref{HomR}).  To simplify
the notations, in what follows we allow each of these quantities to stand either for the
result of the maximization or for the function to maximize, depending on context.

Taking into account that the symplectic eigenvalue for $2\times2$-matrix $V$ is
$\sqrt{\det V}$, we have for the matrices $V_\mathrm{out}$ and $\overline
V_\mathrm{out}$ the relations
\begin{equation}
\nu=\sqrt{o_{u}o_{u_\star}},\qquad\qquad
\overline\nu=\sqrt{\overline o_{u}\overline o_{u_\star}},
\label{Nusesnunuov}
\end{equation}
where
\begin{equation}
\begin{split}
&o_{u}=\eta\,i_{u} + (1-\eta)\,e_{u},\\
&\overline o_{u}=\eta\,(i_{u}+m_{u})+ (1-\eta)\,e_{u},\\
&o_{u_\star}=\eta\,i_{u_\star} + (1-\eta)\,e_{u_\star},\\
&\overline o_{u_\star}=\eta\,(i_{u_\star}+m_{u_\star})+ (1-\eta)\,e_{u_\star}.
\end{split}
\label{eigs_defs}
\end{equation}

\subsection{Heterodyne variables}\label{HetVars}

In the following it will be convenient to introduce the 
\emph{heterodyne environment matrix}
\begin{equation*} 
V_\mathrm{env}^{(\mathrm{het})} := V_\mathrm{env}+\frac{\mathbb{I}_2}{2\,(1-\eta)},
\end{equation*} 
whose eigenvalues are
\begin{equation}
e_{u}^\mathrm{(het)}=e_{u}^{\phantom a}+\frac1{2\,(1-\eta)},\qquad
e_{u_\star}^\mathrm{(het)}=e_{u_\star}^{\phantom a}+\frac1{2\,(1-\eta)}.
\label{EuHets}
\end{equation}
Replacing $V_\mathrm{env}$ by $V_\mathrm{env}^{(\mathrm{het})}$ in the
relations~\eqref{VoutInit} and~\eqref{VoutAv}, one can also define the
``heterodyne version'' of the other matrices:
\begin{align}
&V_\mathrm{out}^{\mathrm{(het)}}\:=
\eta\,V_\mathrm{in}^{\mathrm{(het)}}+(1-\eta)\,V_\mathrm{env}^{\mathrm{(het)}},
\label{VoutInitHet}\\
&\overline V_\mathrm{out}^{\mathrm{(het)}}:=
\eta\left(V_\mathrm{in}^{\mathrm{(het)}}+V_\mathrm{mod}^{\mathrm{(het)}}\right)+
(1-\eta)\,V_\mathrm{env}^{\mathrm{(het)}},
\label{VoutAvHet}
\end{align}
where the eigenvalues of matrices $V_\mathrm{out}$, $\overline V_\mathrm{out}$,
$V_\mathrm{out}^{\mathrm{(het)}}$ and $\overline
V_\mathrm{out}^{\mathrm{(het)}}$ are related as follows:
\begin{equation*}
\begin{split}
o_u^\mathrm{(het)}=o_u^{\phantom a}+\frac12,\qquad\qquad
o_{u_\star}^\mathrm{(het)}=
o_{u_\star}^{\phantom a}+\frac12,\\
\overline o_u^\mathrm{(het)}=
\overline o_u^{\phantom a}+\frac12,\qquad\qquad
\overline o_{u_\star}^\mathrm{(het)}=
\overline o_{u_\star}^{\phantom a}+\frac12.
\end{split}
\end{equation*}
Then, one can define symplectic eigenvalues in the heterodyne setting (similarly to
Eqs.~\eqref{Nusesnunuov}) by the relations
\begin{equation}
\nu^{(\rm het)}=\sqrt{o_{u}^{\mathrm{(het)}}o_{u_\star}^{\mathrm{(het)}}},\qquad
\overline\nu^{(\rm het)}=
\sqrt{\overline o_{u}^{\mathrm{(het)}}\overline o_{u_\star}^{\mathrm{(het)}}}.
\label{sympleigsHet}
\end{equation}

The average amount of photons $N_\mathrm{env}^{(\rm het)}$ in the heterodyne
environment 
\begin{equation*}
V_\mathrm{env}^{(\mathrm{het})}=
V\left(\EuScript{N}_\mathrm{env}^{(\rm het)},s^{(\rm het)}\right)
\end{equation*}
(see Eq.~\eqref{ind1mode}) can be introduced using the standard
relation~\eqref{indNsm}.  The parameters of the environment matrices
$V_\mathrm{env}=V(\EuScript{N}_\mathrm{env},s)$ and $V_\mathrm{env}^{(\rm het)}$
are related by
\begin{equation}
\left[\EuScript{N}_{\rm env}^{(\rm het)}+\frac12\right]^2=
\left[\EuScript{N}_{\rm env}+\frac12+\frac1{2\,(1-\eta)}\right]^2+
\frac{N_{\rm env}-\EuScript{N}_{\rm env}}{1-\eta},
\label{NenvHet}
\end{equation}
\begin{equation}
s^\mathrm{(het)}=\;
\frac12\ln\frac{1+(1-\eta)(2\,\EuScript{N}_\mathrm{env}+1)\,e^s\phantom{a}}
{1+(1-\eta)(2\,\EuScript{N}_\mathrm{env}+1)\,e^{-s}},\phantom{1}
\label{sHet}
\end{equation}
\begin{equation}
s=\frac12\ln
\frac{1-(1-\eta)\left(2\,\EuScript{N}_\mathrm{env}^\mathrm{(het)}+1\right)\,
e^{s^\mathrm{(het)}}\phantom{a}}
{1-(1-\eta)\left(2\,\EuScript{N}_\mathrm{env}^\mathrm{(het)}+1\right)
e^{-s^\mathrm{(het)}}},
\label{sHet2s}
\end{equation}
\begin{equation}
N_\mathrm{env}^\mathrm{(het)}=N_\mathrm{env}+\frac1{2\,(1-\eta)}.
\label{MenvHet}
\end{equation}
In particular, for thermal environment $V(\EuScript N_\mathrm{env},0)$ we have
$N_\mathrm{env}^\mathrm{(het)}=\EuScript N_\mathrm{env}^\mathrm{(het)}$ and
\begin{equation*}
\EuScript N_\mathrm{env}^\mathrm{(het)}=\EuScript N_\mathrm{env}+\frac1{2\,(1-\eta)}.
\end{equation*}
Notice, that the heterodyne environment $V_\mathrm{env}^{(\rm het)}$ is squeezed if
and only if $V_\mathrm{env}$ is squeezed.

The quantities with upper index ``(het)'' defined in this subsection will allow
us to simplify the relations for the heterodyne rate. We shall refer to them as
\emph{heterodyne variables}. The latter, which are eigenvalues will be also
called \emph{heterodyne eigenvalues} to distinguish them from \emph{standard}
eigenvalues (of $V_\mathrm{in}$, $V_\mathrm{env}$, $V_\mathrm{out}$, etc.).

\subsection{Heterodyne and homodyne rates}\label{TransRates1D}

Let us consider the homodyne rate~\eqref{HomR}.  It corresponds to a measurement
of the $u_\star$-quadrature, which is the less noisy according to the convention
$e_u>e_{u_\star}$ (obviously, there is no difference in the choice of quadrature
if $e_u=e_{u_\star}$).  Such a choice gives higher rate in comparison with the
measurement of $u$-quadrature.  In what follows (see Subsec.~\ref{TheSolStages}), 
it will be shown that this case corresponds to eigenvalue $m_u=0$ be
optimal for homodyne rate.  In explicit form it is
\begin{equation}
R^\mathrm{(hom)}=
\frac12\log_2\frac{\overline o_{u_\star}}{o_{u_\star}},
\label{Hom1D}
\end{equation}
which coincides with $\log_2\left(\overline\nu/\nu\right)$ if $m_u=0$. This
property gives rise to the relation (see Eqs.~\eqref{HolDef0app}
and~\eqref{Nusesnunuov})
\begin{equation} 
R^\mathrm{(hom)}=\underline{C}^{(\mathrm{log})}
\label{HomLim}
\end{equation} 
if optimal $m_u$ is also zero for logarithmic approximation to
capacity\footnote{It will be shown in Subsec.~\ref{TheThirdStage}, that if
$m_u=0$ is optimal for one of the quantities $\underline C$ and $\underline
C^{(\log)}$, then it is optimal also for the other. Then, remember (see
Subsec.~\ref{EstimClassCapacity}) that optimal eigenvalues for $\underline
C^{(\log)}$ and $\underline C^{(0)}$ are always the same by definition.} (we
equalize the quantities $R^\mathrm{(hom)}$ and $\underline{C}^{(\mathrm{log})}$
for the same channel parameters).  This holds true for small values of $N$ (see
Eq.~\eqref{Nthr23} below).  Thus, in this case the homodyne rate coincides with
the logarithmic approximation to capacity.

Analogously, Eq.~\eqref{HetR} gives
\begin{equation} 
R^\mathrm{(het)}=
\log_2\frac{\overline\nu^{(\rm het)}}{\nu^{(\rm het)}},
\label{Het1D}
\end{equation} 
for heterodyne measurement (see Eqs.~\eqref{sympleigsHet}), \emph{i.e.} the
heterodyne rate is equal to the logarithmic approximation to capacity calculated
with $V_\mathrm{env}^{(\rm het)}$.  One can also get for a fixed~$s$ that 
\begin{equation}
\lim_{\EuScript{N}_\mathrm{env}\to\infty}R^{(\rm het)}=\underline{C}^{(\log)}.
\label{HetLim}
\end{equation} 
Eqs.~\eqref{HomLim} and~\eqref{HetLim} define the values of parameters $s$ and
$\EuScript{N}_\mathrm{env}$ for which the rates approach the capacity (the
comparison between capacity, homodyne and heterodyne rates was discussed earlier
in~\cite{GG2}).

The simple form of~\eqref{Het1D} explains why the description of heterodyne rate
using heterodyne variables introduced in Subsec.~\ref{HetVars} is the most
natural one.  Keep in mind, that the heterodyne rate can be described using both
approaches: as standard variables used for capacity and homodyne rate, or as
heterodyne variables. Despite we shall usually work with heterodyne variables,
sometimes standard variables will be used.

As far as quantities~\eqref{HetR} and~\eqref{Het1D} are identical as functions
of input and modulation eigenvalues, the latter do not depend on representation
(type of variables) used for $R^\mathrm{(het)}$. It means that upper indices
``(het)'' written for input and modulation eigenvalues are used only to indicate
that they are optimal for heterodyne rate (to distinguish from those optimal for
capacity and homodyne rate). However, indices ``(het)'' written for environment,
output and average output eigenvalues indicate both different variables used and
optimality for heterodyne rate. Loosely speaking, $i_u^\mathrm{(het)}=i_u$ and
$m_u^\mathrm{(het)}=m_u$, while $e_u^\mathrm{(het)}\neq e_u$,
$o_u^\mathrm{(het)}\neq o_u$, $\overline o_u^\mathrm{(het)}\neq\overline o_u$
($u\in\{q,p\}$) \emph{as abstract variables}, but in our convention \emph{all}
of them are different, because $i_u^\mathrm{(het)}$ and $m_u^\mathrm{(het)}$ are
used \emph{only} for heterodyne case and are optimal for it\footnote{Writing,
e.g.  $i_u^\mathrm{(het)}=i_u$ would be misleading, as $i_u$ are those
eigenvalues optimal for capacity, but not optimal for heterodyne rate. This is
less problematic for homodyne rate, because its optimal eigenvalues coincide in
some cases with that of the logarithmic approximation to capacity (see
Eq.~\eqref{HomLim}) and therefore can be treated as a particular case of
eigenvalues optimal for capacity.}.

Below, we shall usually write the relations for capacity and then explain which
replacements should be applied to get analogous relations for rates.  These
replacements can be some of the following:
\begin{align}
&i_u^{\phantom{1}}\to i_u^\mathrm{(het)},&
&i_{u_\star}^{\phantom{1}}\to i_{u_\star}^\mathrm{(het)},
\label{Repli}\\ 
&m_u^{\phantom{1}}\to m_u^\mathrm{(het)},&
&m_{u_\star}^{\phantom{1}}\to m_{u_\star}^\mathrm{(het)},
\label{Replm}\\ 
&e_u^{\phantom{1}}\to e_u^{\mathrm{(het)}},&
&e_{u_\star}^{\phantom{1}}\to e_{u_\star}^{\mathrm{(het)}},
\label{Reple}\\
&o_u^{\phantom{1}}\to o_u^{\mathrm{(het)}},&
&o_{u_\star}^{\phantom{1}}\to o_{u_\star}^{\mathrm{(het)}},
\label{Replo}\\ 
&\overline o_u^{\phantom{1}}\to\overline o_u^{\mathrm{(het)}},&
&\overline o_{u_\star}^{\phantom{1}}\to\overline o_{u_\star}^{\mathrm{(het)}},
\label{Replovo}\\
&N_\mathrm{env}^{\phantom{1}}\to N_\mathrm{env}^{\mathrm{(het)}},&
&\EuScript N_\mathrm{env}^{\phantom{1}}\to\EuScript N_\mathrm{env}^{\mathrm{(het)}},
\label{ReplN}
\end{align}
\begin{align}
&\nu\to\nu^{\mathrm{(het)}},&
&\overline\nu\to\overline\nu^{\mathrm{(het)}},
\label{Replnu}\\ 
&g_1\to\frac1{\ln2},&
&g_2\to-\frac1{\ln2}.
\label{Replg}
\end{align}
Each of the above numbered lines specifies two replacements.  However, only
those replacements, which correspond to \emph{explicit} variables of the
relation (subjected to replacements) must be applied.  Finally, when discussing
about the rates, if we refer to relations written for the capacity, we should first
apply the proper replacements.

\subsection{Optimization problem}\label{OptProbl}

The optimization problem for the heterodyne rate can be formulated as follows.
One needs to find the matrices $V_\mathrm{in}^{\mathrm{(het)}}$ and
$V_\mathrm{mod}^{\mathrm{(het)}}$ (see Eqs.~\eqref{VoutInitHet}
and~\eqref{VoutAvHet}), which provide the maximum for the function~\eqref{Het1D} and
satisfy the energy constraint
\begin{equation}
\frac12\Tr\left(\overline V_\mathrm{in}^{\mathrm{(het)}}\right)=N+\frac12,
\label{enconstrHet}
\end{equation}
where 
\begin{equation*}
\overline V_\mathrm{in}^{\mathrm{(het)}}=
V_\mathrm{in}^{\mathrm{(het)}}+V_\mathrm{mod}^{\mathrm{(het)}}.
\end{equation*}
By substituting Eq.~\eqref{VoutInitHet} into Eq.~\eqref{indNsm} written for
$\overline V_\mathrm{out}^{\mathrm{(het)}}$ and taking into account the energy
constraint~\eqref{enconstrHet}, we get the amount of photons in the average output state
\begin{equation}
\overline N_\mathrm{out}^{(\mathrm{het})}=
\eta N+(1-\eta)N_\mathrm{env}^{(\mathrm{het})}.
\label{Nout1D}
\end{equation}
Analogously, for the case of capacity the relations~\eqref{VoutAv}
and~\eqref{enrestr} give
\begin{equation}
\overline N_\mathrm{out}=\eta N+(1-\eta)N_\mathrm{env}.
\label{Nout1Dcap}
\end{equation}

Notice, that theorems~\ref{InpPurTheor} and~\ref{InpPurTheor2} allow us to
exclude the variables $i_{u_\star}$ and $i_{u_\star}^\mathrm{(het)}$ from the
optimization problems due to the purity of the input states: 
\begin{equation*}
i_ui_{u_\star}=
\frac14,\qquad\qquad i_u^\mathrm{(het)}i_{u_\star}^\mathrm{(het)}=\frac14.
\end{equation*}
Then, the optimization problems for the single-mode channel
can be formulated as follows.  One needs to find the maxima of functions (see
definitions~\eqref{HGaussian}, \eqref{HolDef}, \eqref{Hom1D} and~\eqref{Het1D})
\begin{align}
&\underline{C}=g\left(\overline\nu-\frac12\right)-g\left(\nu-\frac12\right),
\label{defcap}\\
&R^{(\mathrm{het})}=
\log_2\overline\nu^\mathrm{(het)}-
\log_2\nu^\mathrm{(het)},
\label{defhet}
\end{align}
\begin{equation}
R^{(\mathrm{hom})}=\frac12\bigl[\log_2\overline o_{u_\star}-\log_2o_{u_\star}\bigr],
\label{defhom}
\end{equation}
over the variables $i_u$, $m_u$, $m_{u_\star}$ in the case of $\underline{C}$
and $R^{(\mathrm{hom})}$, and over the variables $i_u^\mathrm{(het)}$,
$m_u^\mathrm{(het)}$, $m_{u_\star}^\mathrm{(het)}$ in the case of
$R^{(\mathrm{het})}$, taking into account the constraints
\begin{align}
&i_u>0,
\label{ium0}\\
&m_u,m_{u_\star}\geqslant0,\\
&i_u+\frac1{4i_u}+m_u+m_{u_\star}=2N+1,
\label{1useConstr}
\end{align}
in the case of $\underline{C}$ and $R^{(\mathrm{hom})}$, and the
constraints~\eqref{ium0}-\eqref{1useConstr} after the replacements~\eqref{Repli}
and~\eqref{Replm} in the case of $R^{(\mathrm{het})}$.  
In Subsec.~\ref{TheThirdStage} and~\ref{TheSecondStage} we shall solve it using
Lagrange multipliers method.

It is interesting to note that the relations for symplectic
eigenvalues~\eqref{Nusesnunuov} and~\eqref{sympleigsHet} allow the
capacity~\eqref{defcap} and heterodyne rate~\eqref{defhet} to be represented as
\begin{align}
&\underline{C}=g\left(\overline{\EuScript{N}}_\mathrm{out}\right)-
g\left(\EuScript{N}_\mathrm{out}\right),
\label{defcapNout}\\
&R^{(\mathrm{het})}=
\log_2\left(\overline{\EuScript{N}}_\mathrm{out}^\mathrm{(het)}+\frac12\right)-
\log_2\left(\EuScript{N}_\mathrm{out}^\mathrm{(het)}+\frac12\right),
\label{defhetNout}
\end{align}
where $\overline{\EuScript{N}}_\mathrm{out}$, $\EuScript{N}_\mathrm{out}$,
$\overline{\EuScript{N}}_\mathrm{out}^\mathrm{(het)}$ and
$\EuScript{N}_\mathrm{out}^\mathrm{(het)}$ are the amounts of thermal photons for
the states $\overline V_\mathrm{out}$, $V_\mathrm{out}$, $\overline
V_\mathrm{out}^\mathrm{(het)}$ and $V_\mathrm{out}^\mathrm{(het)}$,
respectively.

\subsection{The solution stages}\label{TheSolStages}

Let us consider the capacity and the homodyne rate.  In Subsec.~\ref{TheThirdStage}
and~\ref{TheSecondStage} it will be shown that all the solutions of Lagrange
equations, associated to the optimization problem stated in
Subsec.~\ref{OptProbl}, give positive~$i_u$, that is $m_u,m_{u_\star}\geqslant0$
are the only inequalities to satisfy. This allows us to classify the solutions
depending on the amount of positive optimal $m$-eigenvalues.  The following
terminology is used for this purpose.

\bigskip

\begin{definition}\label{stagesdef}
The solution belongs to the \emph{first stage} if the optimal $m_u$,
$m_{u_\star}$ are both equal to zero, to the \emph{second stage} if the optimal
$m_u$, $m_{u_\star}$ are one equal to zero and the other is positive, and to the
\emph{third stage} if the optimal $m_u$, $m_{u_\star}$ are both positive.
\end{definition}

\bigskip

As far as $R^{(\mathrm{hom})}$ does not depend on $m_u$, due to the
condition~\eqref{1useConstr} the maximum is achieved for $m_u=0$, which shows the
absence of the third stage in homodyne rate.  In other words, energy $N$ should
not be wasted in the quadrature unused for information transmission.

The first stage holds if and only if capacity is equal to zero, which can only
be if $N=0$ (if $N\neq0$ one can always get non-zero capacity and rates by
taking $i_u=\frac12$, $m_u=m_{u_\star}=N$).  In particular, Eq.~\eqref{1useConstr}
applied for the first stage gives $i_u=\frac12$.  The same consideration holds true
also for the homodyne rate.

\bigskip

\begin{propos}\label{EuNeqEustGen}
Given \mbox{$\overline o_u\neq\overline o_{u_\star}$}
in the second stage, the eigenvalues \mbox{$m_u=0$} and \mbox{$m_{u_\star}>0$} are
optimal for capacity\footnote{This proposition holds for both cases
$e_u>e_{u_\star}$ and $e_u<e_{u_\star}$.} if and only if \mbox{$\overline
o_u>\overline o_{u_\star}$}.
\end{propos}

\bigskip

\begin{IEEEproof}
Suppose that $m_{u_\star}=0$ and $m_u>0$ are optimal in the case of
\mbox{$\overline o_u>\overline o_{u_\star}$}.  The energy
constraint~\eqref{1useConstr} is preserved by the change of variables
$m_u'=m'_{u_\star}=m_u/2$, $i'_u=i_u$. The new variables do not change the
second term in Eq.~\eqref{defcap} but increase the first term\footnote{The area
of a rectangle with fixed perimeter is higher if the lenth of sides differs
less. In the considered case $\overline o'_u+\overline o'_{u_\star}= \overline
o_u+\overline o_{u_\star}$ but $|\overline o'_u-\overline
o'_{u_\star}|<|\overline o_u-\overline o_{u_\star}|$.  In addition, $g$ is
monotonically increasing and concave function.}.  Thus, they give higher maximum
for capacity.  Similarly, one can prove that $m_u=0$ and
\mbox{$m_{u_\star}>0$} are not optimal if \mbox{$\overline o_u<\overline
o_{u_\star}$}.  Hence, the proposition is proved by contradiction. 
\end{IEEEproof}

\bigskip

\begin{propos}\label{EuNeqEust}
If $e_u>e_{u_\star}$, then in the second stage \mbox{$m_u=0$} and
$m_{u_\star}>0$ are optimal for capacity.
\end{propos}

\bigskip

\begin{IEEEproof}
The proof is reported in Appendix~\ref{EuNeqEustProof}.
\end{IEEEproof}

\bigskip

It follows from propositions~\ref{EuNeqEustGen} and~\ref{EuNeqEust} that
the case of \mbox{$e_u>e_{u_\star}$} requires
$o_u=\overline o_u$ and
\begin{equation}
o_u\geqslant\overline o_{u_\star}>o_{u_\star}
\label{ineqs2ndStage}
\end{equation}
in the second stage.

Similar consideration gives $\overline o_u=\overline o_{u_\star}$ in the third
stage (by supposing $\overline o_u\neq \overline o_{u_\star}$ one can always
redistribute the energy $N$ among $m$-eigenvalues so to decrease the difference
$|\overline o_u-\overline o_{u_\star}|$ thus giving higher maximum for
capacity). Taking into account Eq.~\eqref{1useConstr}, we get in this case
\begin{equation}
\overline o_u=\overline o_{u_\star}=
\eta\left(N+\frac12\right)+(1-\eta)\left(N_\mathrm{env}+\frac12\right).
\label{auequation}
\end{equation}
The equality $\overline o_u=\overline o_{u_\star}$ is equivalent to the equation
\mbox{$\overline{\EuScript N}_\mathrm{out}=\overline N_\mathrm{out}$}, where the
latter is given by Eq.~\eqref{Nout1Dcap}.  Thus, for the third stage, the first
term in the relation~\eqref{defcapNout} is already found.

Notice, that the above considerations for the capacity (including
definition~\ref{stagesdef}, propositions~\ref{EuNeqEustGen}
and~\ref{EuNeqEust}, Eqs.~\eqref{ineqs2ndStage} and~\eqref{auequation}) hold
also for the heterodyne rate if the replacements \eqref{Repli}--\eqref{ReplN}
and $g\to\log_2$ are applied, and if Eqs.~\eqref{Nout1D}, \eqref{defhet}
and~\eqref{defhetNout} are mentioned instead of Eqs.~\eqref{Nout1Dcap},
\eqref{defcap} and \eqref{defcapNout}, respectively.  Below the solutions for
the third and the second stages are presented.

\subsection{The third stage}\label{TheThirdStage}

In the case of the third stage, the Lagrange multipliers method applied to the
function $\underline{C}$ with the constraint~\eqref{1useConstr} leads to the
following system of equations (see definition of $g_k$ in Eq.~\eqref{gkfun}):
\begin{align}
&\frac{\partial L}{\partial i_{u}}=\frac\eta2\biggl[
g_1(\overline\nu)\left(\frac1{\overline o_u}-
\frac1{4i_u^2\overline o_{u_\star}^{\phantom{2}}}\right)\nonumber\\
&\quad\:\:\:\,-g_1(\nu)\left(\frac1{o_u}-
\frac1{4i_u^2o_{u_\star}^{\phantom{2}}}\right)\biggr]-
\varkappa\left[1-\frac1{4i_u^2}\right]=0,
\label{Eq1Cap}\\
&\frac{\partial L}{\partial m_{u}}\;\,=
\frac\eta2\frac{g_1(\overline\nu)}{\overline o_u}-\varkappa=0,
\label{Eq2Cap}\\
&\frac{\partial L}{\partial m_{u_\star}}=
\frac\eta2\frac{g_1(\overline\nu)}{\overline o_{u_\star}}-\varkappa=0,
\label{Eq3Cap}
\end{align}
where the Lagrange function is
\begin{equation*}
L=\underline{C}-\varkappa\left(i_u+\frac1{4i_u}+m_u+m_{u_\star}-2N-1\right),
\end{equation*}
with $\varkappa$ the Lagrange multiplier. 

Eqs.~\eqref{Eq2Cap} and~\eqref{Eq3Cap} give $\overline o_u=\overline
o_{u_\star}$ which was obtained before from qualitative considerations.  By
substituting Eqs.~\eqref{Eq2Cap} and~\eqref{Eq3Cap} into Eq.~\eqref{Eq1Cap} one
can find that squeezing in input $s_\mathrm{in}$ equals that of environment $s$
and output $s_\mathrm{out}$:
\begin{equation}
\frac{i_{u_\star}}{i_u}=\frac{e_{u_\star}}{e_u}=\frac{o_{u_\star}}{o_u},
\label{squeezequality}
\end{equation}
which allows us to find optimal input eigenvalues
\begin{equation}
i_u=\frac12\sqrt{\frac{e_u}{e_{u_\star}}},\qquad\quad
i_{u_\star}=\frac12\sqrt{\frac{e_{u_\star}}{e_u}}.
\label{inputEigs}
\end{equation}
Thus, given the environment state $V_\mathrm{env}=V(\EuScript
N_\mathrm{env},s)$, the optimal input state is $V_\mathrm{in}=V(0,s)$.
Combining Eq.~\eqref{auequation} with~\eqref{inputEigs} one can obtain optimal
$m$-eigenvalues
\begin{equation}
\begin{split}
m_{u_{\phantom{\star}}}=N+\frac12-i_{u_{\phantom{\star}}}+
\frac{1-\eta}\eta\left(N_\mathrm{env}+\frac12-e_{u_{\phantom{\star}}}\right),\\
m_{u_\star}=N+\frac12-i_{u_\star}+\frac{1-\eta}\eta\left(N_\mathrm{env}+\frac12-e_{u_\star}\right).
\label{cueigs}
\end{split}
\end{equation}

In order to get analogous relations for the heterodyne rate, the
replacements~\eqref{Repli}-\eqref{ReplN} and~\eqref{Replg} must be applied to
Eqs.~\eqref{Eq1Cap}-\eqref{Eq3Cap} and \eqref{squeezequality}-\eqref{cueigs}.
In particular, it gives
\begin{equation*}
s_\mathrm{in}^\mathrm{(het)}=s^\mathrm{(het)}=
s_\mathrm{out}^\mathrm{(het)}
\end{equation*}
and $V_\mathrm{in}^\mathrm{(het)}=V(0,s^\mathrm{(het)})$.  Notice, that
Eqs.~\eqref{EuHets} and~\eqref{MenvHet} give
\begin{align*}
&N_\mathrm{env}^\mathrm{(het)}+\frac12-e_{u_{\phantom{\star}}}^\mathrm{(het)}=
N_\mathrm{env}+\frac12-e_u,\\
&N_\mathrm{env}^\mathrm{(het)}+\frac12-e_{u_\star}^\mathrm{(het)}=
N_\mathrm{env}+\frac12-e_{u_\star}
\end{align*}
for the relations~\eqref{cueigs}.

Finally, the explicit relations for capacity and heterodyne rate in the third
stage read
\begin{align}
&\underline{C}=g\bigl[\eta N+(1-\eta)N_\mathrm{env}\bigr]-
g\bigl[(1-\eta)\EuScript{N}_\mathrm{env}\bigr],
\label{CC1useSimple}\\
&R^{(\mathrm{het})}=
\log_2\left[\eta N+(1-\eta)N_{\rm env}^{(\rm het)}+\frac12\right]
\nonumber\\
&\qquad\qquad\qquad\qquad
-\log_2\left[(1-\eta)\EuScript{N}_\mathrm{env}^{(\rm het)}+\frac12\right],
\label{HR1useSimple}
\end{align}
where Eq.~\eqref{HR1useSimple} becomes
\begin{equation}
R^{(\mathrm{het})}=
\log_2\left[1+\frac{\eta N}{1+(1-\eta)\EuScript{N}_\mathrm{env}}\right]
\label{HR1useSimpleNenv0}
\end{equation}
for the case of thermal nonsqueezed environment.  The
relation~\eqref{CC1useSimple} originally was found in~\cite{NJP} and generalizes
that obtained for lossy bosonic channel with vacuum environment~$g(\eta
N)$~\cite{GG2} and, later, with thermal nonsqueezed environment~\cite{GM}. In
turn, Eq.~\eqref{HR1useSimple} generalizes the relation for the heterodyne rate,
$\log_2(1+\eta N)$, found in~\cite{CavesDrummond} for vacuum environment (see
also discussion in~\cite{GG2}).  

By comparing Eqs.~\eqref{HolDef0app},~\eqref{defcap} and~\eqref{CC1useSimple} we
get the logarithmic approximation to the capacity\footnote{Remember, that according
to Eq.~\eqref{zeroth} $g(v)\approx\log_2\left(v+\frac12\right)$.}
\begin{multline}
\underline{C}^{(\log)}=\log_2\left[\eta N+(1-\eta)N_{\rm env}+\frac12\right]\\
-\log_2\left[(1-\eta)\EuScript{N}_\mathrm{env}+\frac12\right],
\label{Clog1use}
\end{multline}
which coincides with the heterodyne rate~\eqref{HR1useSimple} after the
replacements~\eqref{ReplN} (it follows from Eqs.~\eqref{NenvHet}
and~\eqref{MenvHet} that the limits of the ratios
$N_\mathrm{env}^\mathrm{(het)}/N_\mathrm{env}$ and
$\EuScript{N}_\mathrm{env}^\mathrm{(het)}/\EuScript{N}_\mathrm{env}$ for
$\EuScript{N}_\mathrm{env}\to\infty$ are equal to one).  Thus, the
limit~\eqref{HetLim} actually holds.  Notice, that eigenvalues~\eqref{inputEigs}
and~\eqref{cueigs} are optimal also for the quantity $\underline{C}^{(\log)}$,
therefore we have $\underline C\equiv\underline C^{(0)}$ in the third stage.

In the case of pure environment, the capacity~\eqref{CC1useSimple} 
can be written as
\begin{equation}
\underline C=g(\overline N_\mathrm{out}),
\label{ClCapNL}
\end{equation}
where $\overline N_\mathrm{out}$ is given by Eq.~\eqref{Nout1Dcap}. The form of
the relation~\eqref{ClCapNL} provides the most natural generalization of the
noiseless channel capacity $g(N)$.  Thus, in the third stage the capacity of the
channel with pure environment is completely defined by the average amount of photons
contained in the channel (\emph{i.e.} in the system ``environment plus input''),
where probability weights $\eta$ and $1-\eta$ specify the contribution of input
and environment states into the channel capacity.

Previously it was proved (see proposition~\ref{EuNeqEust}) that
$m_{u_\star}\neq0$ is optimal for the chosen  convention ($e_u>e_{u_\star}$),
therefore the third stage holds if $m_u>0$. This is the case for the capacity (for
the quantities $\underline C$, $\underline C^{(\log)}$ and $\underline C^{(0)}$)
if the amount $N$ of input photons is higher than the threshold
\begin{equation}
N_{2\to3}=
i_u-\frac12-\frac{1-\eta}\eta\left(N_{\rm env}+\frac12-e_u\right),
\label{Nthr23}
\end{equation}
where $i_u$ is defined by the first of Eqs.~\eqref{inputEigs}.  It is equivalent
to the restriction $s<s_{2\to3}^{\phantom{1}}$ for given values of $\eta$, $N$ and $\EuScript
N_\mathrm{env}$, where
\begin{equation*}
s_\mathrm{2\to3}^{\phantom{1}}=
-\ln\Bigl[\sqrt{1+\phi_0+\bigl(N+1/2\bigr)^2\phi_0^2}-
\bigl(N+1/2\bigr)\phi_0\Bigr]
\end{equation*}
with
\begin{equation}
\phi_0:=\frac{\eta}{(1-\eta)\left(\EuScript N_\mathrm{env}+\frac12\right)}.
\label{phi0}
\end{equation}
Notice, that the quantity $s_\mathrm{2\to3}^{\phantom{1}}$ has the limits
\begin{align}
&\lim_{\eta\to1}\,s_\mathrm{2\to3}^{\phantom{1}}=\ln(2N+1),
\label{s23eta1}\\
&\lim_{\eta\to0}\,s_\mathrm{2\to3}^{\phantom{1}}=0.
\nonumber
\end{align}
The threshold~\eqref{Nthr23} holds also for the heterodyne rate if the
replacements~\eqref{Repli} and $N_{2\to3}^{\phantom{1}}\to
N_{2\to3}^\mathrm{(het)}$ are applied, where $i_u^\mathrm{(het)}$ expressed
through standard eigenvalues reads
\begin{equation}
i_u^\mathrm{(het)}=\frac12\sqrt{\frac{1+2\,(1-\eta)\,e_u\phantom{i}}{1+2\,(1-\eta)\,e_{u_\star}}}.
\label{iu3sthetStEigs}
\end{equation}

The threshold $N_{2\to3}$ is a nonnegative number which equals zero only for the
vacuum environment. As far as the third stage holds only if $N>N_{2\to3}$ and
the first stage holds for only $N=0$, the second stage must correspond to values
$0<N\leqslant N_{2\to3}$. Thus, the type of solution increases its stage in
sequence starting from the first stage and ending to the third one if $N$ grows
from zero to infinity.  This explains the origin of the adopted term ``stage''.
Also, it can be interpreted as ``the third stage is always the most preferable
if energy $N$ is sufficiently high, otherwise the second stage should be taken,
and the first stage holds if only both the third and the second stage fail to
satisfy the constraint''. This mnemonic rule, although trivial for the
single-mode channel, will be useful when applied to the multi-mode memory
channel.  The above consideration is also valid for the heterodyne threshold
$N_{2\to3}^\mathrm{(het)}$.

Similarly to the quantity $s_\mathrm{2\to3}^{\phantom{1}}$, 
given the values of $s$, $N$ and $\EuScript N_\mathrm{env}$, the
relation for transmissivity $\eta_{2\to3}^{\phantom{1}}$ 
corresponding to transition from second to third stage can be written as
\begin{equation}
\eta_{2\to3}^{-1}=
1-\frac{\;\;\;\,N+\frac12-i_u}{N_\mathrm{env}+\frac12-e_u},
\label{Eta23}
\end{equation}
where $0<s\leqslant\ln(2N+1)$ and $i_u$ is defined by the first of Eqs.~\eqref{inputEigs}.
Taking into account that $s_\mathrm{in}=s$, $s\geqslant0$, the limit~\eqref{s23eta1}
and monotonicity of $\eta_{2\to3}^{\phantom{1}}$ with respect to $s$, one can
see that 
\begin{equation}
\frac12\leqslant i_u\leqslant N+\frac12,
\label{iuforeta23}
\end{equation}
where higher values of $i_u$
correspond to higher values of $\eta_{2\to3}^{\phantom{1}}$.
The transmissivity $\eta_{2\to3}^{\phantom{1}}$ is plotted vs $s$ in Fig.\ref{NLfig}-left.

\subsection{The second stage}\label{TheSecondStage}

One can show that the Lagrange equations for the capacity (and heterodyne rate)
in the second stage can be obtained from the
system~\eqref{Eq1Cap}-\eqref{Eq3Cap} by substituting $m_u=0$
($m_u^\mathrm{(het)}=0$) in all equations and by removing Eq.~\eqref{Eq2Cap}
corresponding to derivative with respect to $m_u$ ($m_u^\mathrm{(het)}$). 
This is because unknown variables enter in the Lagrange equations as linear combinations.
For the homodyne rate the Lagrange equations are the same as for
the capacity in the second stage if the replacement~\eqref{Replg} is applied.

Then, solving the Lagrange equations for the homodyne rate one can find the
ratio
\begin{equation}
\frac{i_{u_\star}}{i_u}=\frac{o_{u_\star}}{\overline o_{u_\star}},
\label{squeezequality2st}
\end{equation}
which also holds for the heterodyne rate after replacements \eqref{Repli},
\eqref{Replo} and~\eqref{Replovo}. For the capacity the Lagrange equations give
a \emph{mode transcendental equation} on $i_u$
\begin{equation}
\mathcal{F}(i_u)=0,
\label{ModeTrEq1use}
\end{equation}
where
\begin{equation}
\mathcal{F}:=
g_1(\overline\nu)\left[\frac1{o_u}-\frac1{\overline o_{u_\star}}\right]
-g_1(\nu)\left[\frac1{o_u}-\frac1{4i_u^2o_{u_\star}^{\phantom{2}}}\right].
\label{Ffun}
\end{equation}
Notice, that Eq.~\eqref{ModeTrEq1use}
results to Eq.~\eqref{squeezequality2st} if the $g_1$-function is taken to
zeroth-order approximation (\emph{i.e.} \mbox{$g_1\equiv(\ln2)^{-1}$}).
Remember, that in the second stage the optimal
eigenvalues for $\underline C^\mathrm{(\log)}$,
$\underline C^\mathrm{(0)}$ and $R^\mathrm{(hom)}$ (see
Eqs.~\eqref{HolDef0app} and~\eqref{HomLim})
are the same. They follow from Eq.~\eqref{squeezequality2st} solved for
the variable $i_u$ and equal to
\begin{align}
&m_u=0,
\label{iuapproxmu}\\
&m_{u_\star}=2N+1-i_u-\frac1{4i_u},
\label{iuapproxmust}\\
&i_u=\frac12\left[\sqrt{1+(2N+1)\phi+(\phi/2)^2}-\phi/2\right],
\label{iuapproxiu}\\
&i_{u_\star}=\frac1{4i_u},
\label{iuapproxiust}
\end{align}
where
\begin{equation}
\phi=\frac\eta{1-\eta}e_{u_\star}^{-1}
\label{Phi2ndStage}
\end{equation}
is equal to $\phi_0$ (see Eq.~\eqref{phi0}) in the case of thermal environment
($s=0$).  The exact values of the optimal eigenvalues for the capacity $\underline C$ 
are given by Eqs. \eqref{ModeTrEq1use}, \eqref{iuapproxmu},
\eqref{iuapproxmust} and~\eqref{iuapproxiust}.  As far as eigenvalues
\eqref{iuapproxmu}--\eqref{iuapproxiust} are optimal for the quantity
$\underline C^\mathrm{(0)}$, below we will call them as \emph{the zeroth-order
solution} (or \emph{the zeroth-order eigenvalues}) for capacity.  Thus,
similarly to the third stage, in the second stage the quantity $\underline
C^\mathrm{(0)}$ is also expressed in an explicit form.  In turn, the condition
$m_{u_\star}>0$ (see Eq.~\eqref{iuapproxmust}) restricts the admissible region
for $i_u$ to the interval
\begin{equation*}
N+\frac12-\sqrt{N^2+N}<i_u<N+\frac12+\sqrt{N^2+N}.
\end{equation*}
The optimal eigenvalues for the heterodyne rate are given by the same
relations \eqref{iuapproxmu}--\eqref{iuapproxiust}
if the replacements~\eqref{Repli}, \eqref{Replm} and
\begin{equation}
\phi\to\phi^\mathrm{(het)},
\label{Replphi}
\end{equation}
with
\begin{equation}
\phi^\mathrm{(het)}=\frac\eta{1-\eta}\left[e_{u_\star}^\mathrm{(het)}\right]^{-1}
\label{Phi2ndStageHet}
\end{equation}
are applied.

By comparing Eq.~\eqref{defhom} with Eq.~\eqref{squeezequality2st} one can
get the homodyne rate
\begin{align}
R^\mathrm{(hom)}&=\log_2(2i_u)
\nonumber\\
&=\log_2\left[\sqrt{1+(2N+1)\phi+(\phi/2)^2}-\phi/2\right].
\label{SndStF}
\end{align}
Remember, that $R^\mathrm{(hom)}\equiv\underline C^\mathrm{(log)}$ (see
Eq.~\eqref{HomLim}) in the second stage.  Then, similarly, by comparing
Eqs.~\eqref{defhet} and~\eqref{squeezequality2st} we get the same
relation~\eqref{SndStF} for the heterodyne rate if the
replacements~\eqref{Repli} and~\eqref{Replphi} are applied to it.

The first-order approximation for mode transcendental
equation~\eqref{ModeTrEq1use} can be obtained by replacing the function $g_1$
with its first-order approximation~\eqref{g1v1stOrdApp}.  Since
Eq.~\eqref{ModeTrEq1use} cannot be exactly solved within this approximation, we
will solve it in the neighborhood of the zeroth-order solution
\eqref{iuapproxmu}--\eqref{iuapproxiust} as linear perturbation. In particular,
by denoting input zeroth-order eigenvalue~\eqref{iuapproxiu} as~$i_u^{(0)}$ and
substituting $i_u$ with $i_u^{(0)}+\varepsilon_{u}$ in the first-order approximation of
Eq.~\eqref{ModeTrEq1use}, we get a linear equation
for small deviation $\varepsilon_u$. Its solution is
\begin{equation}
\varepsilon_{u}=\frac{\eta\,\overline o_{u_\star}o_ui_{u}^{(0)}m_{u_\star}
(\overline o_{u_\star}-o_u)}
{2\bigl[\eta^2(o_u^2+\overline o_{u_\star}^2-
\overline o_{u_\star}o_u)\,i_{u}^{(0)}m_{u_\star}-
\overline o_{u_\star}^2o_u^2(12\nu^2+1)\bigr]},
\label{FirstOrd1use}
\end{equation}
whose variables are the zeroth-order eigenvalues. Thus, we have found \emph{the
first-order solution}\footnote{Similarly, another first-order
solution can be obtained if exact relation
for $g_1$ function is used instead of approximation~\eqref{g1v1stOrdApp}.} 
$i_u^{(1)}=i_u^{(0)}+\varepsilon_u$.  Remember that in the second stage, 
by virtue of Eqs.~\eqref{iuapproxmu}, \eqref{iuapproxmust} and~\eqref{iuapproxiust},
the only degree of freedom is represented by~$i_u$.
Hence, it is sufficient to specify its value in order to have the complete solution
of the optimization problem.

Similarly to the quantity $\underline C^{(0)}$ whose variables are the
zeroth-order eigenvalues \eqref{iuapproxmu}--\eqref{iuapproxiust}, the
first-order solution has to be substituted into the exact\footnote{Notice, that in
order to get the zeroth-order and the first-order approximate solutions we
replaced the exact $g$-function and its derivatives (everywhere in optimization
problem) by their zeroth-order and the first-order approximations,
respectively.} relation~\eqref{defcap} instead of its first-order approximation
which was used to derive the first-order eigenvalues.  Otherwise, the loss in
accuracy becomes significant. In particular, although input and modulation
eigenvalues calculated through exact and approximate approaches essentially
differ each other, they give rise to almost equal values for capacity.  This can
be explained by the fact that the quantity~\eqref{defcap}, considered as a
function of only one unknown variable\footnote{The other input and modulation
eigenvalues have to be expressed through $i_u$ using Eqs.~\eqref{iuapproxmu},
\eqref{iuapproxmust} and~\eqref{iuapproxiust}.} $i_u$, has zero derivative in
the neighborhood of its optimal value (\emph{i.e.} the deviation of $i_u$ affects
maximum of the capacity only in the second order).  The quantity~\eqref{defcap}
considered as a function of the first-order eigenvalues below will be called
\emph{the first-order approximation to capacity} $\underline C^{(1)}$.

The homodyne rate $\log_2\sqrt{1+4\eta N}$ was found in~\cite{CavesDrummond} by
supposing both the environment and input states to be vacuum (see also
discussion in~\cite{GG2}). Indeed, it can be obtained without solving the
optimization problem, by substituting in Eq.~\eqref{defhom}
$i_u=i_{u_\star}=e_u=e_{u_\star}=1/2$ and $m_{u_\star}=2N$ as it follows from
the constraint~\eqref{1useConstr}.  However, since optimal input state is never
vacuum according to Eq.~\eqref{iuapproxiu}, that rate holds (approximately) only
if the value of $N$ is close to zero.

\subsection{$\overline o_{u_\star}$-representation}

We have solved the problem of finding the optimal eigenvalues $i_u$,
$i_{u_\star}$, $m_u$ and $m_{u_\star}$ for given values of $e_u$, $e_{u_\star}$,
$\eta$ and $N$. It is interesting to note, that the eigenvalue $\overline
o_{u_\star}$ can be used as the equivalent replacement\footnote{This fact
will be used in Sec.~\ref{MemoryChannelsSec} for discussing memory channels.}
of the quantity $N$. In fact, Eq.~\eqref{auequation} makes it evident in the
third stage. Let us show this also for the second stage. Combining
Eqs.~\eqref{eigs_defs}, \eqref{iuapproxmust}, \eqref{iuapproxiust}
and~\eqref{Phi2ndStage} one can get the relation
\begin{equation}
2N+1=i_u+\frac{\overline o_{u_\star}}{\eta}-\frac1{\phi}.
\label{Noustovrel}
\end{equation}
By substituting it into Eq.~\eqref{iuapproxiu} and then solving
the latter for $i_u$ one can obtain
\begin{equation}
i_u=\frac12\left[
\sqrt{(\phi/4)^2+\overline o_{u_\star}\phi/\eta}-\phi/4\,\right].
\label{iuapproxIUO}
\end{equation}
Hence, Eq.~\eqref{SndStF} can be equivalently rewritten through variable 
$\overline o_{u_\star}$ as
\begin{equation}
R^\mathrm{(hom)}=
\log_2\left[
\sqrt{(\phi/4)^2+\overline o_{u_\star}\phi/\eta}-\phi/4\,\right],
\label{SndStFou}
\end{equation}
which coincides with the quantity $\underline C^\mathrm{(log)}$ in the second
stage.  Notice, that eigenvalue $m_{u_\star}$ can be expressed through
$\overline o_{u_\star}$ as
\begin{equation}
m_{u_\star}=\frac{\overline o_{u_\star}}\eta-\frac1\phi-\frac1{4i_u}.
\label{iuapproxMUstO}
\end{equation}
Thus, the eigenvalues~\eqref{iuapproxmu}, \eqref{iuapproxiust},
\eqref{iuapproxIUO} and~\eqref{iuapproxMUstO} are optimal for the quantities
$R^\mathrm{(hom)}$, $\underline C^\mathrm{(log)}$ and $\underline C^{(0)}$ in
the second stage and expressed through the quantity $\overline o_{u_\star}$
instead of $N$. Eqs.  \eqref{Noustovrel}--\eqref{iuapproxMUstO} hold also for
the heterodyne rate if the replacements \eqref{Repli}, \eqref{Replm},
\eqref{Replovo} and~\eqref{Replphi} are applied.

Similarly, the mode transcendental equation~\eqref{ModeTrEq1use}
also does not depend
on $N$ if eigenvalue $\overline o_{u_\star}$ is assumed to be a known constant.
In this case the admissible region for the eigenvalue $i_u$ (root of
Eq.~\eqref{ModeTrEq1use}) can be estimated using inequalities $\overline\nu>\frac12$
and $m_{u_\star}>0$ (see Eq.~\eqref{iuapproxMUstO}), which can be rewritten as
\begin{equation*}
i_u>\frac1\eta\left[\frac1{4\,\overline o_{u_\star}}-(1-\eta)\,e_u\right]
\end{equation*}
and
\begin{equation*}
i_u>\frac14\left[\frac{\overline o_{u_\star}}\eta-\frac1\phi\right]^{-1},
\end{equation*}
respectively.  Analogously to Eq.~\eqref{FirstOrd1use}, 
by expressing $N$ through $\overline o_{u_\star}$ in Eq.~\eqref{ModeTrEq1use}
and using approximation~\eqref{g1v1stOrdApp} one can get the first
order solution $i_u^{(1)}=i_u^{(0)}+\varepsilon_u$ in terms of $\overline
o_{u_\star}$. In this case $\varepsilon_u$ is given by the relation
\begin{multline}
\varepsilon_u=(\overline o_{u_\star}-o_u)
(\overline o_{u_\star}-o_{u_\star})\,o_u^{\phantom{1}}i_u^{(0)}\\
\times\Bigl\{\Bigl(2\,\bigl[o_u^2+\overline o_{u_\star}^2-
(o_u+o_{u_\star})\,\overline o_{u_\star}\bigr]+
\bigl[12\,o_u^2+1\bigr]\nu^2\Bigr)i_u^{(0)}\eta\\
-2\left[12\,\nu^2+1\right]o_u^2\,\overline o_{u_\star}^{\phantom{1}}\Bigr\}^{-1},
\label{FirstOrdNuses}
\end{multline}
whose variables are the zeroth-order eigenvalues~\eqref{iuapproxiust},
\eqref{iuapproxIUO} and~\eqref{iuapproxMUstO}. Notice, that despite the
equations~\eqref{iuapproxiu} and~\eqref{iuapproxIUO} are equivalent (one can be
obtained from another), this is not the case for relations~\eqref{FirstOrd1use}
and~\eqref{FirstOrdNuses}.

\subsection{Noiseless channel}\label{NoiseLessChannel}

Let us demonstrate the above results on the particular case of noiseless
(\emph{i.e.} ideal) channel ($\eta$=1).  Its capacity equals \mbox{$C=g(N)$}.
The optimal eigenvalues for its homodyne rate can be found from
Eqs.~\eqref{squeezequality2st}, \eqref{iuapproxmust}
and~\eqref{iuapproxiust} by substituting $\eta=1$, which gives
\begin{align}
&i_u=N+\frac12,
\label{noiselessiu}\\
&m_{u_\star}=N\left(1+\frac1{2N+1}\right)=\sinh(\ln(2N+1))).
\label{noiselesscust}
\end{align}
The optimal eigenvalues for its heterodyne rate can be obtained from
Eqs.~\eqref{cueigs} and~\eqref{iu3sthetStEigs} by substituting $\eta=1$,
which results in
$i_u^\mathrm{(het)}=i_{u_\star}^\mathrm{(het)}=\frac12$ and 
$m_u^\mathrm{(het)}=m_{u_\star}^\mathrm{(het)}=N$. 
Hence, we have $N_{2\to3}^\mathrm{(het)}=0$ (see Eq.~\eqref{Nthr23}), 
\emph{i.e.} the second stage does not exist in this case.

The relations~\eqref{HR1useSimpleNenv0} and~\eqref{SndStF} applied to 
the noiseless channel give the inequalities~\cite{GG2}
\begin{equation}
R^\mathrm{(het)}<R^\mathrm{(hom)}<C,
\label{epicfail}
\end{equation}
where $R^\mathrm{(hom)}=\log_2(2N+1)$ and $R^\mathrm{(het)}=
\log_2(N+1)$~\cite{CavesDrummond}. It means that both heterodyne and homodyne
rates never achieve the capacity for finite $N$ even for the noiseless
channel\footnote{It is shown in~\cite{CavesDrummond} that the capacity
of the noiseless channel can be achieved by using Fock states for encoding and
photon counting measurement for decoding.}.
In particular, for large values of $N$ inequalities~\eqref{epicfail} read
\begin{equation*}
\log_2N<\log_2N+1<\log_2N+\frac1{\ln2},
\end{equation*}
where the difference between the rates and the capacity disappears in the limit
$N\to\infty$.  In addition,
both capacity and rates of the noiseless channel are
always higher than theirs values in the presence of losses (environment),
\emph{i.e.} when $\eta<1$.

\begin{figure}[t]
\begin{center}
\includegraphics[scale=1]{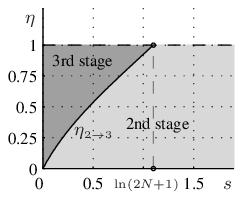}\qquad\includegraphics[scale=1]{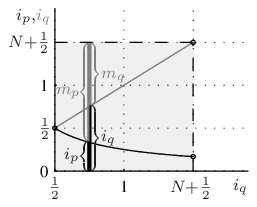}
\caption{On the left, two regions of the plane $(s,\eta)$ corresponding to the third
(darker background) and the second stage for $N=1$ and
$\EuScript N_\mathrm{env}=1$. The regions are separated by the curve
$\eta_{\scriptscriptstyle 2\to3}$
(see Eq.~\eqref{Eta23}) plotted vs $s$ (black curve), it splits the whole
plane into two regions marked with different grey-scale backgrounds (the darker
background corresponds to third stage). The curve $\eta_{\scriptscriptstyle
2\to3}$ reaches the value $\eta=1$ at squeezing value $s=\ln(2N+1)$. The
horizontal dashed line plotted for $\eta=1$ corresponds to noiseless channel. On the
right, the eigenvalues $i_p$ and $i_q$ for the noiseless channel are plotted vs
$i_q$ for $N=1$.  The area of the square of grey color is equal to
$\overline\nu^2=\left(N+\frac12\right)^2$ which defines the capacity
$g\bigl(\sqrt{\overline\nu^2}-\frac12\bigr)=g(N)$. Depicted braces show that each value
of $i_q$ corresponds to two different methods to distribute the energy
$N+\frac12$ between input and modulation quadratures.}
\label{NLfig}
\end{center}
\end{figure}

One can also notice 
that despite optimal input and modulation eigenvalues are unique for
heterodyne and homodyne rate, this is not the case for noiseless channel
capacity.  The latter has infinite amount of solutions, which can be shown as
follows.  At first, since the theorem~\ref{InpPurTheor} holds also for $\eta=1$,
the optimal input state must be pure. At second, any pure input state zeros the
second term in Holevo-$\chi$ quantity. As far as the area of a rectangle with fixed
perimeter is maximal if and only if rectangle's sides are equal (see proof of
proposition~\ref{EuNeqEust}), we have the system of equations
\begin{equation}
\begin{split}
i_u+m_{u_{\phantom{\star}}}=N+\frac12&,\\
\frac1{4i_u}+m_{u_\star}=N+\frac12&,
\end{split}
\label{NLoptInpMod}
\end{equation}
where the energy restriction~\eqref{1useConstr} is the ``perimeter''.  Thus, by
taking any input eigenvalue from the interval
\begin{equation}
\frac1{4\left(N+\frac12\right)}\leqslant i_u\leqslant N+\frac12,
\label{NLoptInp}
\end{equation}
and obtaining the eigenvalues $i_{u_\star}$, $m_u$, $m_{u_\star}$ from the
relations~\eqref{iuapproxiust}, \eqref{NLoptInpMod} we arrive at the same value
of capacity $g(N)$.

Taking into account that the capacity is symmetric over
quadratures and considering as usual only positive values of $s_\mathrm{in}$,
we can parametrize the interval~\eqref{NLoptInp} as (see similarity with
Eq.~\eqref{iuforeta23})
$
s_\mathrm{in}=\sigma\ln(2N+1),
$
where $\sigma\in\bigl[0,1\bigr]$.  Notice, that $\sigma=1$
is the only solution corresponding to second stage in this interval. 
Then, the optimal eigenvalues can be expressed as functions
of $\sigma$ as
\begin{equation}
\begin{split}
&i_{u_{\phantom{\star}}}=\frac12(2N+1)^\sigma,\\
&i_{u_\star}=\frac12(2N+1)^{-\sigma},\\
&m_{u_{\phantom{\star}}}=N+\frac12\,\bigl[1-(2N+1)^\sigma\bigr],\\
&m_{u_\star}=N+\frac12\left[1-(2N+1)^{-\sigma}\right],
\end{split}
\label{musigma}
\end{equation}

The set of eigenvalues optimal for noiseless channel are plotted
in Fig.\ref{NLfig}-right. The black point at the left part of the graph corresponds to well-known
solution $i_u=i_{u_\star}=\frac12$, $m_u=m_{u_\star}=N$ which is particular case
of $\sigma=0$ in Eqs.~\eqref{musigma} . Two black points at the
right part of the graph correspond to solution~\eqref{noiselessiu},
\eqref{noiselesscust} following from Eqs.~\eqref{musigma} for $\sigma=1$.

Let us consider how the solution (\textit{i.e.} the optimal input and
modulation eigenvalues) changes if noise in the channel disappears 
($\eta\to1$). The loci $(s,\eta)$ corresponding to different
stages are shown in Fig.\ref{NLfig}-left. One can see, that if $s$ belongs to
the interval $(0,\ln(2N+1))$, then by increasing $\eta$ from $0$ to $1$ we always 
change the second stage to the third one. As far as $s_\mathrm{in}=s$ in the
third stage, the solutions for different values of $s$ for noisy channel tend to different
solutions for noiseless channel and remain in the third stage. These
solutions of noiseless channel correspond to the interval
$\sigma\in\bigl[0,1\bigr)$.
Then, all solutions for $s\geqslant\ln(2N+1)$ of noisy channel tend to the same solution
of the noiseless channel which corresponds to the second stage and to $\sigma=1$.

\subsection{Universal limit}

Let us analyze the behavior of the capacity and rates in the limit of infinite
environment squeezing ($s\to\infty$) if channel parameters $\eta$, $N$ and
$\EuScript{N}_\mathrm{env}$ are fixed.  Notice, that only the second stage is
possible in this case according to Eq.~\eqref{Nthr23}.  By substituting
eigenvalues~\eqref{iuapproxmu}, \eqref{iuapproxmust} and~\eqref{iuapproxiust}
into mode transcendental equation~\eqref{ModeTrEq1use} and then solving it for
$i_u$ in the case of $s\to\infty$ one can get the result~\eqref{noiselessiu}.
Thus, the eigenvalues maximizing the capacity in the limit of \mbox{$s\to\infty$} are
the same as for the noiseless channel and given by Eqs.~\eqref{noiselessiu}
and~\eqref{noiselesscust}.  Substituting them into Eqs.~\eqref{Nusesnunuov} one
can see that both symplectic eigenvalues $\nu$ and $\overline\nu$ tend to
infinity if $s\to\infty$. This allows us to use the logarithmic approximation to
the capacity to find the limit.  Hence, by comparing Eqs.~\eqref{HomLim},
\eqref{SndStF} and~\eqref{noiselessiu} we obtain the result~\cite{NJP}
\begin{equation}
\lim_{s\rightarrow\infty}\underline{C}(s,\eta,N,\EuScript{N}_\mathrm{env})=
\log_2(2i_u)=\log_2(2N+1),
\label{Climit}
\end{equation}
which will be called below as \emph{the universal limit}.

As far as $\lim_{\phi\to\infty}i_u$ (see Eq.~\eqref{iuapproxiu}) gives the
relation~\eqref{noiselessiu}, the limit~\eqref{Climit} also holds for homodyne
rate $\lim_{s\to\infty}R^{(\mathrm{hom})}$~\cite{CosmoRates}.  Analogously,
taking into account that
$\lim_{s\to\infty}\phi^\mathrm{(het)}=2\eta$
(see Eq.~\eqref{Phi2ndStageHet}),
we get the limit
\begin{equation}
\lim_{s\to\infty}R^{(\mathrm{het})}=
\log_2\left[\sqrt{1+2(2N+1)\eta+\eta^2}-\eta\,\right].
\label{HetLimitFinal}
\end{equation}

Notice, that the limiting value~\eqref{Climit} of the capacity equals the homodyne rate in
the case of perfect (noiseless) channel (see Eq.~\eqref{epicfail}).  This fact can be
understood by considering that, for $s\to\infty$, the quadrature $u$ becomes infinitely
noisy while the quadrature $u_\star$ becomes noiseless.  Thus, by encoding the information
in the quadrature $u_*$, the information transmission becomes noiseless.

\subsection{Concavity of solution}\label{ConcavityOfSolution}

The concavity over $N$ for the capacity and rates will be essential below for
discussing multiple channels uses. It is also the important property allowing to
show additivity of the capacity and rates for the memoryless channels.

Let us show the concavity of the function $\underline C(N)$.
In the second stage, the latter can be represented as $\underline C(N,i_u(N))$,
therefore 
the first derivative with respect to $N$ is
\begin{equation*}
\frac{d{}^{}\underline{C}}{dN}=\frac{\partial{}^{}\underline{C}}{\partial N}+
\frac{\partial{}^{}\underline{C}}{\partial{}^{}i_u}\frac{\partial{}^{}i_u}{\partial N}.
\end{equation*}
However, since only the eigenvalues maximizing $\underline{C}$ are of interest,
we have
${\partial\underline{C}}/{\partial{}^{}i_u}=(\eta/2)\mathcal{F}=0$ (see
definition~\eqref{Ffun}), therefore
\begin{equation*}
\frac{d{}^{}\underline{C}}{dN}=\frac{\partial{}^{}\underline{C}}{\partial N}.
\end{equation*}
Then, one can show that for all values of $N$ and for both second and
third stages
\begin{equation}
\frac{d{}^{}\underline{C}}{dN}=\frac\eta{\overline o_{u_\star}}g_1(\overline\nu)>0,
\label{dCdNGen}
\end{equation}
which proves that $\underline C(N)$ is a monotonically increasing function of its
argument. Notice, that
\begin{equation}
\max_{N}\frac{\partial{}^{}\underline{C}}{\partial N}=\lim_{N\to0}
\frac{\partial{}^{}\underline{C}}{\partial N}\leqslant\infty,
\label{DerThrRestr}
\end{equation}
where equality is achieved only by the pure environment state ($e_u=e_{u_\star}=1/2$).

It is shown in Appendix~\ref{App2ndDer} that
\begin{equation}
\frac{d^2\underline C}{d{}^{}N^2}<0.
\label{SecDirNegCap}
\end{equation}
Then, we deduce from Eqs.~\eqref{dCdNGen} and~\eqref{SecDirNegCap}
that the function $\underline C(N)$ is concave on the whole
region of $N\in[0,\infty)$.
Thus, the single-mode (one-shot) capacity for fixed values of $e_{u},e_{u_\star}$ and $\eta$
can be considered as the concave function: 
\begin{equation}
{N\longrightarrow\boxed{\underline{C}=
\underline{C}(N)}\longrightarrow\underline{C}},
\label{blackbox1use}
\end{equation}
\emph{i.e.} as a ``blackbox'' returning the value of $\underline{C}$ upon
``input'' $N$ while respecting the concavity property.  

The derivative~\eqref{dCdNGen} holds also for rates if the replacement~\eqref{Replg} is
applied. Besides it, for the heterodyne rate the replacements~\eqref{Replovo}
and~\eqref{Replnu} must be applied. The concavity of both rates and logarithmic
approximation to capacity can be deduced from explicit relations~\eqref{HR1useSimple},
\eqref{Clog1use} and~\eqref{SndStF}.  Hence, both heterodyne and homodyne rates are also
concave functions which can be treated in the same ``blackbox'' form.

\subsection{$\lambda$-representation}\label{AlternativeRepr}

As far as function $\underline C(N)$ is concave and monotonically increasing,
the value of the derivative~\eqref{dCdNGen} can be used as the equivalent
replacement for the amount of photons $N$ granted for the channel input.  Such
approach below will be called the \mbox{\emph{$\lambda$-representation}} to
distinguish it from the standard approach using the quantity $N$
(\emph{$N$-representation}).  Thus, we can specify an input energy for capacity
using
\begin{equation}
\lambda(N):=
\frac{\partial\underline C}{\partial N}=\frac\eta{\overline o_{u_\star}}g_1(\overline\nu).
\label{ETlambda}
\end{equation} 
Eq.~\eqref{ETlambda} can be equivalently
rewritten\footnote{Here we use the property: if $g'(v)=y$, then
$v=1/(e^{y\ln2}-1)$.} in the form of Planck
distribution\footnote{Similar result was obtained in
Ref.~\cite{CavesDrummond} for a number-state channel, where the optimal
photon-number distribution is Planck distribution parametrized by a Lagrange
multiplier.}
\begin{equation}
\overline{\EuScript N}_\mathrm{out}=\frac1{e^{\,\omega/T}-1},
\label{genPlanck}
\end{equation}
where $\overline{\EuScript N}_\mathrm{out}=\overline\nu-1/2$, the
``temperature'' 
$T:=\eta/(\lambda\ln2)$
and ``frequency''
$\omega:=\overline\nu/\overline o_u$. We will also use the ``temperature'' for
the ideal channel
\begin{equation}
T_1:=\left(\lambda\ln2\right)^{-1}
\label{Tdef}
\end{equation}
obtained from the relation for $T$ with $\eta=1$.

In the third stage $\omega=1$ and $\overline{\EuScript
N}_\mathrm{out}=\overline N_\mathrm{out}$ (see Subsec.~\ref{TheSolStages}),
\emph{i.e.} the quantities $\lambda$ and $\eta$ completely define the average amount
of photons~\eqref{Nout1Dcap} contained in channel and, if the environment is
pure, its capacity (see Eq.~\eqref{ClCapNL}).  Moreover, the
dependence $N(\lambda)$ given by Eq.~\eqref{genPlanck} is expressible in
explicit form:
\begin{equation}
N=\frac1\eta\left[
\frac1{e^{\,1/T}-1}-(1-\eta)N_\mathrm{env}\right].
\label{Planck3st}
\end{equation}

Let us now consider the second stage. Following~\cite{NewNoiseChannel} one can
substitute $g_1(\overline\nu)=\overline o_{u_\star}\lambda/\eta$ (see
Eq.~\eqref{ETlambda}) in the relation~\eqref{ModeTrEq1use}. That leads to
\begin{equation}
\omega=\sqrt{1+\frac{\eta\,g_1(\nu)}\lambda\left[
\frac1{o_u}-\frac1{4i_u^2o_{u_\star}^{\phantom{1}}}\right]},
\label{omega2st}
\end{equation}
where we used the relation $\omega^2=\overline o_{u_\star}/o_u$ 
(remember, that in the second stage we have $\overline o_u=o_u$). Then, by
substituting Eq.~\eqref{omega2st} and the relation 
$\overline\nu=\omega o_u$ in Eq.~\eqref{genPlanck} we
get a transcendental equation which relates $\lambda$ and $i_u$.
Hence, Eq.~\eqref{genPlanck} (after all substitutions) becomes 
the mode transcendental equation~\eqref{ModeTrEq1use} 
written in $\lambda$-representation. If the value of
$i_u$ is found for a given value of $\lambda$, the input energy $N$ reads
\begin{equation}
N=\frac12\left[\frac{o_u\,\omega^2}\eta-\frac1\phi+i_u-1\right],
\label{N2stLamRepr}
\end{equation}
which is the relation~\eqref{Noustovrel} with $\overline
o_{u_\star}=o_u\omega^2$. Thus, in any representation ($N$-, 
$\overline o_{u_\star}$- or $\lambda$-representation) we
have to solve only a single transcendental equation to find all variables.

Similarly to the threshold value $N_{2\to3}$ defined by
Eq.~\eqref{Nthr23}, one can consider the threshold $N_{1\to2}=0$ which
defines the amount of photons corresponding to the transition from first to
second stage. These thresholds in the $\lambda$-representation will be denoted by
$\lambda_{2\to3}$ and $\lambda_{1\to2}$ and can be obtained as follows.

The threshold $\lambda_{1\to2}$ is the limit of $\lambda$ for $N\to0$.
Remember, that $N=0$ is the case of the first stage with optimal eigenvalues
$i_u=i_{u_\star}=\frac12$ and $m_u=m_{u_\star}=0$ (see Subsec.~\ref{TheSolStages}).
Then, the convention $e_u>e_{u_\star}$ means for the first stage that $u$ is the
quadrature corresponding to $m_u=0$ for infinitesimal non-zero values of $N$.
Hence, the general relation~\eqref{ETlambda} gives
\begin{equation}
\lambda_{1\to2}\equiv\lim_{N\to0}\lambda=
{\frac\eta{o_{u_\star}}g_1(\nu)}=
\eta\sqrt{\frac{o_u}{o_{u_\star}}}g'\left(\nu-\frac12\right),
\label{Thr12Gen}
\end{equation}
where the input eigenvalues are those of vacuum. Analogously,
\begin{align}
&\lambda_{2\to3}\equiv\lambda\left(N_{2\to3}\right)=
{\eta\,g_1(\overline\nu)}/\overline\nu=
\nonumber\\
&\qquad\quad\:\:\,\eta\,g'\left[\eta\,\left(i_u-\frac12\right)+
\left(1-\eta\right)\left(e_u-\frac12\right)\right],
\label{Thr23Gen}
\end{align}
where $i_u$ is given by~\eqref{inputEigs}.

\bigskip

\begin{propos}\label{NoiseEffectPropos}
The function $\lambda_{1\to2}(e_u,e_{u_\star})$ is monotonically decreasing over
each of its arguments.
\end{propos}

\bigskip

\begin{IEEEproof}
The dependence $\lambda_{1\to2}(e_u)$ is proportional to the function $g_1(\nu)$,
and the dependence $\lambda_{1\to2}(e_{u_\star})$ is proportional to the function
$g_1(\nu)/\nu^2$.  Both these functions are monotonically decreasing over the
argument $\nu$. In turn, $\nu$ is monotonically increasing over $o_u$ and
$o_{u_\star}$ which are linear functions of $e_u$ and $e_{u_\star}$,
respectively. Taking into account that the composition of monotonically decreasing
and monotonically increasing functions is monotonically decreasing, the
proposition is proved.
\end{IEEEproof}

\bigskip

Using the zeroth-order approximation for the $g_{1}$-function in Eq.~\eqref{ETlambda}, 
one can consider the quantity
\begin{equation}
\lambda^{(0)}=\frac\eta{\overline o_{u_\star}\ln2},
\label{lambda0}
\end{equation}
which will play the role of $\lambda$ for both\footnote{Our purpose
is to get (as much as possible)
analytical relation for capacity in multi-mode setting discussed in
Sec.~\ref{MemoryChannelsSec}. If the first-order approximation for
$g_1$-function is used (see Eq.~\eqref{g1v1stOrdApp}), 
then the inversion of the dependence 
$\lambda(N)$ given by Eq.~\eqref{ETlambda} gives rise to algebraic equation of
high order,
therefore
we use the quantity $\lambda^{(0)}$ also to derive $\underline C^{(1)}$.} 
the approximated quantities $\underline C^{(0)}$ and $\underline
C^{(1)}$. Analogously to Eq.~\eqref{Tdef} we will use the notation
$$
T_1^{(0)}:=\left(\lambda^{(0)}\ln2\right)^{-1}.
$$ 
Then, the thresholds $\lambda^{(0)}_{1\to2}$ and
$\lambda^{(0)}_{2\to3}$ can be defined like the
quantities~\eqref{Thr12Gen} and~\eqref{Thr23Gen}.  

Similarly to capacity (the derivatives $dR^\mathrm{(hom)}/dN$ and
$dR^\mathrm{(het)}/dN$ were defined in Subsec.~\ref{ConcavityOfSolution}) 
one can introduce the quantities
\begin{align}
&\lambda^{(\mathrm{hom})}:=\frac{dR^\mathrm{(hom)}}{dN}
=\frac\eta{\overline o_{u_\star}\ln2},
\label{ETlambdaHom}\\
&\lambda^{(\mathrm{het})}:=\frac{dR^\mathrm{(het)}}{dN}
=\frac\eta{\overline o_{u_\star}^{\mathrm{(het)}}\ln2}
\label{ETlambdaHet}
\end{align} 
for homodyne and heterodyne rates, respectively. Their threshold values
will be denoted as $\lambda^{(\mathrm{hom})}_{1\to2}$,
$\lambda^{(\mathrm{het})}_{1\to2}$ and $\lambda^{(\mathrm{het})}_{2\to3}$. 
The ``temperatures'' for rates can be defined analogously to
Eq.~\eqref{Tdef} as
\begin{align}
T_1^\mathrm{(hom)}:=\left[\lambda^\mathrm{(hom)}\ln2\right]^{-1},\quad
T_1^\mathrm{(het)}:=\left[\lambda^\mathrm{(het)}\ln2\right]^{-1}
\label{Tdefhomhet}
\end{align}
Then, in the third stage 
the quantities $N$ and $\lambda^{(\mathrm{het})}$ are related by equation
\begin{equation*}
N=T_1^\mathrm{(het)}-\frac{1-\eta}\eta N_\mathrm{env}-\frac1\eta.
\end{equation*}

It follows from Eqs.~\eqref{Noustovrel} and~\eqref{iuapproxIUO} that 
in the second stage $N$
depends on $\lambda^{(0)}$ (for capacity $\underline C^{(0)}$) as 
\begin{equation}
N=\frac12\bigl[T_1^{(0)}-\phi^{-1}+i_u-1\bigr],
\label{lambdaToN2st}
\end{equation}
where
\begin{equation}
i_u=\frac12\left[\sqrt{(\phi/4)^2+\phi\,T_1^{(0)}}-\phi/4\right].
\label{iuapproxIUOlambda}
\end{equation}
Notice the similarity between Eqs.~\eqref{N2stLamRepr} and~\eqref{lambdaToN2st}.
In fact, the first term in Eq.~\eqref{N2stLamRepr} is equal to $\overline
o_{u_\star}/\eta$, which can be rewritten as (see Eq.~\eqref{ETlambda})
$(g_1(\overline\nu)/\eta)\,T\ln2$. The latter is equal to $T_1$ if the
replacement~\eqref{Replg} is applied and $\eta$ is set to $1$.  Taking into
account the definitions \eqref{ETlambdaHom}--\eqref{Tdefhomhet} one can see that
Eqs.~\eqref{lambdaToN2st} and~\eqref{iuapproxIUOlambda} hold also for rates if
$T_1^{(0)}$ is replaced by $T_1^{(\mathrm{hom})}$ or $T_1^{(\mathrm{het})}$,
$\phi$ is given by Eq.~\eqref{Phi2ndStage} or replaced by $\phi^\mathrm{(het)}$
(see Eq.~\eqref{Phi2ndStageHet}), for homodyne and heterodyne rate,
respectively.

\begin{figure}[t]
\begin{center}
\includegraphics[scale=1]{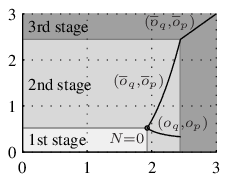}\quad\quad\includegraphics[scale=1]{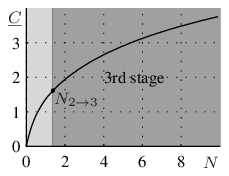}
\caption{On the left, the loci $(o_q(N),o_p(N))$ (bottom curve) 
and $(\overline o_q(N),\overline o_p(N))$ (top curve)
for values of $N\in(0,3)$ are plotted. The values of other parameters are
$\EuScript N_{\rm env}=s=1$, $\eta=0.6$.  
Different grey-scale backgrounds indicate the parameters' regions corresponding 
to different stages for given curves
(the higher the stage is, the darker the color is). 
The first stage is a single point at
$N=0$, where $(o_q,o_p)=(\overline o_q,\overline o_p)$. In the third stage the
locus $(o_q,o_p)$ is mapped into a single point (situated at the
border between the second and the third stages) for all values of $N$, since $V_{\rm in}$ does not
depend on $N$. In turn, the locus $(\overline o_q,\overline
o_p)$ in the third stage is the line $\overline o_p=\overline o_q$. On the
right, the quantity $\underline C$ is plotted vs $N$. One can see that the
dependence $\underline C(N)$ is actually concave. The values of
$N\in(0,N_{2\to3}]$ (marked using light grey color) corresponds to the second
stage, and the values of $N\in(N_{2\to3},10)$ (marked with dark grey color)
corresponds to the third stage.}
\label{CofNNG}
\end{center}
\end{figure}

\subsection{Stage transition and quantum water filling}

\begin{figure}[t]
\begin{center}
\includegraphics[scale=1]{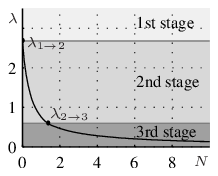}\quad\quad\includegraphics[scale=1]{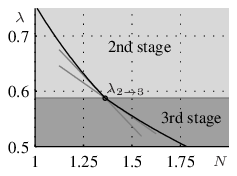}
\caption{The quantity $\lambda$ is plotted vs $N$ (the right is a magnification of the stage transition point
$\lambda_{2\to3}$). The values of other parameters are $s=1$, $\eta=0.7$,
$\EuScript N_\mathrm{env}=0.5$. 
One can see that $\lambda_{1\to2}=\lambda(N=0)<\infty$. 
On the right, the left and right tangents
are plotted at the point of $\lambda_{2\to3}=\lambda(N_{2\to3})$ 
(the left and right derivatives are
different as it follows from Eqs.~\eqref{SndDerReduced} and~\eqref{2ndDer3st} in
Appendix~\ref{App2ndDer}).}
\label{1dlambdaNG}
\end{center}
\end{figure}

Finally, let us discuss the point of stage transition. As far as different
stages correspond to solutions of different systems of Lagrange equations, it is
natural that some properties (e.g. smoothness, see
Eqs.~\eqref{SndDerReduced} and~\eqref{2ndDer3st} in
Appendix~\ref{App2ndDer}) are violated at this point. In fact, this can be seen
from Fig.\ref{CofNNG}-left, where the loci $(\overline o_q,\overline o_p)$ and
$(o_q,o_p)$ are plotted for different values of $N$ and fixed values of
$s,\eta,\EuScript N_\mathrm{env}$.  The dependence of $\overline o_p$ vs
$\overline o_q$ given by the locus $(\overline o_q,\overline o_p)$ has a kink in
the point of transition from second to third stage. Similarly, the
function $\lambda(N)$ has a kink and the function $d\lambda/dN=d^2{}^{}\underline{C}/dN^2$ is
discontinuous at this point (see Fig.\ref{1dlambdaNG}).
However, the function $\underline C(N)$ is smooth at the point of stage
transition, because its derivative~\eqref{dCdNGen} is continuous (see
Fig.\ref{CofNNG}-right). 

In the third stage we have the
equality~\eqref{auequation}, which can be written as
\begin{equation*}
\eta\,(i_u+m_u)+(1-\eta)\,e_u=\eta\,(i_{u_\star}+m_{u_\star})+(1-\eta)\,e_{u_\star}.
\end{equation*}
It means that the energy spent for modulation is distributed between
quadratures in a way to
equalize the eigenvalues of the state $\overline V_\mathrm{out}$. This type of
solution is typical for optimization problems and it appears also for classical
channels~\cite{CoverThomas}, where it was called ``water filling''.
Later such solution was shown to hold for some parameters also for quantum
channel with additive noise~\cite{NewNoiseChannel,HSH,SchaferKarpov},
where it was called ``quantum water filling''. For the case of lossy channel
this type of solution was presented in~\cite{NJP}.

Quite generally one can call ``quantum water filling'' all types of solutions
for the optimal distribution of input energy between quadratures.  It will
be shown later in Sec.~\ref{MemoryChannelsSec} for memory channels that the
input energy has to be distributed between many modes.  In addition, all modes
belonging to the third stage must possess equal average number of photons 
$\overline N_\mathrm{out}$, and for all of them equality $\overline
o_u=\overline o_{u_\star}$ must hold.  Furthermore, if almost all modes are in the
third stage, the solution can be interpreted as a small perturbation of water
filling. Thus, the term ``quantum water filling'' used for all types of
solutions underlines the ``physical'' meaning of the performed optimization.

\section{Role of channel parameters}\label{CritParsSec}

In this section we discuss the dependence from parameters of 
capacity and rates found in Sec.~\ref{MLChannel} (\emph{i.e.} for single channel
use). Apart from characterizing the one-shot capacity this study is 
also relevant for the case of multiple channel uses and additivity problem
discussed below in Sec.~\ref{MemoryChannelsSec}.

It is evident that both capacity and rates must be monotonic functions of
parameters $\eta$, $N$ and $\EuScript{N}_\mathrm{env}$. In fact, higher
transmissivity and input energy cannot result to less capacity or rates from
physical point of view. In
addition, it was explicitly shown in Subsec.~\ref{ConcavityOfSolution} that
both capacity and rates are monotonic concave functions of $N$. 

In turn, monotonic dependence of capacity from $\EuScript{N}_\mathrm{env}$ can be
shown as follows. Given the value
$\EuScript{N}_\mathrm{env}'>\EuScript{N}_\mathrm{env}$ the lossy channel for the
parameters $s$, $\eta$ and $\EuScript{N}_\mathrm{env}'$ can be
represented as a channels composition $\mathcal G_\mathrm{N}\circ\mathcal
G_\mathrm{L}$, where $\mathcal G_\mathrm{L}$ is a lossy channel with parameters $s$,
$\eta$, $\EuScript{N}_\mathrm{env}$ and $\mathcal G_\mathrm{N}$ is an additive
(classical) noise channel (see Eqs.~\eqref{VoutInitAN} and~\eqref{VoutAvAN}) with
environment matrix 
\begin{equation*}
V_\mathrm{env}=(1-\eta)(\EuScript{N}_\mathrm{env}'-\EuScript{N}_\mathrm{env})
\left(
\begin{array}{cc}
e^s&0\\
0&e^{-s}
\end{array}
\right).
\end{equation*}
Since the capacity of the composition of two channels cannot exceed that of each
individual channel, we deduce that the capacity is non-increasing
function of $\EuScript{N}_\mathrm{env}$. Furthermore,
the following \emph{environment purity theorem} states
that the optimal $\EuScript{N}_\mathrm{env}$ is zero:

\bigskip

\begin{theorem}\label{EnvPurTheor}
The maximum of capacity on the set of environment states $\{V_\mathrm{env}\}$
whose elements have the same average amount of photons $N_\mathrm{env}$
is achieved on pure environment state, \emph{i.e.}
$e_{u}e_{u_\star}=1/4.$
\end{theorem}

\bigskip

\begin{IEEEproof}
Proof is given in Appendix~\ref{EnvPurTh}.
\end{IEEEproof}

\bigskip

Extension of this theorem to the case of rates is straightforward.

Thus, the only parameter which can make capacity and rates non-monotonic is
the environment squeezing $s$. In this section we investigate this
non-monotonic dependence. Below, the subsections~\ref{RoleOfInpAndEnvSqSubsec}
and~\ref{RoleOfTransForCapSubsec} are mainly devoted to definitions, properties
and numerical results on channel parameters, while the other subsections contain
analytical results justifying the numerics.

\begin{figure}[t]
\begin{center}
\includegraphics[scale=1]{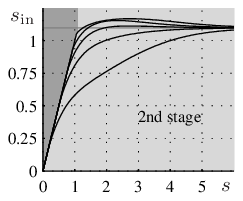}\quad\quad\includegraphics[scale=1]{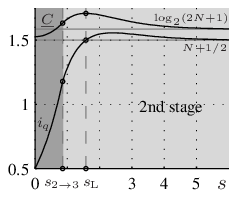}
\caption{On the left, the optimal input squeezing
$s_\mathrm{in}$ is plotted vs $s$, for values of $\eta$ going from $0.1$ (bottom
curve) to $0.9$ (top curve) with step $0.2$. The values of the other parameters
are $N=1$, $\EuScript N_{\rm env}=0$. On the right, both the capacity
$\underline C$ and optimal input eigenvalue $i_q$ are plotted vs $s$. The
value of the other parameters are $\EuScript N_\mathrm{env}=0$, $N=1$, $\eta=0.6$.}
\label{cofs_ropts}
\end{center}
\end{figure}

\begin{figure}[t]
\begin{center}
\includegraphics[scale=1]{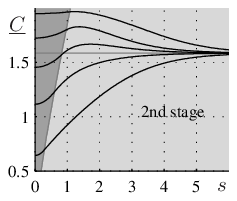}\quad\quad\includegraphics[scale=1]{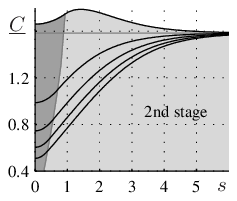}
\caption{On the left, capacity $\underline{C}$ 
vs $s$, for values of $\eta$ going from $0.15$ (bottom
curve) to $0.95$ (top curve) with step $0.2$. The values of the other parameters
are $N=1$, $\EuScript N_{\rm env}=0$. On the right, capacity
$\underline C$ vs $s$ for values of $\EuScript N_\mathrm{env}$ going from 0 (top curve) 
to 4 (bottom curve) with step 1. The values of the other parameters
are $\eta=0.7$, $N=1$.}
\label{cofs_family}
\end{center}
\end{figure}

\begin{figure}[t]
\begin{center}
\includegraphics[scale=1]{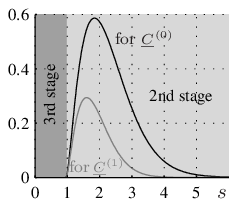}\quad\quad\includegraphics[scale=1]{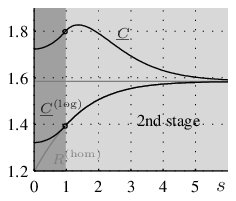}
\caption{On the left, the quantities 
$\bigl[(\underline C-\underline C^{(0)})/\underline C\bigr]\times100\%$
(black) and
$\bigl[(\underline C-\underline C^{(1)})/\underline C\bigr]\times100\%$ (grey)
are plotted vs $s$ for
$\eta=0.75$, $N=1$, $\EuScript N_\mathrm{env}=0$. On the right, the
quantities $\underline{C}$, $\underline{C}^{(\log)}$ and $R^\mathrm{(hom)}$ are
plotted vs $s$ for $\eta=0.75$, $N=1$, $\EuScript N_\mathrm{env}=0$.}
\label{cLogGraphs}
\end{center}
\end{figure}

\subsection{Role of input and environment squeezing}\label{RoleOfInpAndEnvSqSubsec}

Using the representation~\eqref{ind1mode} for input covariance matrix
$V_\mathrm{in}=V(\EuScript{N}_\mathrm{in},s_\mathrm{in})$, one can relate the
optimal degree of input squeezing $s_\mathrm{in}$ to the degree of environment
squeezing $s$. It follows from Eqs.~\eqref{squeezequality}
and~\eqref{iuapproxiu} that $s_\mathrm{in}=s$ for the third stage
(for $\underline C$, $\underline C^{(0)}$, $\underline C^{(1)}$ and $\underline C^{(\log)}$) and
\begin{equation}
s_\mathrm{in}=\ln\left[\sqrt{1+(2N+1)\phi+\phi^2/4}-\phi/2\right]
\label{ropts1use}
\end{equation}
for the second stage (for $\underline C^{(0)}$ and $\underline C^{(\log)}$). 
Analogously,
it follows from Eq.~\eqref{squeezequality} that
$s_\mathrm{in}=s^{(\mathrm{het})}$ (see Eq.~\eqref{sHet}) for the heterodyne
rate in the third stage.  In the second stage both homodyne and heterodyne rates
result to the same relation~\eqref{ropts1use}, where the
replacement~\eqref{Replphi} must be applied for the heterodyne case. At the
transition point between different stages there is a \emph{kink} in the function
$s_\mathrm{in}(s)$ (see Fig.\ref{cofs_ropts}-left).  It reflects the fact that different stages
correspond to solution of different systems of equations. The dependence
$i_q(s)$ is shown in Fig.\ref{cofs_ropts}-right (this is discussed in the 
following subsections in a more detailed way).

\begin{figure}[t]
\begin{center}
\includegraphics[scale=1]{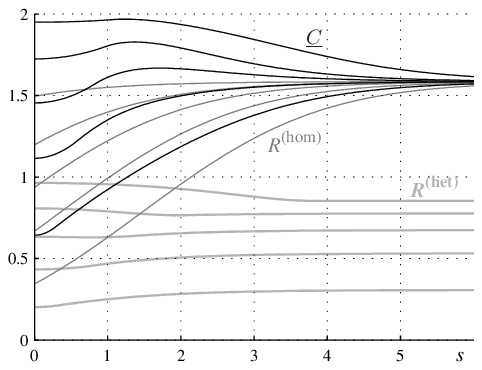}
\end{center}
\caption{Classical capacity $\underline{C}$ (solid curves), homodyne
$R^\mathrm{(hom)}$ (thin grey curves) and heterodyne  $R^\mathrm{(het)}$ (bold grey
curves) rates vs $s$, for values of $\eta$ going from $0.15$ (bottom curve) to
$0.95$ (top curve) with step $0.2$. The values of the other parameters are
$N=1$, $\EuScript N_{\rm env}=0$.}
\label{ratesFig}
\end{figure}

\begin{figure}[t]
\begin{center}
\includegraphics[scale=1]{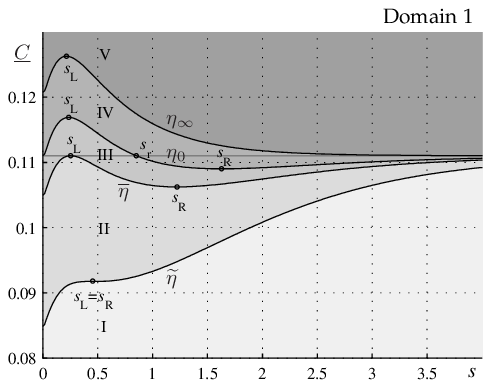}
\caption{
The dependence $C(s)$ for channel parameters $N$ and $\EuScript N_\mathrm{env}$
belonging to the \emph{first domain} ($N=0.04$, $\EuScript N_\mathrm{env}=0.001$)
is plotted for values of transmissivity $\eta=0.28$, $0.359$, $0.384$ and
$0.424$ (from bottom to top), which approximately correspond to border values between different
\emph{regimes} ($\widetilde\eta$, $\overline\eta$, 
$\eta_{\scriptscriptstyle 0}$ and $\eta_\infty$,
respectively). Regimes are indicated with roman numbers (I--V) and different
gray scale colors. Any curve $C(s)$ corresponding to a particular regime 
would completely lie in the area with the background color corresponding to that regime.}
\label{porogi_domain_1}
\end{center}
\end{figure}

\begin{figure}[t]
\begin{center}
\includegraphics[scale=1]{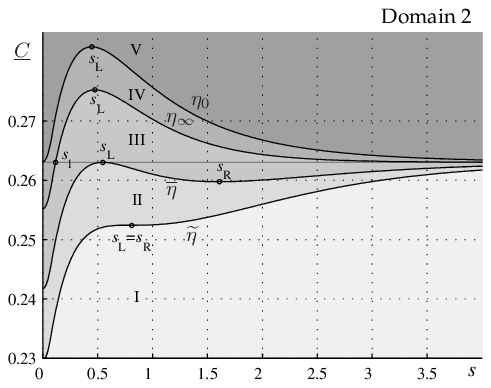}
\caption{
The dependence $C(s)$ for channel parameters $N$ and $\EuScript N_\mathrm{env}$
belonging to the \emph{second domain} ($N=0.1$, $\EuScript N_\mathrm{env}=0$)
is plotted for values of transmissivity $\eta=0.369$, $0.394$, $0.423$, $0.44$ 
(from bottom to top), which approximately correspond to border values between different
\emph{regimes} ($\widetilde\eta$, $\overline\eta$, $\eta_\infty$ and $\eta_{\scriptscriptstyle 0}$,
respectively).  Regimes are indicated with roman numbers (I--V) and different
gray scale colors. Any curve $C(s)$ corresponding to a particular regime 
would completely lie in the area
with the background color corresponding to that regime.}
\label{porogi_domain_2}
\end{center}
\end{figure}

\begin{figure}[t]
\begin{center}
\includegraphics[scale=1]{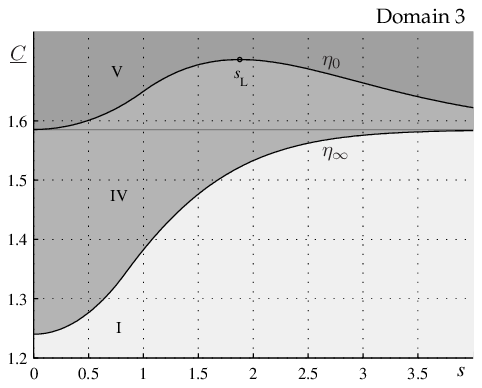}
\caption{
The dependence $C(s)$ for channel parameters $N$ and $\EuScript N_\mathrm{env}$
belonging to the \emph{third domain} ($N=1$, $\EuScript N_\mathrm{env}=1$)
is plotted for values of transmissivity $\eta=0.808$ (bottom curve) and
$0.919$ (top curve), which approximately correspond to border values between different
\emph{regimes} ($\eta_\infty$ and $\eta_{\scriptscriptstyle 0}$, respectively).  
Regimes are indicated with roman numbers (I, IV and V: other regimes do not
exist in the third domain) and different
gray scale colors. Any curve $C(s)$ corresponding to particular regime 
would completely lie in the area
with the background color corresponding to that regime.}
\label{porogi_domain_3}
\end{center}
\end{figure}

\begin{figure}[t]
\begin{center}
\includegraphics[scale=1]{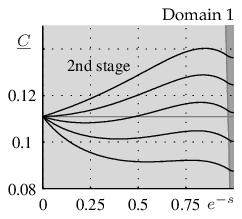}\quad\hspace{7pt}\includegraphics[scale=1]{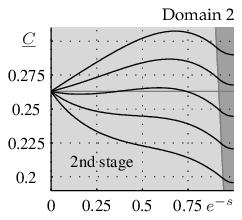}\\
\hspace{7pt}\includegraphics[scale=1]{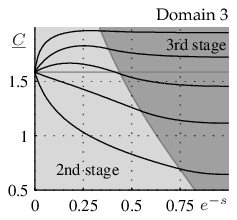}\quad\hspace{7pt}\includegraphics[scale=1]{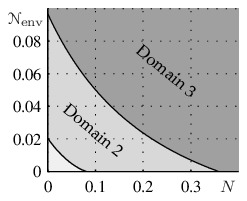}
\caption{The dependence $\underline C(e^{-s})$ for different values of $\eta$
and fixed values of $N$ and $\EuScript N_\mathrm{env}$. In particular, 
$N=0.04$ and $\EuScript N_\mathrm{env}=0$ (corresponding to the first domain), 
$\eta$ goes from $0.275$ (bottom) to $0.475$ (top) with step $0.05$ (top-left figure);
$N=0.1$ and $\EuScript N_\mathrm{env}=0$ (corresponding to the second domain), 
$\eta$ goes from $0.3$ (bottom) to $0.5$ (top) with step $0.05$ (top-right figure); 
$\EuScript N_\mathrm{env}=0$ and $N=1$ (corresponding to the third domain), 
$\eta$ goes from $0.15$ (bottom) to $0.95$ (top) with step $0.2$ (bottom-left figure). 
The parts corresponding to different backgrounds belongs to
different stages (lighter color states for second stage, and darker color states for
third stage). Bottom-right:
the loci $(N_0,\EuScript N_\mathrm{env,0})$ (bottom curve) and 
$(\widetilde N,\widetilde{\EuScript N}_\mathrm{env})$ (top curve), corresponding
to transitions between different channel domains. 
The points $(N,\EuScript N_\mathrm{env})$ belonging to the area between these curves
correspond to the domains indicated with different grey scale backgrounds
colors.}
\label{expScaleCofs}
\end{center}
\end{figure}

\begin{figure}[t]
\begin{center}
\includegraphics[scale=1]{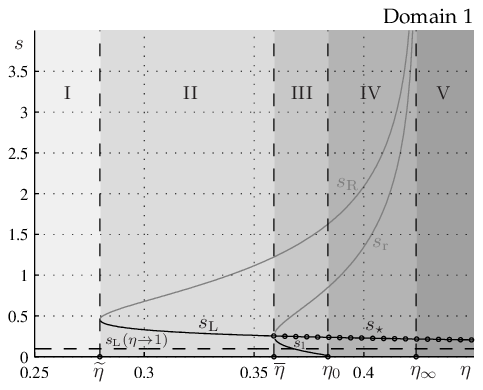}
\caption{The dependence of the quantities 
$s_\mathrm{r}$, $s_\mathrm{l}$, $s_\mathrm{R}$, $s_\mathrm{L}$ and $s_\star$
vs $\eta$ for capacity $\underline C$. The value of the other
parameters are $N=0.04$ and $\EuScript N_\mathrm{env}=0.001$ 
(corresponding to the first domain). Different
gray scale backgrounds corresponds to transmissivities $\eta$ from different
regimes (indicated with roman numbers). Vertical asymptotes are plotted for the
critical transmissivities, and the horizontal asymptote shows the limit
$\lim_{\eta\to1}s_\mathrm{L}$. The part of the curve $s_\mathrm{L}(\eta)$ coinciding
with $s_\star(\eta)$ is shown with dots. At the point $\eta=\overline\eta$ the
quantity  $s_\star$ jumps to infinity and for all values of
$\eta\leqslant\overline\eta$ is equal to infinity.}
\label{chan1sq}
\end{center}
\end{figure}

\begin{figure}[t]
\begin{center}
\includegraphics[scale=1]{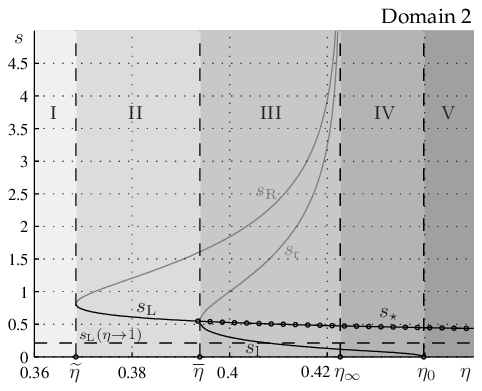}
\caption{
The dependence of the quantities 
$s_\mathrm{r}$, $s_\mathrm{l}$, $s_\mathrm{R}$, $s_\mathrm{L}$ and $s_\star$
vs $\eta$ for capacity $\underline C$. The value of the other
parameters are $N=0.1$ and $\EuScript N_\mathrm{env}=0$ 
(corresponding to the second domain). Different
gray scale backgrounds corresponds to transmissivities $\eta$ from different
regimes (indicated with roman numbers). Vertical asymptotes are plotted for the
critical transmissivities, and the horizontal asymptote shows the limit
$\lim_{\eta\to1}s_\mathrm{L}$. The part of the curve $s_\mathrm{L}(\eta)$ coinciding
with $s_\star(\eta)$ is shown with dots. At the point $\eta=\overline\eta$ the
quantity  $s_\star$ jumps to infinity and for all values of
$\eta\leqslant\overline\eta$ is equal to infinity.}
\label{chan2sq}
\end{center}
\end{figure}

\begin{figure}[t]
\begin{center}
\includegraphics[scale=1]{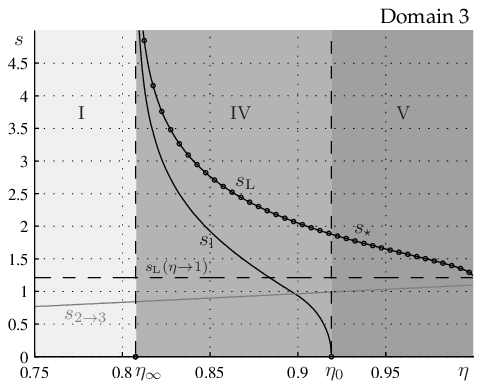}
\caption{
The dependence of the quantities 
$s_\mathrm{r}$, $s_\mathrm{l}$, $s_\mathrm{R}$, $s_\mathrm{L}$, $s_\star$
and $s_{2\to3}$
vs $\eta$ for capacity $\underline C$. The value of the other
parameters are $N=1$ and $\EuScript N_\mathrm{env}=1$ 
(corresponding to the third domain). Different
gray scale backgrounds corresponds to transmissivities $\eta$ from different
regimes (indicated with roman numbers, the second and the third regimes do not
exist). Vertical asymptotes are plotted for the
critical transmissivities, and the horizontal asymptote shows the limit
$\lim_{\eta\to1}s_\mathrm{L}$. The part of the curve $s_\mathrm{L}(\eta)$ coinciding
with $s_\star(\eta)$ is shown with dots ($s_\star$ coincides with
$s_\mathrm{L}$ on the whole region of transmissivities where $s_\mathrm{L}$ is
defined, \emph{i.e.} for $\eta\in(\eta_\infty,1)$, and $s_\star$ equals infinity if
$\eta\in(0,\eta_\infty)$. At the point $\eta=\overline\eta$ the
quantity  $s_\star$ together with $s_\mathrm{L}$ asymptotically tends to
infinity. Also notice, that $s_{2\to3}\neq s_\mathrm{L}$ in the limit
$\eta\to1$.}
\label{chan3sq}
\end{center}
\end{figure}

\begin{figure}[t]
\begin{center}
\includegraphics[scale=1]{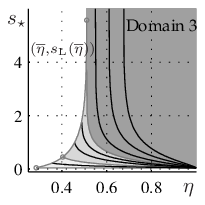}\quad\quad\includegraphics[scale=1]{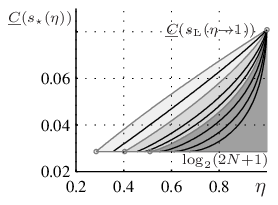}
\caption{Optimal environment squeezing $s_\star$ (left) and capacity
$\underline{C}(s_\star(\eta))$ (right) are plotted  vs $\eta$ for values of
$\EuScript N_\mathrm{env}$ equal to $0$, $0.0055$, $0.0165$, $0.0413$, $0.066$,
$0.0898$, $0.1403$, $0.2393$ and $0.3879$ (from bottom to top at left, and from
top to bottom at right).  The value of the other parameter is $N=0.01$.  Curves,
corresponding to different backgrounds belongs to different domains (darker
background corresponds to higher domain). The curves corresponding to transition
between different domains are plotted using grey color. At the left: each curve
belonging to the first or the second domain jumps to infinity at some finite
value and equals infinity for all transmissivities to the left of that jump.
Then, curves, corresponding to the third domain tend asymptotically to infinity
and are equal to infinity for all values of $\eta$ which are to the
left of that asymptote.  The whole region occupied by the finite dependences
$s_\star(\eta)$ is bounded to the left by the locus
$(\overline\eta,s_\mathrm{L}(\overline\eta))$. The values of
$s_\star(\min\eta)<\infty$ corresponding to borders between different domains
are indicated with grey points.  It is interesting to note that the area
corresponding to the second domain is bounded by finite value from the top,
\emph{i.e.} for the values of $N$ and $\EuScript N_\mathrm{env}$ corresponding
to transition from second to third domain, the value of $s_\star$ is
still finite at the point $\eta=\overline\eta$.  The area occupied by curves is
bounded from the bottom by the curve $s_\star(\eta)$ for $\EuScript
N_\mathrm{env}=0$ which is not zero. 
At the right: the whole region occupied by family of possible curves
$\underline{C}(s_\star(\eta))$ is bounded from the bottom by the
limit~\eqref{Climit}. Here we plotted capacity corresponding to finite values of
$s_\star$ (for small values of $\eta$ we have $s_\star=\infty$, therefore each
curve $\underline{C}(s_\star(\eta))$ is equal to $\log_2(2N+1)$ -- this is not
plotted).  By increasing $\eta$ from zero to one, we reach the point of
$\eta=\overline\eta$ (for the first and second domain) or $\eta=\eta_\star$ (for
the third domain) where the curve $\underline{C}(s_\star(\eta))$ is detached from
horizontal line $\log_2(2N+1)$. And finally, when $\eta$ tends to $1$, all
curves $\underline{C}(s_\star(\eta))$ tend to the same value
$\lim_{\eta\to1}\underline C(s_\mathrm{L}(\eta))$. 
One can see from numerics that this value does not depend on $\EuScript N_\mathrm{env}$.}
\label{sNenv_Eta}
\end{center}
\end{figure}

\begin{figure}[t]
\begin{center}
\includegraphics[scale=1]{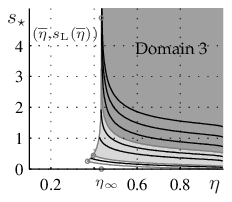}\quad\includegraphics[scale=1]{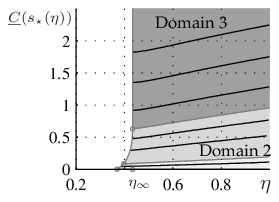}
\caption{
Optimal environment squeezing $s_\star$ (left) and capacity
$\underline{C}(s_\star(\eta))$ (right) are plotted  vs $\eta$ for values of $N$ 
equal to $10^{-6}$, $0.0149$, $0.0298$, $0.1127$, $0.1956$, $0.2755$, $0.4443$, $0.7760$, $1.2734$
(from bottom to top for both: left and right graphs).  
The value of the other parameter is $\EuScript N_\mathrm{env}=0.01$.
Curves, corresponding to different backgrounds belongs to different domains
(darker background corresponds to higher number of domain). The curves
corresponding to transition between different domains are plotted using grey
color. At the left: each curve belonging to the first or the second domain jumps
to infinity at some finite value and equals infinity for all transmissivities
which are to the left of that value. 
Then, curves, corresponding to the third domain tend asymptotically to
infinity (when $\eta$ tends to $\eta_\infty$ from the right)
and are equal to infinity to the left of that asymptote. 
The whole region occupied by the finite
dependences $s_\star(\eta)$ is bounded to the left by the locus
$(\overline\eta,s_\mathrm{L}(\overline\eta))$. The values of $s_\star(\min\eta)<\infty$
corresponding to borders between different domains are
indicated with grey points. It is interesting to note that the area
corresponding to the second domain is bounded by the finite value from the top,
\emph{i.e.} for the values of $N$ and $\EuScript N_\mathrm{env}$ corresponding
to transition from the second to the third domain, the value of $s_\star$ is
still finite at the point $\eta=\overline\eta$. 
The area occupied by curves is
bounded from the bottom by the curve $s_\star(\eta)$ for $N\to0$ 
which is not zero. 
One can see that in the third domain 
the value of $\eta=\eta_\infty$ is the same for all curves. This is in fact in
agreement with the analytical result (obtainied in subsequent
subsections) that $\eta_\infty$ does not depend on $N$.
At the right: the whole region occupied by family of possible curves
$\underline{C}(s_\star(\eta))$ is bounded from the bottom by the zero.
Here we plotted the capacity corresponding to finite values of
$s_\star$ (for small values of $\eta$ we have $s_\star=\infty$, therefore each
curve $\underline{C}(s_\star(\eta))$ should be continued horizontally to the
left being at the same level as in left border 
-- this is not plotted).}
\label{sN_Eta}
\end{center}
\end{figure}

\begin{figure}[t]
\begin{center}
\includegraphics[scale=1]{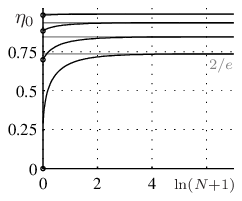}\quad\quad\includegraphics[scale=1]{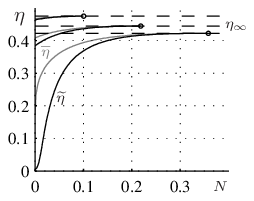}
\includegraphics[scale=1]{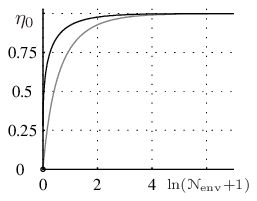}\quad\includegraphics[scale=1]{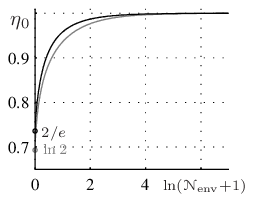}
\caption{Top-left: the quantity $\eta_{\scriptscriptstyle 0}$ is plotted vs $\ln(N+1)$ for the values
of $\EuScript N_\mathrm{env}$ equal to $0$, $0.2$, $1$, $10$ (from bottom to top).
Top-right: the quantitues $\widetilde\eta$ (black solid), $\overline\eta$
(grey) and $\eta_\infty$ (horizontal dashed black lines)
are plotted vs $N$ for the values  $\EuScript N_\mathrm{env}$ equal to $0$,
$0.02$, $0.05$ (from bottom to top for all curves). The points where
$\widetilde\eta$ and $\overline\eta$ touch the line corresponding to
$\eta_\infty$ are indicated with bold points.
Bottom-left: function $\lim_{N\to0}\eta_{\scriptscriptstyle 0}$ vs $\ln(\EuScript N_\mathrm{env}+1)$ 
as exact (black curve, see Eq.~\eqref{ETA0N0}) and approximate (grey curve, see
Eq.~\eqref{ETA0N0approx}) quantities.
Bottom-right: function $\lim_{N\to\infty}\eta_{\scriptscriptstyle 0}$ 
as exact (black curves, Eq.~\eqref{NInf}) 
and approximate (grey curve, see Eq.~\eqref{lambert}) quantities. The values
of all the quantities for all the graphs for zero argument are shown by bold
point.}
\label{eta0ref}
\end{center}
\end{figure}

\begin{figure}[t]
\begin{center}
\includegraphics[scale=1]{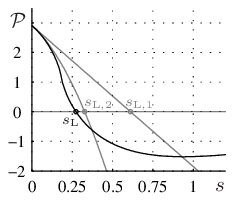}\quad\quad\includegraphics[scale=1]{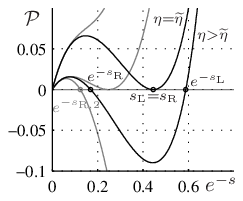}
\includegraphics[scale=1]{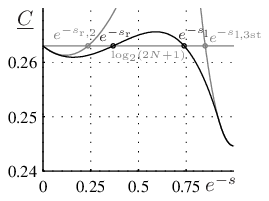}\quad\includegraphics[scale=1]{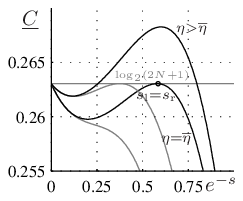}
\caption{Top-left (analytical method of estimation of $s_\mathrm{L}$): 
quantity $\mathcal P$ vs $s$ together with its linea and
quadratic approximations at the point $s=0$. The value of parameters are
$\eta=0.8$, $N=0.1$, $\EuScript N_\mathrm{env}=0$.
Top-right (analytical method of estimation of $\widetilde\eta$):
quantity $\mathcal P$ vs $e^{-s}$ for the values
$\eta=0.3985$ (bottom black curve, its quadratic and cubic approximations at point
$e^{-s}=0$ through partial Taylor sum -- grey color) and $\eta=0.3685$ (top black curve and
its third approximation at point $e^{-s}=0$ -- grey color). The value of other
parameters are $N=0.1$, $\EuScript N_\mathrm{env}=0$.
The points, where quadratic and linear approximations corss the line $\mathcal P=0$ defines
quadratic and liner approximations for the corresponding $s$-quantities:
approximations $s_\mathrm{L,2}$ and $s_\mathrm{L,1}$ for the quantity
$s_\mathrm{L}$, and approximation $s_\mathrm{R,2}$ for the quantity $s_\mathrm{R}$.
Similarly, the value of transmissivity at which cubic approximation has two
roots (touches the line $\mathcal P=0$) defines approximation for
$\widetilde\eta$.
Bottom-left (method of estimation of $s_\mathrm{l}$ and $s_\mathrm{r}$): 
$\underline C$ vs $e^{-s}$ for the values of $\eta=0.4$, $N=0.1$ and $\EuScript N_\mathrm{env}=0$.
Bottom-right (method of estimation of $\overline\eta$): $\underline C$ vs
$e^{-s}$ for the values of $\eta=0.3939$ (bottom black curve) and
$\eta=0.4062$ (top black curve). The value of other parameters are $N=0.1$, $\EuScript N_\mathrm{env}=0$.
For the top black curve its cubic approximation gives an estimation for the
quantity $\overline\eta$.}
\label{sLestref}
\end{center}
\end{figure}

\begin{figure}[t]
\begin{center}
\includegraphics[scale=1]{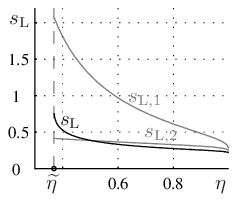}\quad\quad\includegraphics[scale=1]{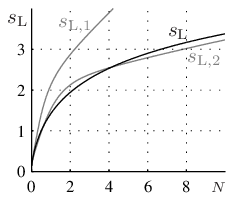}
\includegraphics[scale=1]{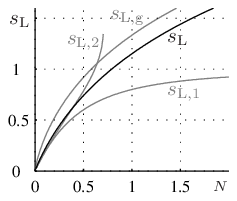}\quad\quad\includegraphics[scale=1]{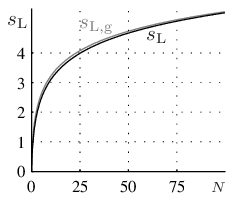}
\caption{Top: the dependence of the quantities $s_\mathrm{L}$, $s_\mathrm{L,1}$ and
$s_\mathrm{L,2}$ vs $\eta$ (left) and vs $N$ (right). The value the other
parameters are: $N=0.1$ and $\EuScript N_\mathrm{env}=0$ (left),
$\eta=0.9$ and $\EuScript N_\mathrm{env}=0.1$ (right).
Bottom (left and right): the quantity
$\lim_{\eta\to1}s_\mathrm{L}$ vs $N$ (the value of $\EuScript N_\mathrm{env}$
was set to zero, but numerically this limit does not depend on $\EuScript N_\mathrm{env}$).
To the left it is plotted together with the approximations $s_\mathrm{L,1}$,
$s_\mathrm{L,2}$, $s_\mathrm{L,g}=\frac23g(N)$, and to the right it is plotted
together only with $s_\mathrm{L,g}=\frac23g(N)$ (see Eqs.~\eqref{sLlim} and~\eqref{sLlim2}). 
Approximations  $s_\mathrm{L,1}$ and $s_\mathrm{L,2}$ are plotted only for those
values of the argument where they are applicable.}
\label{sLdepref}
\end{center}
\end{figure}

The capacity $\underline{C}$ found by the exact analytical
solution is shown in Fig.\ref{cofs_family} for fixed $N$ as function of~$s$
and for different values of $\eta$ (at left) and $\EuScript N_\mathrm{env}$ (at
right).  One can see that the squeezed environment
($s\neq 0$) may result to capacity enhancement. This phenomenon shows similarity
with the improvement of the signal to noise ratio achieved by squeezed vacuum
injection in an optical wave-guide tap~\cite{tap}.  The highest enhancement
occurs at either finite value of~$s$ or at $s\to\infty$ depending on the value
of~$\eta$.  In any case, the capacity in the limit of large~$s$ becomes only
function of the energy constraint~$N$ (see Eq.~\eqref{Climit}) explaining why
all curves $\underline{C}(s)$ flow together to the same value\footnote{This
behavior originally was observed in~\cite{PRA} for the capacity of particular
memory channel found as maximum over a small subset of Gaussian states.} 
when $s\rightarrow\infty$. Similarly,
$s_\mathrm{in}$ tends to the value~\eqref{noiselessiu} for $s\to\infty$ 
because it follows from Eq.~\eqref{ropts1use} (see also Fig.\ref{cofs_ropts}). 

The difference among quantities $\underline C^{(0)}$, $\underline C^{(1)}$
and $\underline C$ is shown in Fig.\ref{cLogGraphs}-left. In
Fig.\ref{cLogGraphs}-right the quantities $\underline C$, $\underline
C^{(\log)}$ and $R^\mathrm{(hom)}$ are shown together. One can see that
$\underline C^{(\log)}$ coincides with homodyne rate in the second stage (see
Eq.~\eqref{HomLim}).

The rates $R^{(\mathrm{hom})}$ and $R^{(\mathrm{het})}$ together with the exact
solution for capacity are shown in Fig.\ref{ratesFig} for fixed $N$ as functions
of~$s$ and for different values of $\eta$. One can see that in the second stage
both rates are monotonically growing functions of $s$ which is in agreement with
the Eq.~\eqref{SndStF}. In the third stage the heterodyne rate may be
non-monotonic achieving its minimum. As it can be seen from Fig.\ref{ratesFig},
the optimal heterodyne rate is achieved at either $s\to\infty$ or $s=0$.
Analytical description of this behavior is given in Subsec.\ref{critparhetsec}.
A similar (to Fig.\ref{ratesFig}) family of curves can be obtained if
$R^\mathrm{(het)}$ or $R^\mathrm{(hom)}$ is plotted versus $s$
for different values of $\EuScript{N}_\mathrm{env}$ and fixed $\eta$.  One can
see that the universal limit~\eqref{Climit} holds also for this case.

Despite the behavior shown in Fig.\ref{cofs_family} is the most typical, there are
parameters values giving more complicated dependence for $\underline{C}(s)$ (all possible
cases are plotted at Figs.\ref{porogi_domain_1}, \ref{porogi_domain_2},
\ref{porogi_domain_3} and~\ref{expScaleCofs}). In particular, the capacity may have both
minimum and maximum each of them attained at finite environment squeezing $0<s<\infty$.
Such behavior and the parameters related with its description are discussed in the
following subsections.

\subsection{Role of transmissivity for capacity}\label{RoleOfTransForCapSubsec}

It was shown in Fig.\ref{cofs_family} that both $\eta$ and $\EuScript
N_\mathrm{env}$ can be chosen to parametrize the family of curves
$\underline C(s)$.  In order to completely characterize how capacity depends on
squeezing we will use $\eta$. All ``qualitative'' possibilities for the
dependence $\underline C(s)$ are shown in Figs.\ref{porogi_domain_1},
\ref{porogi_domain_2}, \ref{porogi_domain_3} and~\ref{expScaleCofs}.  
Such a dependence can be interpreted as crossing different \emph{regimes} by increasing $\eta$
from zero to one. In turn, the set of regimes depend on the \emph{domain} which
$N$ and $\EuScript N_\mathrm{env}$ values belong to (see
Fig.\ref{expScaleCofs}-bottom-right). Let us consider this behavior in more detail
(in the relations below the argument of $\underline{C}$ is assumed to be $s$).

Let us define the specific values of squeezing and transmissivity
in a formal way. First we notice that from numerical calculations it results:

\bigskip

\begin{propos}\label{Only2ExtremaCofs}
The function $\underline C(s)$ may have at maximum two extrema for the values of
squeezing $0<s<\infty$. 
\end{propos}

\bigskip

Evidently, if the function $\underline C(s)$ has two extrema, then one of them
must be maximum and the other minimum. They can be formally defined as
follows.

\bigskip

\begin{definition}\label{sLsRDef}
The finite positive value of squeezing $s$ will be denoted by $s_\mathrm{L}$ or
$s_\mathrm{R}$ if the function $\underline C(s)$ has its local maximum or minimum
at that values, respectively:
\begin{align*}
&\underline C'(s_\mathrm{L})=0,&\underline C''(s_\mathrm{L})<0,\\
&\underline C'(s_\mathrm{R})=0,&\underline C''(s_\mathrm{R})>0.
\end{align*}
\end{definition}

\bigskip

One of our purpose is to study the value of squeezing giving highest capacity,
which can be defined as written below.

\bigskip

\begin{definition}\label{sOptDef}
The value of squeezing $s$ will be denoted by $s_\star$ and called
\emph{optimal} if it corresponds to global maximum of the function $\underline
C(s)$: 
\begin{equation*}
\underline C(s_\star)=\max_{0\leqslant s\leqslant\infty}\underline C(s).
\end{equation*}
\end{definition}

\bigskip

Finite values of squeezing providing the same value of capacity as infinite 
squeezing can exist:

\bigskip

\begin{definition}\label{slsrDef}
The finite positive value of squeezing $s$ will be denoted by $s_\mathrm{l}$ 
($s_\mathrm{r}$) if both the value of function $\underline C(s)$ at that
squeezing coincides with the value at infinity and $\underline C(s)$ is
increasing (decreasing) function at this point:
\begin{align*}
&\underline C(s_\mathrm{l})=\underline C(\infty),&\underline C'(s_\mathrm{l})>0,\\
&\underline C(s_\mathrm{r})=\underline C(\infty),&\underline C'(s_\mathrm{r})<0.
\end{align*}
\end{definition}

\bigskip

As far as $\underline C(s)$ belongs to the third stage for small values of $s$
and $\underline C(s)$ is always increasing function of $s$ in the third stage,
the extremum corresponding to the smallest value of squeezing must be the
maximum.  Hence, the minimum must correspond to higher value of squeezing which
exists only if the maximum does. This also makes the function $\underline C(s)$
increasing in $s_\mathrm{l}$ and decreasing in $s_\mathrm{r}$. Thus, if both
extrema exist, we have $s_\mathrm{L}<s_\mathrm{R}$ and
$s_\mathrm{l}<s_\mathrm{r}$ which explains the notations introduced in the
definitions~\ref{sLsRDef} and~\ref{slsrDef}). 
In the general case it follows from numerical results that the function
$\underline C(s)$ in the interval $0<s<\infty$ can be one of the following:
\begin{itemize}
\item Monotonic function without stationary points.
\item Monotonic function with single saddle-point.
\item Function with one maximum. 
\item Function with one maximum and one minimum.
\end{itemize}
In order to study the case with the saddle-point we define the following
transmissivity:

\bigskip

\begin{definition}\label{SaddlePointDef}
The transmissivity $\eta$ will be denoted by $\widetilde\eta$ and 
called \emph{saddle-point transmissivity}
if $\underline C(s)$ has saddle-point in some finite positive value of
squeezing:
\begin{equation*}
\exists\;\widetilde s\in(0,\infty)\;|\;
\underline C'(\widetilde s)=\underline C''(\widetilde s)=0.
\end{equation*}
\end{definition}

\bigskip

It follows from the results of numerical study that $\widetilde\eta$ exists if
and only if the following transmissivity does:

\bigskip

\begin{definition}\label{OverlEtaDef}
The transmissivity $\eta$ will be denoted by $\overline\eta$ and 
called \emph{$\overline\eta$-transmissivity}
if a finite positive value of squeezing exists such that
$\underline C(s)$ has maximum at that squeezing and the value of maximum is 
equal to $C(\infty)$:
\begin{equation*}
\exists\;\overline s\in(0,\infty)\;|\;
\underline C(\overline s)=\underline C(\infty),\;
\underline C'(\overline s)=0,\;\underline C''(\overline s)<0.
\end{equation*}
\end{definition}

\bigskip

Then, by considering the behavior of $\underline C(s)$ for zero and infinite
squeezings the following definitions can be introduced.

\bigskip

\begin{definition}\label{Eta0Def}
The transmissivity $\eta$ will be denoted by $\eta_0$ and 
called \emph{$\eta_0$-transmissivity}
if the values of $\underline C(s)$ for zero and infinite squeezing coincide:
$\underline C(0)=\underline C(\infty)$.
\end{definition}

\bigskip

\begin{definition}\label{EtaInfDef}
The transmissivity $\eta$ will be denoted by $\eta_\infty$ and 
called \emph{$\eta_\infty$-transmissivity}
if $\underline C'(\infty)=0$, \emph{i.e.} in neighborhood of infinite squeezing 
$\underline C(s)$ is decreasing for $\eta>\eta_\infty$ and increasing
for $\eta<\eta_\infty$.
\end{definition}

\bigskip

One can note that some properties (e.g. saddle-point) can be observed only for
particular ``domains'' of the parameters $N$ and $\EuScript N_\mathrm{env}$,
which requires to introduce further classification. It follows from the results
of numerical study that the following definitions allow to divide the quadrant
$(N>0$, $\EuScript N_\mathrm{env}\geqslant0)$ into three non-overlapping
\emph{domains} (see Fig.\ref{expScaleCofs}-bottom-right), thus providing
consistent classification of all possible cases.

\bigskip

\begin{definition}\label{Domain12Def}
The parameters $(N,\EuScript N_\mathrm{env})$ 
belong to the \emph{first} or to the \emph{second domain} if $\eta_0<\eta_\infty$
or $\eta_0>\eta_\infty$, respectively.
\end{definition}

\bigskip

\begin{definition}\label{Domain3Def}
The parameters $(N,\EuScript N_\mathrm{env})$ 
belong to the \emph{third domain} if the function $\underline C(s)$ has at
maximum one extremum in the interval $0<s<\infty$ for all values of
transmissivity $\eta\in(0,1)$.
\end{definition}

\bigskip

These domains correspond to the following relations between transmissivities:
\begin{itemize}
\item \emph{First domain:} $\widetilde\eta<\overline\eta<\eta_0<\eta_\infty$.
\item \emph{Second domain:} $\widetilde\eta<\overline\eta<\eta_\infty<\eta_0$.
\item \emph{Third domain:} $\eta_\infty<\eta_0$ ($\widetilde\eta$ and
$\overline\eta$ do not exist).
\end{itemize}

In order to characterize the transitions from one domain to another 
we will use the following definitions.

\bigskip

\begin{definition}\label{Supercr12Def}
The value of $N$ will be denoted by $N_0$ and called
\emph{supercritical} for a given value of $\EuScript N_\mathrm{env}$, if the point
$(N_0,\EuScript N_\mathrm{env})$ corresponds to the transition from 
first to second domain.  Similarly, the value of $\EuScript N_\mathrm{env}$
will be denoted by $\EuScript N_\mathrm{env,0}$ and called
\emph{supercritical} for a given value of $N$, if the point
$(N,\EuScript N_\mathrm{env,0})$ corresponds to transition from first to second domain. 
\end{definition}

\bigskip

\begin{definition}\label{Supercr23Def}
The value of $N$ will be denoted by $\widetilde N$ and called
\emph{supercritical} for a given value of $\EuScript N_\mathrm{env}$, if the point
$(\widetilde N,\EuScript N_\mathrm{env})$ corresponds to transition from
second to third domain.  Similarly, the value of $\EuScript N_\mathrm{env}$
will be denoted by $\widetilde{\EuScript N}_\mathrm{env}$ and called
\emph{supercritical} for a given value of $N$, if the point
$(N,\widetilde{\EuScript N}_\mathrm{env})$ corresponds to transition from
second to third domain. 
\end{definition}

\bigskip

\begin{definition}\label{SupercrFunDef}
The function $f(N,\EuScript N_\mathrm{env})=0$ will be denoted by $f_0$ and called
\emph{supercritical} if it corresponds to the boundary between the first and
second domain. Similarly, the function $f(N,\EuScript N_\mathrm{env})=0$ will
be denoted by $\widetilde f$ and called \emph{supercritical} if it corresponds
to the boundary between the second and third domain.
\end{definition}

\bigskip

As far as the boundary between domains characterize the critical parameters
(transmissivities), e.g. appearance of some critical parameters or the relations
between them, the term ``supercritical'' was used in the
definitions~\ref{Supercr12Def}, \ref{Supercr23Def} and~\ref{SupercrFunDef}.  One
can also say that supercritical parameters are those critical parameters which
characterize the other critical parameters.

The mnemonic rule to remember the notations used for the critical and
supercritical parameters is the following. The quantities $\eta_0$ and
$\eta_\infty$ are defined by considering the behavior of capacity at the points
of zero and infinite squeezing, therefore these values are used as subscripts.
The supercritical values $N_0$ and $\EuScript N_\mathrm{env,0}$ correspond to
transition between the domains which have different relations between $\eta_0$
and $\eta_\infty$, therefore subscript zero is used. The
$\widetilde\eta$-transmissivity corresponds to the case when $\underline C(s)$ 
decays into maximum to the left and minimum to the right if transmissivity
$\eta$ is slightly above the value $\widetilde\eta$, \emph{i.e.} $\underline
C(s)$ forms a ``wave'' in such case. This explains the usage of tilde sign. The
transmissivity $\overline\eta$ corresponds to the case when the curve
$\underline C(s)$ ``touches'' the upper line $\underline C(s)=\log_2(2N+1)$,
therefore overlining is used. Finally, the transition from the third to the
second domain corresponds to the appearance of the quantity $\widetilde\eta$,
\emph{i.e.} ``wave'' behavior of the curve $\underline C(s)$, therefore the tilde
sign is used for supercritical parameters $\widetilde N$ and
$\widetilde{\EuScript N}_\mathrm{env}$.

Thus, we have defined four critical
transmissivities ($\widetilde\eta$, $\overline\eta$, $\eta_0$ and $\eta_\infty$) 
and four specific values of squeezing ($s_\mathrm{R}$, $s_\mathrm{L}$,
$s_\mathrm{r}$ and $s_\mathrm{l}$). Similarly, by considering
the family of functions $\underline C(s)$ parametrized by $\EuScript N_\mathrm{env}$ (for
fixed values of $\eta$ and $N$) or by $N$ (for fixed values of $\eta$ and
$\EuScript N_\mathrm{env}$), the corresponding critical values for environment
thermal or input photons can be considered, respectively. All these approaches
can be generalized and considered as particular cases of \emph{critical
functions}. The latter are functions of the form $\delta(\eta,N,\EuScript
N_\mathrm{env})=0$, where any of the parameters $\eta$, $N$ and $\EuScript
N_\mathrm{env}$ is critical if the others are considered to be constants. In
particular, by assuming $N$ and $\EuScript N_\mathrm{env}$ to be constants and
using notations for critical functions similarly to transmissivities, we get
the following relations:
\begin{align*}
&\widetilde\delta\,(\widetilde\eta,N,\EuScript N_\mathrm{env})=0,\\
&\overline\delta\,(\overline\eta,N,\EuScript N_\mathrm{env})=0,\\
&\delta_0(\eta_0,N,\EuScript N_\mathrm{env})=0,\\
&\delta_\infty(\eta_\infty,N,\EuScript N_\mathrm{env})=0.
\end{align*}

Now that we have introduced all necessary definitions we can discuss how $\underline
C(s)$ is varying with the increasing of $\eta$ from zero to one. As one can see
from Figs.\ref{porogi_domain_1}, \ref{porogi_domain_2}, \ref{porogi_domain_3}
and~\ref{expScaleCofs} it passes in sequence the following five \emph{regimes}:
\begin{enumerate}
\item[I.]
\emph{$0<\eta\leqslant\widetilde\eta$} (for the third domain one can consider 
$\eta_\infty$ instead of $\widetilde\eta$).

Capacity is monotonically increasing function of
$s\in\mathbb{R}_{+}$ and tends to its universal limit~\eqref{Climit} from the
bottom. Optimal squeezing~$s_\star$ is equal to~$\infty$. In particular, when
$\eta=\widetilde\eta$, capacity has its saddle-point for the value of squeezing
$s=s_\mathrm{L}=s_\mathrm{R}$. 
\item[II.]
$\widetilde\eta<\eta\leqslant\overline\eta$ (this regime does not exist for the third
domain).

The saddle-point decays into two extrema -- 
the capacity maximum to the left at the point of $s=s_\mathrm{L}$ and the
capacity minimum to the right at the point of $s=s_\mathrm{R}$, where it is
$$\underline{C}(0)<\underline{C}(s_\mathrm{R})<
\underline{C}(s_\mathrm{L})<\underline{C}(s_\star)=\underline{C}(\infty).$$
Higher values of $\eta$ correspond to lower values $s_\mathrm{L}$ and to higher values of
$s_\mathrm{R}$. Thus, despite we still have $s_\star=\infty$, the value of
$s=s_\mathrm{L}$ could
be more preferable because it is finite.  When $\eta=\overline\eta$, the local
maximum at the point $s=s_\mathrm{L}$ reaches the value of global maximum:
$$\underline{C}(s_\mathrm{L})=\underline{C}(s_\star)=\underline{C}(\infty)=\log_2(2N+1).$$
\item[III.]
$\overline\eta<\eta<\eta_0$ or  
$\overline\eta<\eta\leqslant\eta_\infty$ for the first and the second domains,
respectively (this regime does not exist for the third
domain). 

Optimal squeezing $s_\star$ becames finite and equal to $s_\mathrm{L}$. Two values of
squeezing $s_\mathrm{l}$ and $s_\mathrm{r}$ providing the same capacity as in the universal limit appear:
$$\underline{C}(s_\mathrm{l})=\underline{C}(s_\mathrm{r})=\underline{C}(\infty).$$
Higher values of $\eta$ correspond to higher $s_\mathrm{r}$ and lower $s_\mathrm{l}$.
Capacity approaches its universal limit from the bottom:
$\underline{C}(0\ll s<\infty)<\underline{C}(s=\infty)$.
\begin{itemize}
\item \emph{The first domain.} With the increasing of $\eta$ the value of $s_\mathrm{l}$
is decreasing. It tends to zero when $\eta$ tends to $\eta_0$ and then disappears (does not exist
for $\eta\geqslant\eta_0$). The global capacity minimum for $\eta\to\eta_0-0$ is
achieved at the value $s_\mathrm{R}$.
\item \emph{The second domain.} When $\eta\to\eta_\infty$ both values of 
$s_\mathrm{r}$ and $s_\mathrm{R}$ tend to infinity, \emph{i.e.}
$$\underline{C}(s_\mathrm{r})=\underline{C}(s_\mathrm{R})=\underline{C}(\infty)$$
and one of capacity extrema disappears. 
The global capacity minimum is still achieved at $s=0$,
therefore any squeezed environment is still more preferable.
\end{itemize}
\item[IV.]
$\eta_0\leqslant\eta\leqslant\eta_\infty$ (for the first domain) or
$\eta_\infty<\eta\leqslant\eta_0$ (for the second and the third domains). 
\begin{itemize}
\item \emph{The first domain.}
The capacity has two minima at values of zero and $s=s_\mathrm{R}$,
where $\underline{C}(0)>\underline{C}(s_\mathrm{R})$. When $\eta$ tends to
$\eta_\infty$, 
both $s_\mathrm{r}$ and $s_\mathrm{R}$ tend to infinity and the right extremum
of $\underline C(s)$ disappears. 
\item \emph{The second and the third domains.}
The capacity has two minima at values of zero and
infinite squeezing, where $\underline{C}(0)<\underline{C}(\infty)$. Starting
from this regime it will have only one extremum for finite non-zero values of
$s$ which is maximum at $s=s_\mathrm{L}$. When $\eta$ reaches $\eta_0$, we have 
\begin{equation*}
\underline{C}(0)=\underline{C}(\infty)=\log_2(2N+1).
\end{equation*}
\end{itemize}
\item[V.]
$\eta>\eta_\infty$ (for the first domain) or $\eta>\eta_0$ (for the second and
the third domains). 

The global capacity minimum is at infinite squeezing, \emph{i.e.}
$\underline{C}(\infty)<\underline{C}(0)$.
\end{enumerate}

The notion of regime can be also clarified by considering specific values of
squeezing as functions of transmissivity for fixed values of $N$ and
$\EuScript N_\mathrm{env}$. In fact, one can see that these values of squeezing 
appear and disappear at some critical values of transmissivity which can also
correspond to asymptotic lines (see Figs.\ref{chan1sq}, \ref{chan2sq}
and~\ref{chan3sq}). In particular, the optimal squeezing equals
\begin{equation}
s_\star=
\begin{cases}
\infty,&\mbox{if}\quad0<\eta<\eta'.\\
s_\mathrm{L},&\mbox{if}\quad\eta'<\eta<1,
\end{cases}
\label{sstar}
\end{equation}
where $\eta'=\overline\eta$ for the case of first and second domains and
$\eta'=\eta_\infty$ for the case of third domain. Moreover, the optimal
squeezing asymptotically tends to infinity for the case of third domain, but
discontinuously jumps to infinity for first and second domains.
This transition behavior of optimal squeezing is shown in
Fig.\ref{sNenv_Eta}-left and Fig.\ref{sN_Eta}-left, where $s_\star<\infty$ is plotted
as function of $\eta$ for different values of $\EuScript N_\mathrm{env}$ and $N$
respectively. The capacity corresponding to these finite values of $s_\star$ is plotted
in Fig.\ref{sNenv_Eta}-right and Fig.\ref{sN_Eta}-right, respectively.

Finally, let us consider how critical transmissivities depend on $N$ and
$\EuScript N_\mathrm{env}$. One can see that $\eta_\infty$ does not depend on
$N$ and has non-trivial minimum for $\EuScript N_\mathrm{env}=0$ (see
Fig.\ref{eta0ref}-top-right). Then, both $\widetilde\eta$ and $\overline\eta$
(we have always $\widetilde\eta<\overline\eta<\eta_\infty$) are monotonically
growing functions of $N\in(0,\widetilde N)$ which disappear for the values of
$N\geqslant\widetilde N$ and tend to $\eta_\infty$ if $N$ tends to $\widetilde
N$ from the left. Notice, that the values of $\widetilde\eta$ and
$\overline\eta$ do not tend to zero for $N\to0$ if $\EuScript N_\mathrm{env}>0$.

The $\eta_0$-transmissivity is plotted vs $\ln(N+1)$ for different values of
$\EuScript N_\mathrm{env}$ in Fig.\ref{eta0ref}-top-left. One can see that
$\eta_0$ has non-trivial limits for the values of $N$ tending to zero and
infinity. These limits are plotted in Fig.\ref{eta0ref}-left and
Fig.\ref{eta0ref}-right, respectively. It is interesting to note that the
quantity $\eta_0(N\to\infty)$ also has non-trivial minimum.

The next subsections will be devoted to analytical estimation of critical and
supercritical parameters as well as to estimation of the specific values of
squeezing. This will eventually allow us to prove most of their properties
discussed in this subsection.

\subsection{Stationary points for capacity}\label{critparcapsec}

Let us consider the quantities $s_\mathrm{L}$, $s_\mathrm{R}$, $\delta_\infty$
and $\widetilde\delta$ analytically.

The critical function $\delta_\infty$ which characterize the behavior of the channel in
the neighborhood of infinite environment squeezing $s\to\infty$ can be found as
follows.  We first note that only eigenvalues maximizing $\underline{C}$ are of
interest, therefore it is $\partial\underline{C}/\partial
i_u=(\eta/2)\mathcal{F}=0$, where $\mathcal{F}=0$ is the mode transcendental
equation~\eqref{ModeTrEq1use}. This allows us to simplify the derivative over $s$
as
\begin{equation*}
\frac{d\underline{C}}{ds}=
\frac{\partial\underline{C}}{\partial s}+
\frac{\partial\underline{C}}{\partial i_u}\frac{\partial i_u}{\partial s}
=\frac{\partial\underline{C}}{\partial s}.
\end{equation*}
Its asymptotic behavior is
\begin{equation*}
\frac{\partial\underline{C}}{\partial s}=
\frac{\left[2N+1-(2N+1)^{-1}\right]\delta_\infty}
{\eta\,(1-\eta)\left(\EuScript N_\mathrm{env}+\frac12\right)\ln2}e^{-s},
\end{equation*}
where
\begin{equation}
\delta_\infty=(1-\eta)^2\left(\EuScript N_\mathrm{env}+\frac12\right)^2-\frac1{12}.
\label{Delta}
\end{equation}
Thus, $\eta$ and $\EuScript N_\mathrm{env}$ are the only parameters which define
how capacity tends to its universal limit~\eqref{Climit}. In particular, given
a value of $\EuScript N_\mathrm{env}$, the capacity tends to this limit
from the top if $\eta>\eta_\infty$ and from the bottom if $\eta<\eta_\infty$,
where \mbox{$\eta_\infty$-transmissivity} can be found from the relation
\begin{equation}
(1-\eta_\infty)\left(\EuScript N_\mathrm{env}+\frac12\right)=\frac1{\sqrt{12}}.
\label{CritPars}
\end{equation}
In particular, for the vacuum environment it is
\begin{equation*}
\eta_\infty=1-\frac1{\sqrt{3}}\approx0.42265.
\end{equation*}

Analogously, if the value of $\eta$ is fixed, the capacity tends
to the universal limit~\eqref{Climit} from the top or bottom
depending on the value of $\EuScript N_\mathrm{env}$, which
follows from Eq.~\eqref{CritPars}.  Consequently, this
value plays a role similar to critical transmissivity if the
family of curves $\underline C(s)$ parametrized by $\EuScript
N_\mathrm{env}$ for fixed $\eta$ and $N$ is considered (compare
the curves in Fig.\ref{cofs_family}-left and
Fig.\ref{cofs_family}-right).  It is interesting to note that
this effect also exists for additive noise Gaussian channel where
the quantity $\EuScript N_{\mathrm{env}}$ has the same meaning
and its critical value equals
$1/\sqrt{12}$~\cite{SchaferKarpovSPIE}. Thus, the critical
parameters and the behavior shown in Fig.\ref{cofs_family} may be
relevant for a general Gaussian channel.

In order to specify the region where environment squeezing increases the
capacity, in the following we estimate the values of $s$ corresponding to
extrema of function $\underline{C}(s)$. Let us consider the system of equations
$\partial\underline{C}/\partial s=0$, \mbox{$\partial\underline{C}/\partial
i_u=0$} taken for the eigenvalues maximizing $\underline{C}$ and belonging to
the second stage (extremum cannot be in the third stage since
$\partial\underline{C}/\partial s\neq0$ according to Eq.~\eqref{CC1useSimple}).
Its solution results to the value $i_u=N+1/2$ and the value of $s$ defined by
the equation
\begin{equation}
\mathcal P(s)=0,
\label{TrUr}
\end{equation}
where
\begin{equation*}
\mathcal P=\frac{g_1(\overline\nu)}{\overline\nu^2}\sinh s-
\frac{g_1(\nu)}{\nu^2}\frac{\sinh(s-\ln(2N+1))}{2N+1}.
\end{equation*}
Thus, we have the same value of $i_u$ for both local extrema of $\underline
C(s)$ and the point of $s=\infty$.

Solving Eq.~\eqref{TrUr} in neighborhood of zero or infinite values of $s$ one
can estimate both its roots ($s_\mathrm{L}$ and $s_\mathrm{R}$).
In particular, after the expansion of
Eq.~\eqref{TrUr} in powers of $e^{-s}$ in the neighborhood of $s\to\infty$, where
terms higher than the second order are neglected, it takes the form
$z_2(e^{-s})=0$ with $z_2(e^{-s})=be^{-2s}+ce^{-s}$ ($b$ and $c$ are some
constants).  The function $z_2(e^{-s})$ is the partial sum for Laurent series of
the function $Z(e^{-s})$, where $Z=0$ is the equation~\eqref{TrUr}. 
Both functions $z_2(e^{-s})$ and $Z(e^{-s})$ are concave in the neighborhood of
$s\to\infty$ (see Fig.\ref{sLestref}-top-right), 
therefore their nontrivial\footnote{The trivial solution which we
imply is $e^{-s}=0$ corresponding to $s=\infty$.} roots are close each other.
The latter property explains why the approximation $z_2(e^{-s})=0$ is applicable
and leads to the result
\begin{equation}
s_\mathrm{R}=\ln\frac{[1+(2N+1)^2]\left[\delta_\infty^2+\frac1{180}\right]-
\frac{\eta^2}6\left[N+\frac12\right]^2}
{\eta\,(1-\eta)\left(N+\frac12\right)
\left(\EuScript N_\mathrm{env}+\frac12\right)\delta_\infty},
\label{sR}
\end{equation}
where critical function $\delta_\infty$ is given by
Eq.~\eqref{Delta} and characterizes the ``criticality''
of the given channel parameters (their vicinity to the transition point).
Notice, that according to estimation~\eqref{sR} we have
$$\lim_{\eta\to\eta_\infty}s_\mathrm{R}=\lim_{\delta_\infty\to0}s_\mathrm{R}=\infty.$$

Analogously, considering the next order approximation for
Eq.~\eqref{TrUr}, one can construct the function
$z_3(e^{-s})=ae^{-3s}+be^{-2s}+ce^{-s}$ and find the condition
when both nontrivial roots of the equation $z_3(e^{-s})=0$
coincide.  This is the case of $s_\mathrm{L}=s_\mathrm{R}$ (both
$s_\mathrm{L}$ and $s_\mathrm{R}$ are taken from approximation
$z_3$, see Fig.\ref{sLestref}),
\emph{i.e.} the saddle-point of the curve $\underline{C}(s)$
where both the derivatives $\partial\underline{C}/ds$ and
$\partial^2\underline{C}/ds^2$ equal zero. 
In Subsec.\ref{SuperCritSubSec} this approach will be used in order to
provide analytical estimation of the saddle-point transmissivity
$\widetilde\eta$.

Similarly, expanding Eq.~\eqref{TrUr} in powers of $s$ in the neighborhood of
$s=0$ we get an equation of the form $as^3+bs^2+cs=0$ ($a$, $b$ and
$c$ are some constants depending on channels parameters), whose
nontrivial root is an estimation for the left extremum
$s_\mathrm{L}$ (see Subsec.~\ref{AppsL} for its value and derivation).

Analyzing the equation $\underline C(s)=0$ instead of
Eq.~\eqref{TrUr} and applying the same method (expansion in
powers of $e^{-s}$ in the neighborhood of $s=\infty$ and in powers
of $s$ in the neighborhood of $s=0$) one can estimate both left
($s_\mathrm{l}$) and right ($s_\mathrm{r}$) roots. In particular,
one can get the relation
$$
s_\mathrm{r}=s_\mathrm{R}-\ln2.
$$
Estimation of
$s_\mathrm{l}$ is given in Subsec.~\ref{Appsl}.  The case when
$s_\mathrm{l}$ and $s_\mathrm{r}$ (considered for this
approximation) coincide corresponds to
$\overline\eta$-transmissivity, which is estimated below in
Subsec.\ref{SuperCritSubSec}.

\subsection{Estimation of $s_\mathrm{L}$}\label{AppsL}

Let us estimate the quantity $s_\mathrm{L}$.
Our purpose is to solve Eq.~\eqref{TrUr} in the neighborhood of $s=0$ taking
into account that all eigenvalues in extrema points are known. At first, notice
that squares of symplectic eigenvalues as functions of $s$ in the extrema points read
\begin{align*}
\nu_\mathrm{e}^2(s)&=
\EuScript Q\left[\frac\eta2,(1-\eta)\left(\EuScript N_\mathrm{env}+\frac12\right),
s-\ln(2N+1)\right],
\nonumber\\
\overline\nu_\mathrm{e}^2(s)&=
\EuScript Q\left[\eta\left(N+\frac12\right),(1-\eta)\left(\EuScript
N_\mathrm{env}+\frac12\right),s\right],
\nonumber
\end{align*}
where
$$
\EuScript Q(a,b,\varphi):=a^2+b^2+2ab\cosh(\varphi).
$$
Below we use the notations $\nu_{\mathrm{e},0}=\nu_\mathrm{e}(0)$ and
$\overline\nu_{\mathrm{e},0}=\overline\nu_\mathrm{e}(0)$.  Let us define the
function
\begin{equation*}
\EuScript Y(x_1,x_2)=\frac{x_1}{\nu_\mathrm{e,0}^2}+\\
\frac{x_2}{\left(\nu_\mathrm{e,0}^2-\frac14\right)g_1(\nu_\mathrm{e,0})}
\end{equation*}
and introduce the following notations:
\begin{align*}
&\EuScript{X:}=\frac\eta2(1-\eta)\left(\EuScript N_\mathrm{env}+\frac12\right),\\
&\cn:=\cosh(\ln(2N+1)),\\
&\sn:=\sinh(\ln(2N+1)).
\end{align*}
By representing Eq.~\eqref{TrUr} as $\mathcal P(s)=as^2+bs+c=0$ we can find
both linear (supposing $a=0$) and quadratic ($a\neq0$) approximations. They
result as estimations of squeezing in left extremum of $C(s)$ (denoted as
$s_\mathrm{L,1}$ for linear approximation and $s_\mathrm{L,2}$ for quadratic one):
\begin{align*}
&s_\mathrm{L,1}^{-1}=\EuScript{K}_1,
&s_\mathrm{L,2}^{-1}=
\frac12\left(\sqrt{\EuScript{K}_1^2-2\EuScript{K}_2}+\EuScript{K}_1\right),
\end{align*}
where
\begin{align}
&\EuScript{K}_1=\frac{\cn}{\sn}-\EuScript X\EuScript Y(1,1)\,\sn-
\frac{(2N+1)}{\sn}\frac{\nu_\mathrm{e,0}^2}{\overline\nu_\mathrm{e,0}^2}
\frac{g_1(\overline\nu_\mathrm{e,0})}{g_1(\nu_\mathrm{e,0})},
\label{K1}\\
&\EuScript{K}_2=1-3\EuScript X\EuScript Y(1,1)\,\cn+
\sn^2\EuScript X^2\EuScript Y\left(3\EuScript Y(1,1),\frac2{\nu_\mathrm{e,0}^2-\frac14}\right).
\label{K2}
\end{align}
Approximation $s_\mathrm{L,1}$ is applicable only if $\EuScript K_1>0$ and $s_\mathrm{L,2}$ is
applicable only if $\EuScript K_1^2>2\EuScript K_2$ (it is equivalent to
$N<\bigl(\sqrt{2}\bigr)^{-1}$ if $\eta\to1$). These regions of applicability follow from
the condition that proper equations must have their roots positive.

Let us consider the limit $\lim_{\eta\to1}s_\mathrm{L}$. First, note that both second and
third terms in relations Eq.~\eqref{K1} and~\eqref{K2} disappear when
$\eta\to1$, therefore we have
\begin{equation}
\lim_{\eta\to1}s_\mathrm{L,1}^{\phantom{1}}=\frac{\sn}{\cn}=
\frac{2N(N+1)}{2N(N+1)+1}=\frac1{1+\phi_{s_\mathrm{L}}},
\label{sLlim}
\end{equation}
where 
\begin{equation*}
\phi_{s_\mathrm{L}}=\frac1{2N(N+1)}.
\end{equation*}
The quantity $\lim_{\eta\to1}s_\mathrm{L,1}$, in turn, tends to $1$ for $N\to\infty$
and to zero for $N\to0$.  Analogously, 
\begin{equation}
\lim_{\eta\to1}s_\mathrm{L,2}^{-1}=
\frac12\Bigl(1+\phi_\mathrm{s_L}+\sqrt{\phi_\mathrm{s_L}^2+2\phi_{s_\mathrm{L}}-1}\;\Bigr).
\label{sLlim2}
\end{equation}
Notice, that the approximations~\eqref{sLlim} and~\eqref{sLlim2} do not depend on thermal photons
$\EuScript N_\mathrm{env}$.
This is an argument in support of the behavior observed numerically in Fig.\ref{sNenv_Eta}
for the exact limit.

The dependence of $s_\mathrm{L}$ and its approximation from
parameters is shown in Fig.\ref{sLdepref}.  One can see that
$s_\mathrm{L}$ is monotonically decreasing function of $\eta$,
which indeed has non-trivial limit for $\eta\to1$.

\subsection{Estimation of $\eta_0$}\label{Appeta0}

As it follows from a definition~\ref{Eta0Def}
the transmissivity $\eta_0$ is given by the equation
\begin{equation}
\mathcal O(\eta)=0,
\label{ETA0}
\end{equation}
where
\begin{multline}
\mathcal O(\eta) := 
g\bigl[\eta N+(1-\eta){\EuScript N}_\mathrm{env}\bigr]\\
-g\bigl[(1-\eta){\EuScript N}_\mathrm{env}\bigr]
-\log_2(2N+1).
\label{ETA0expanded}
\end{multline}
Note that for $s=0$ we have the case of the third stage
and $N_\mathrm{env}=\EuScript N_\mathrm{env}$.

We can have the following cases: $\eta_0<\eta_\infty$ (see Fig.\ref{porogi_domain_1}), 
$\eta_0>\eta_\infty$ (see Fig.\ref{porogi_domain_2}) and
$\eta_0=\eta_\infty$. 
The latter case corresponds to transition from first to second domain and 
defines the locus $(N,\EuScript N_\mathrm{env})$ where
equality $\eta_0=\eta_\infty$ holds. One can see from Fig.\ref{expScaleCofs}-bottom-right that 
the limit value $\EuScript N_\mathrm{env}=\EuScript N_\mathrm{env,0}$, 
which still can have $\eta_0=\eta_\infty$ is achieved at $N=0$. However, since Eq.~\eqref{ETA0}
is satisfied by any values of $\eta$ and $\EuScript N_\mathrm{env}$ if $N=0$, we have to solve it
for the limit $N\to0$. By expanding Eq.~\eqref{ETA0} over $N$ we get equation
$N\partial\mathcal O/\partial N=0$. Then, by substituting $N=0$ into $\partial\mathcal O/\partial N=0$
we get the equation 
\begin{equation}
\eta_0\ln\left(1+\frac1{(1-\eta_0)\EuScript N_\mathrm{env}}\right)=2.
\label{ETA0N0}
\end{equation}
The joint solution of the system of Eqs.~\eqref{ETA0N0}, \eqref{CritPars} and
\mbox{$\eta_0=\eta_\infty$}
results to the equation
\begin{equation*}
\left(1-\frac1{\sqrt{3}\,\bigl(2\EuScript N_\mathrm{env}+1\bigr)}\right)
\ln\left(1+\frac{\sqrt{3}\,\bigl(2\EuScript N_\mathrm{env}+1\bigr)}{\EuScript N_\mathrm{env}}\right)=2.
\end{equation*}
Its solution is the supercritical value
$$\EuScript N_\mathrm{env,0}(N=0)\approx0.0204.$$

\subsubsection{Limit values of $\EuScript N_\mathrm{env}$}

For high values of $\EuScript N_\mathrm{env}$ Eq.~\eqref{ETA0} has its asymptotic
behavior given by the relation
\begin{equation*}
\log_2\frac{\eta N+(1-\eta)\,\EuScript N_\mathrm{env}+\frac12}{(1-\eta)\,\EuScript N_\mathrm{env}+\frac12}=
\log_2(2N+1),
\end{equation*}
from which one can get
\begin{equation*}
(1-\eta)(1+2\EuScript N_\mathrm{env})=0,
\end{equation*}
{\it i.e.}
\begin{equation}
\lim_{\EuScript N_\mathrm{env}\to\infty}\eta_0=1.
\label{eta0Ne0exact}
\end{equation}

In the case of pure ({\it i.e.} $\EuScript N_\mathrm{env}=0$) environment $\eta_0$ is
equal to (see Eq.~\eqref{ETA0})
\begin{equation}
\eta_0=\frac1Ng^{-1}\bigl[\log_2(2N+1)\bigr].
\label{eta0Nenv0}
\end{equation}
In turn, by supposing $\eta_0=\eta_\infty$ we get the equation for $N$
\begin{equation*}
\frac1Ng^{-1}\bigl[\log_2(2N+1)\bigr]=1-\frac1{\sqrt{3}},
\end{equation*}
whose solution is supercritical value 
$$N_0(\EuScript N_\mathrm{env}=0)\approx0.0817.$$
Thus, $\eta_0>\eta_\infty$ if $N>N_0(0)$, 
and $\eta_0<\eta_\infty$ if $N<N_0(0)$ (see
examples in Figs.~\ref{porogi_domain_1}, \ref{porogi_domain_2}
and~\ref{porogi_domain_3}). In other
words, if and only if $N\geqslant N_0(0)$, we have 
$$
\forall s\quad C(s,\eta\geqslant\eta_0)\geqslant\log_2(2N+1).
$$ 
In particular, if the environment is pure, $\eta>\eta_0$ and $N\geqslant N_0(0)$,
then the universal limit gives the global minimum for
$C(s)$, and $s_\mathrm{L}$ gives the global maximum:
\begin{align*}
&\min_{0\leqslant s\leqslant\infty}C(s)=\log_2(2N+1),\\
&\max_{0\leqslant s\leqslant\infty}C(s)=C(s_\mathrm{L}).
\end{align*}

\subsubsection{The case $N\to0$}

By considering Eq.~\eqref{ETA0N0}
for \mbox{$\EuScript N_\mathrm{env}\gg0$} we obtain, to linear approximation 
in $\EuScript N_\mathrm{env}^{-1}$, that
\begin{equation}
\lim_{N\to0}\eta_0(\EuScript N_\mathrm{env}\gg0)=\frac{2\EuScript N_\mathrm{env}}{2\EuScript N_\mathrm{env}+1}.
\label{ETA0N0approx}
\end{equation}
Using Eq.~\eqref{eta0Nenv0} we get for pure environment
\begin{equation*}
\lim_{N\to0}\eta_0(\EuScript N_\mathrm{env}=0)=0.
\end{equation*}

\subsubsection{The case $N\to\infty$}

By taking the limit $N\to\infty$ in Eq.~\eqref{ETA0} (we use expansion of
$g$-function) one can get that it is equivalent to
\begin{equation}
\log_2\frac{\eta e}2-g\bigl[(1-\eta)\EuScript N_\mathrm{env}\bigr]=0.
\label{NInf}
\end{equation}
In particular, for pure environment we get
\begin{equation}
\lim_{N\to\infty}\eta_0(\EuScript N_\mathrm{env}=0)=\frac2e.
\label{ExactNenv0Ninf}
\end{equation}

The function $g(x)$ behaves like $-x\log_2x$ for small values of $x$.
Using this property and expanding the logarithm in the first term of
Eq.~\eqref{NInf} in powers of $\varepsilon:=1-\eta$ up to the first order one
can obtain the equation
\begin{equation*}
\ln\frac e2-\varepsilon+\varepsilon \EuScript N_\mathrm{env}\ln\left(\varepsilon \EuScript N_\mathrm{env}\right)=0.
\end{equation*}
Its solution gives an estimation of $\lim_{N\to\infty}\eta_0$:
\begin{equation}
\lim_{N\to\infty}\eta_0=
1-\left[\EuScript N_\mathrm{env}W_{-1}\left(e^{-\EuScript N_\mathrm{env}^{-1}}
\ln\frac2e\right)\right]^{-1}\ln\frac2e,
\label{lambert}
\end{equation}
where $W_{-1}$ is $-1$ branch of Lambert $W$ function which is solution of
$W(z)e^{W(z)}=z$ and whose properties are well known~\cite{LambertRef}.
One can show that the approximation~\eqref{lambert} has the limits
\begin{align}
\lim_{\EuScript N_\mathrm{env}\to\infty}\lim_{N\to\infty}\eta_0&=1,\nonumber\\
\lim_{\EuScript N_\mathrm{env}\to0}\lim_{N\to\infty}\eta_0&=\ln2.
\label{AlternativeLimit}
\end{align}
The first limit coincides with the exact value (see Eq.~\eqref{eta0Ne0exact}),
but the second one is different (see Eq.~\eqref{ExactNenv0Ninf}).  Maximal error
of estimation~\eqref{lambert} is about $5\%$ and achieved by $\EuScript N_\mathrm{env}=0$.
As far as $\lim_{N\to\infty}\eta_0$ is monotonic over $\EuScript N_\mathrm{env}$ (see
Eq.~\eqref{lambert}), Eq.~\eqref{ExactNenv0Ninf} gives its minimum:
$$
\lim_{N\to\infty}\eta_0\geqslant\frac2e.
$$

Eq.~\eqref{AlternativeLimit} can be obtained as follows. 
First, note that the following limit holds:
\begin{equation}
\lim_{z\to0-0}\frac{\ln(-z)}{W_{-1}(z)}=1.
\label{SpecLim}
\end{equation}
It can be obtained by applying logarithm to both parts of equation
$$-W_{-1}(z)e^{W_{-1}(z)}=-z,\qquad z<0,$$
and then dividing it on $W_{-1}(z)$.  Notice, that $W_{-1}(z)<0$ for
$z\in\left[-e^{-1},0\right]$ and has the limit
$$\lim_{z\to0-0} W_{-1}(z)=-\infty.$$
Let us define a new variable $x<0$ to be equal to the argument of $W_{-1}$ in
Eq.~\eqref{lambert} and consider the limit of Eq.~\eqref{lambert} for $x\to0-0$
which corresponds to $\EuScript N_\mathrm{env}\to0$. Taking into account
Eq.~\eqref{SpecLim}, we arrive at the result~\eqref{AlternativeLimit}.

The dependence of $\eta_0$ on parameters is shown in
Fig.\ref{eta0ref}.

\subsection{Estimation of $s_\mathrm{l}$}\label{Appsl}

The definition $C(s_\mathrm{l})=\log_2(2N+1)$ results to 
\begin{multline*}
\cosh s_l=\\
\frac{g^{-1}\big[\log_2(2N+1)+g((1-\eta)\,\EuScript N_\mathrm{env})\big]-\eta\left(N+\frac12\right)+\frac12}
{(1-\eta)\left(\EuScript N_\mathrm{env}+\frac12\right)},
\end{multline*}
which for pure environment reads
\begin{equation}
\cosh s_l=1+\frac{2N(\eta_0-\eta)}{1-\eta},
\label{slNenv0}
\end{equation}
where $\eta_0=\eta_0(\EuScript N_\mathrm{env}=0)$ is given by
Eq.~\eqref{eta0Nenv0}. In particular, it is clear from Eq.~\eqref{slNenv0} that 
$$
\lim_{\eta\to\eta_0-0}s_\mathrm{l}=0,
$$
which is in full correspondence with the definition and
properties of $s_\mathrm{l}$.

\subsection{Full channel characterization}

Let us summarize the results that we obtained for channel
characterization.  We started from the point that squeezing $s$
is the only parameter which gives rise to non-monotonic
dependence of capacity $\underline C$. We have analyzed this
behavior for typical values of $N$ and $\EuScript N_\mathrm{env}$
(see Fig.\ref{cofs_family}) and found that $\underline C(s)$ has
maximum in the interval $0<s<\infty$ if $\eta>\eta_\infty$, and
is monotonic otherwise. Then, we have shown that the family of
curves $\underline C(s)$ can be considered also for different
values of $\EuScript N_\mathrm{env}$ and fixed $\eta$. Both these
cases can be described using the parameter $\delta_\infty$ (see
Eq.~\eqref{Delta}).  Thus, we get the pair of parameters
$(\eta_\infty,\EuScript N_\mathrm{env})$ characterizing the
behavior of $\underline C(s)$ in the neighborhood of infinity.
Then, we considered also other critical parameters, namely,
$\widetilde\eta$, $\overline\eta$ and $\eta_0$ by analyzing the
family of curves $\underline C(s)$ for different values of $\eta$
and fixed $\EuScript N_\mathrm{env}$.  However, by considering
the family of curves $\underline C(s)$ for different values of
$\EuScript N_\mathrm{env}$ (or $N$) and fixed $\eta$ one can
also introduce analogous critical parameters as the values of
$\EuScript N_\mathrm{env}$ (or $N$). Hence, we finally have four
triads of critical parameters to characterise the channel. After
that we have analyzed how these critical parameters depend on $N$
and $\EuScript N_\mathrm{env}$ by introducing supercritical
parameters. 

On the other hand, one can also say that critical parameters have
allowed us to split the total space
\mbox{$\bigl(0\leqslant\eta\leqslant1$},\,\mbox{$0\leqslant 
N\leqslant\infty$},\,\mbox{$0\leqslant\EuScript N_\mathrm{env}\leqslant\infty\bigr)$}
into \emph{regimes} with
different properties of the dependence $\underline C(s)$, while
supercritical parameters have allowed us to split the total space
\mbox{$\bigl(0\leqslant N\leqslant\infty$},\,\mbox{$0\leqslant\EuScript
N_\mathrm{env}\leqslant\infty\bigr)$} into \emph{domains} with
different properties of the critical parameters.  Finally, note,
that given the type of domain, regime and stage for parameters
$\eta,N,\EuScript N_\mathrm{env}$, one can qualitatively plot the
family of curves $\underline C(s)$ (for different values of
$\eta$) without numerical calculations and put forward all
important points and extrema of these curves.

This classification completely characterises the role of environment squeezing.
E.g. ``supernonmonotonic'' behavior of $\underline C(s)$ (when it has two
extrema in the interval $0<s<\infty$) is only possible in the first and the
second domains, as in the third domain $\underline C(s)$ has at maximum a single
extremum.  Most of practically interesting channel parameters
belong to the third domain, however, this classification is useful, as it
provides exact conditions {\it when it is so} (expected behavior of $\underline
C(s)$ from the third domain).  The global optimal squeezing $s_\star$ has sudden
jump to infinity at $\overline\eta$ in the first and second domain, but
tends asymptotically to infinity in the third domain.

It is quite nontrivial that despite this difficult classification scheme the
existence of supercritical parameters can be shown analytically (see
Subsec.~\ref{SuperCritSubSec}). Moreover, in some important cases they can be
found exactly and analytically (be expressed through radicals). Thus, despite we
have started from numerical analysis of the dependence $\underline C(s)$, there
are analytical results which support the found properties (see
Subsec.\ref{SuperCritSubSec}).

\subsection{Supercritical parameters}\label{SuperCritSubSec}

First, we have to remember that $\widetilde\eta$ tends to
$\eta_\infty$ when channel
passes from second to third domain (see
Fig.\ref{eta0ref}-top-right).  In particular, the limits
\begin{align}
&\lim_{(N,\EuScript N_\mathrm{env})\to(\widetilde N,\widetilde{\EuScript N}_\mathrm{env})}
\lim_{\eta\to\widetilde\eta^+}s_\mathrm{L}(\eta)=\infty,
\label{EtaLim1}\\
&\lim_{(N,\EuScript N_\mathrm{env})\to(\widetilde{\EuScript N},\widetilde N_\mathrm{env})}
\lim_{\eta\to\widetilde\eta^+}s_\mathrm{R}(\eta)=\infty
\label{EtaLim2}
\end{align}
are supported by numerical calculations (here the notation
``$(N,\EuScript N_\mathrm{env})\to(\widetilde N,\widetilde{\EuScript N}_\mathrm{env})$''
means that we consider the values $(N,\EuScript N_\mathrm{env})$ belonging to
the second domain and tending to the border between second and third
domain).  The relations~\eqref{EtaLim1} and~\eqref{EtaLim2} are equivalent to
the following statement: the value of squeezing corresponding to saddle-point
transmissivity tends to infinity if the values of the channel parameters
($N,\EuScript N_\mathrm{env}$) tend to those from the third domain.
Consequently, in this case the quantities
$e^{-s_\mathrm{L}},e^{-s_\mathrm{R}}$ tend to zero.
Thus, we can say that the transition between second and third domain is
completely characterized by the behavior of the function~\eqref{TrUr} in
the neighborhood of the point $e^{-s}=0$ (remember, that
$e^{-s_\mathrm{L}}$ and $e^{-s_\mathrm{R}}$
are zeros of the function~\eqref{TrUr}). 
Let us now consider the Taylor expansion of
\eqref{TrUr} in the neighborhood of that point. 
To the third order it gives rise to the relation 
\begin{equation}
ae^{-3s}+be^{-2s}+ce^{-s}=0
\label{polynomial}
\end{equation}
which is an approximate form of Eq.~\eqref{TrUr} in the neighborhood of $e^{-s}=0$.
Remember, that the coefficient $c$ is proportional to
$\delta_\infty$ (see Eq.~\eqref{Delta}) and defines the transition from ``undercritical'' to
''uppercritical'' parameters of transmissivity and thermal photons. If we
neglect a constant factor, $c$ is just a denominator of the fraction
under logarithm in $s_\mathrm{R}$ (see Eq.~\eqref{sR}).  
The case when $\widetilde\eta$
disappears corresponds to the case when the function~\eqref{TrUr} has no roots in
the neighborhood of $e^{-s}=0$ except of the point $e^{-s}=0$ itself. As far
as~\eqref{TrUr} in this neighborhood is the polynomial~\eqref{polynomial}, this
condition is equivalent to the statement that this polynomial has no other
extrema except of the point $e^{-s}=0$. It is exactly so if both
$b=c=0$. Thus, by substituting $\delta_\infty=0$ and
$\eta=\eta_\infty$ in the relation $b=0$ (up to a constant factor
$b$ is a numerator in the fraction under logarithm of Eq.~\eqref{sR})
we get
\begin{equation*}
\frac{1+(2N+1)^2}{180}-
\frac16\Biggl[\left[1-\frac1{\sqrt{3}(2\EuScript N_\mathrm{env}+1)}\right]\left[N+\frac12\right]\Biggr]^2=0.
\end{equation*}
This relation between the values of $N$ and $\EuScript N_\mathrm{env}$ is that
defined by the function $\tilde f(\widetilde N,\widetilde{\EuScript
N}_\mathrm{env})=0$, therefore it can be rewritten as (see the parallelism with
relation~\eqref{CritPars})
\begin{equation*}
\left(1-\widetilde\eta_\infty\right)
\left(\widetilde{\EuScript N}_\mathrm{env}+\frac12\right)=\frac1{\sqrt{12}},
\end{equation*}
where the \emph{effective supercritical transmissivity} $\widetilde\eta_\infty$ is
\begin{equation*}
\widetilde\eta_\infty=\sqrt{\frac2{15}\left(\frac1{(2\widetilde N+1)^2}+1\right)}.
\end{equation*}
The quantity $\widetilde{\EuScript N}_\mathrm{env}$ as function of $\tilde N$ was
plotted in Fig.\ref{expScaleCofs}-bottom-right.
Finally, let us write down explicitly the above supercritical values for
the particular important cases:
\begin{align}
&\widetilde{\EuScript N}_\mathrm{env}(\widetilde N=0)
=\frac12\left[\left(\sqrt{3}-\frac2{\sqrt{5}}\right)^{-1}-1\right]
\approx0.0969,
\label{SC1}\\
&\widetilde N(\widetilde{\EuScript N}_\mathrm{env}=0)
=\frac12\left[\sqrt{\frac32+\frac5{2\sqrt{3}}}-1\right]
\approx0.3578
\label{SC2}
\end{align}
where the value~\eqref{SC1} is the maximum amount of thermal photons admissible in
environment which still allows to obtain effects from first and second
domain (e.g., existence of saddle-point transmissivity), and the
value~\eqref{SC2} is the maximum amount of input photons which still allows to
observe the same behavior. These are fundamental constants of lossy bosonic
channel providing its decription on the top level of ``hierarchy of
characterization''.

Remember, that Eq.~\eqref{TrUr} (and hence its
approximation~\eqref{polynomial}) is the derivative of the equation
$\underline C(s)=0$. Therefore, the analogous expansion of equation $\underline
C(s)=0$ in the neighborhood of $s\to\infty$ has the form
\begin{equation}
\mathfrak{A}e^{-3s}+\mathfrak{B}e^{-2s}+\mathfrak{C}e^{-s}+\log_2(2N+1)=0,
\label{polynomialfinal}
\end{equation}
where $\mathfrak{C}=-c$, $\mathfrak{B}=-\frac b2$ and
$\mathfrak{A}=-\frac a3$. Eq.~\eqref{polynomialfinal} allows to interpret
both critical and supercritical parameters in the same framework.  In
particular, zero-order coefficient $\log_2(2N+1)$ is the universal
limit~\eqref{Climit}, zero-equal linear coefficient ($\mathfrak{C}=0$) defines
critical parameter $\eta_\infty$, and if
both linear and quadratic coefficients are zero ($\mathfrak{C}=\mathfrak{B}=0$)
we get supercritical parameters $\widetilde N$ and $\widetilde{\EuScript
N}_\mathrm{env}$. In explicit form they read
\begin{equation*}
\underline C=K_0+K_1\,x+K_2\,x^2+K_3\,x^3,
\end{equation*}
\begin{align*}
&K_0=\log_2(2N+1),\\
&K_1=T_1\,\delta_\infty,\\
&K_2=T_2\left[\bigl(1+M^2\bigr)\left(\delta_\infty^2+\frac1{180}\right)-\frac{\eta^2M^2}{24}\right],\\
&K_3=T_3\,\Biggl[\left(1+M^2+M^4\right)\left(\delta_\infty^3+\frac{\delta_\infty}{60}-\frac1{3780}\right)\\
&\qquad\qquad\quad+\left(1+M^2\right)\left(\frac1{60}-\frac{\delta_\infty}4\right)\frac{\eta^2M^2}4-
\frac{\eta^4M^4}{64}\Biggr],
\end{align*}
where $M:=2N+1$ and
\begin{equation*}
T_j:=\frac{2N(N+1)(-1)^j}
{j\left[\eta\,(1-\eta)\left(N+\frac12\right)\left(\EuScript N_\mathrm{env}+\frac12\right)\right]^j\ln2}
\end{equation*}
with $j=1,2,3$.
The equation (discriminant) \mbox{$K_2^2-4K_1^{\phantom{1}}K_3^{\phantom{1}}=0$} can be rewritten as
\begin{multline*}
\left(1+M^4\right)\left(\frac1{900}+\frac{16}{315}\,\delta_\infty-\frac{14}5\,\delta_\infty^2-156\,\delta_\infty^4\right)\\
+M^2\left(\frac1{450}+\frac{16}{315}\,\delta_\infty-\frac{12}5\,\delta_\infty^2-120\,\delta_\infty^4\right)\\
-\eta^2M^2\left(1+M^2\right)\left(\frac1{60}+\frac45\,\delta_\infty-9\,\delta_\infty^2\right)\\
+\eta^4M^4\left(\frac1{16}+3\,\delta_\infty\right)=0,
\end{multline*}
\begin{equation*}
D=K_1'\,x+K_2'^2\,x^2+K_3'^3\,x^3,
\end{equation*}
\begin{equation*}
K_j'=-\frac j
{\eta\,(1-\eta)\left(N+\frac12\right)\left(\EuScript N_\mathrm{env}+\frac12\right)}K_j.
\end{equation*}
The equation (discriminant) $K_2'^2-4K_1'K_3'=0$ can be rewritten as
\begin{multline*}
\left(1+M^4\right)\left(\frac1{900}+\frac{4}{105}\,\delta_\infty-2\,\delta_\infty^2-108\,\delta_\infty^4\right)\\
+M^2\left(\frac1{450}+\frac4{105}\,\delta_\infty-\frac{8}5\,\delta_\infty^2-72\,\delta_\infty^4\right)\\
-\eta^2M^2\left(1+M^2\right)\left(\frac1{60}+\frac35\,\delta_\infty-6\,\delta_\infty^2\right)\\
+\eta^4M^4\left(\frac1{16}+\frac94\,\delta_\infty\right)=0.
\end{multline*}
Roots of these discriminants provide approximations for the
quantities $\widetilde\eta$ and $\overline\eta$.

Notice, that all of these results (universal limit, critical
and supercritical parameters) are given by exact explicit analytical relations.

In turn, the supercritical parameters $N_0$ and $\EuScript N_\mathrm{env,0}$ are
found in Appendix~\ref{Appeta0}, where the values 
\begin{align*}
&N_0(\EuScript N_\mathrm{env,0}=0)\approx0.0817,\\ 
&\EuScript N_\mathrm{env,0}(N_0=0)\approx0.0204
\end{align*}
are obtained as numerical solutions of a transcendental equations.

\subsection{Critical parameters for heterodyne rate}\label{critparhetsec}

\begin{figure}[t]
\begin{center}
\includegraphics[scale=1]{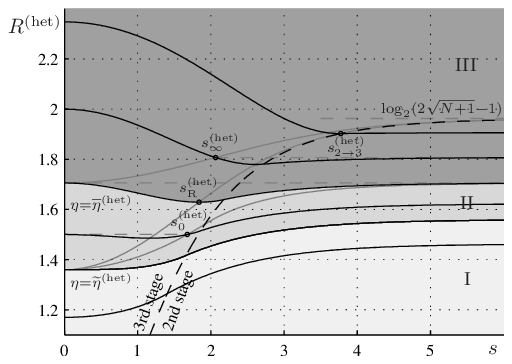}
\caption{Heterodyne rate $R^\mathrm{(het)}$ (black) vs $s$ for the values
of $\eta$ equal to $0.4$, $0.4774$, $0.5356$, $0.623$, $0.75$,
$0.9$ (from bottom to top). The values of other parameters are $N=5$, $\EuScript N_\mathrm{env}=1$. 
The grey curves are the loci
$(s_0^\mathrm{(het)}(\eta),R^\mathrm{(het)}(s_0^\mathrm{(het)}(\eta)))$,
$(s_R^\mathrm{(het)}(\eta),R^\mathrm{(het)}(s_R^\mathrm{(het)}(\eta)))$
and $(s_\infty^\mathrm{(het)}(\eta),R^\mathrm{(het)}(s_\infty^\mathrm{(het)}(\eta)))$
where parameter $\eta$ is varying over whole definitional domain
of the quantities $s_0^\mathrm{(het)}$, $s_R^\mathrm{(het)}$ and
$s_\infty^\mathrm{(het)}$, respectively. Dotted balck curve
devide this quadrant into areas corresponding to different
stages. The curves $R^\mathrm{(het)}(s)$ corresponding to the
same regime have the same gray background color (the higher the regime
the darker the color).}
\label{chanHet}
\end{center}
\end{figure}

\begin{figure}[t]
\begin{center}
\includegraphics[scale=1]{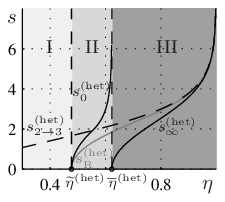}\quad\quad\includegraphics[scale=1]{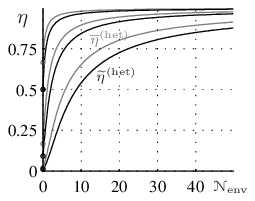}
\caption{Left: The quantities $s_{2\to3}^\mathrm{(het)}$,
$s_0^\mathrm{(het)}$, $s_\infty^\mathrm{(het)}$ (black) and
$s_R^\mathrm{(het)}$ (grey) are plotted vs $\eta$ for $N=5$, $\EuScript N_\mathrm{env}=1$.
Right:
$\overline\eta^\mathrm{(het)}$ and
$\widetilde\eta^\mathrm{(het)}$ are plotted vs $\EuScript
N_\mathrm{env}$ for the values of $N$ equal to $1$, $10$, $100$
(from top to bottom).}
\label{eta0refhet}
\end{center}
\end{figure}

Let us analyze the behavior of the function $R^\mathrm{(het)}(s)$ versus~$s$ 
(below the argument of $R^\mathrm{(het)}$ is assumed to be $s$)
for different values of $\eta$ and fixed $\EuScript N_\mathrm{env}$ (see
Fig.\ref{chanHet}).  By solving Eq. $\partial R^\mathrm{(het)}/\partial s=0$
in the third stage (see Eq.~\eqref{HR1useSimple}), one can show that $R^{(\rm
het)}(s)$ is monotonically increasing function if transmissivity belongs to the
interval\footnote{Notations for critical parameters of heterodyne rate are
chosen to be similar to those for capacity if saddle-point is imagined at $s=0$.}
$0<\eta\leqslant\widetilde\eta^{(\mathrm{het})}$ (we will call
this the \emph{first regime} analogously to capacity), where
\begin{multline}
\widetilde\eta^{(\mathrm{het})}=
\Bigl[\bigl(2\EuScript N_\mathrm{env}+1\bigr)^2+N\\
-\sqrt{N^2+\bigl(2N+1\bigr)\bigl(2\EuScript N_\mathrm{env}+1\bigr)^2}\Bigr]\\
\times\Bigl[2\EuScript N_\mathrm{env}\bigl(2\EuScript N_\mathrm{env}+1\bigr)\Bigr]^{-1},
\label{HetToldeEta}
\end{multline}
which is equal to $(N+1)^{-1}$ in the case of squeezed vacuum state (one needs to
take the limit $\EuScript N_\mathrm{env}\to0$ in Eq.~\eqref{HetToldeEta}). Then,
by equating the heterodyne values taken for $s=0$ and $s=\infty$ (see
Eqs.~\eqref{HR1useSimpleNenv0} and~\eqref{HetLimitFinal}), we obtain
the corresponding transmissivity value
\begin{multline}
\overline\eta^{(\mathrm{het})}=
\Bigr[8\EuScript N_\mathrm{env}\bigl(\EuScript N_\mathrm{env}+1\bigr)+N+2\\
-\sqrt{\bigl(N+2\bigr)^2+
16\,\EuScript N_\mathrm{env}\bigl(\EuScript N_\mathrm{env}+1\bigr)\bigl(N+1\bigr)}\Bigl]\\
\times\Bigl[4\EuScript N_\mathrm{env}\bigl(2\EuScript N_\mathrm{env}+1\bigr)\Bigr]^{-1},
\label{HetOverlEta}
\end{multline}
which becomes $2\,(N+2)^{-1}$ in the case of $\EuScript N_\mathrm{env}=0$. The latter
can be obtained by taking the limit $\EuScript N_\mathrm{env}\to\infty$ in
Eq.~\eqref{HetOverlEta} or by equating the relations $\log_2(1+\eta N)$ and
~\eqref{HetLimitFinal}.

If $\widetilde\eta^{(\mathrm{het})}<\eta\leqslant\overline\eta^{(\mathrm{het})}$
(the \emph{second regime}), one can consider squeezing value
$s^{(\mathrm{het})}_0$ defined by the equality 
$R^{(\rm het)}\bigl(s^{(\mathrm{het})}_0\bigr)=R^{(\rm het)}(0)$. 
In the second stage it equals
\begin{multline*}
s^{(\mathrm{het})}_0=
\ln\Biggl[1+
2\eta N\left[\phi_0^\mathrm{(het)}\right]^{-1}\\
\times\left\{
\left(2\left[\phi_0^\mathrm{(het)}\right]^{-1}+1\right)
\left(2\eta\left[\phi_0^\mathrm{(het)}\right]^{-1}-1\right)
-N\right\}^{-1}
\Biggr],
\end{multline*}
where $\phi_0^\mathrm{(het)}$ is defined similarly to $\phi_0$ (see Eq.~\eqref{phi0})
as the value of $\phi^\mathrm{(het)}$ (see Eq.~\eqref{Phi2ndStageHet}) taken in the point $s=0$.
In explicit form $\phi_0^\mathrm{(het)}$ reads
\begin{equation*}
\phi_0^\mathrm{(het)}=\frac{2\eta}{1+(1-\eta)\bigl(2\EuScript N_\mathrm{env}+1\bigr)}
\end{equation*}
In the third stage $s^{(\mathrm{het})}_0$ is given by the relation
\begin{equation*}
s^{(\mathrm{het})}_0=
\arcosh\bigl[\eta NF_0-1\bigr],
\end{equation*}
where
\begin{equation*}
F_0=
\frac{8\left[N-\left(\eta\left[\phi_0^\mathrm{(het)}\right]^{-1}-1\right)
\left(2\left[\phi_0^\mathrm{(het)}\right]^{-1}+1\right)\right]}
{\eta\left(2\eta\left[\phi_0^\mathrm{(het)}\right]^{-1}-1\right)
\left(2\left[\phi_0^\mathrm{(het)}\right]^{-1}+1\right)^2}.
\end{equation*}
One can show that $s^{(\mathrm{het})}_0\to\infty$ if
$\eta\to\overline\eta^{(\mathrm{het})}-0$, and \mbox{$s^{(\mathrm{het})}_0\to0$} if
$\eta\to\widetilde\eta^{(\mathrm{het})}+0$ (see also Fig.\ref{eta0refhet}).
If the environment is pure ($\EuScript N_\mathrm{env}=0$),
$s^{(\mathrm{het})}_0$ can be rewritten as
\begin{equation*}
s^{(\mathrm{het})}_0=
\ln\frac{(1-\eta)(\eta N+2)}{2-\eta\,(N+2)}
\end{equation*}
in second stage and as
\begin{equation*}
s^{(\mathrm{het})}_0=\arcosh\,
\frac{\eta^2(2N+1)^2-(1-\eta)^2-1}{2\,(1-\eta)}
\end{equation*}
in third stage.

Analogously, if $\overline\eta^{(\mathrm{het})}<\eta\leqslant1$ (the
\emph{third regime}), 
one can consider the quantity $s^{(\mathrm{het})}_\infty$,
such that $R^{(\rm het)}\bigl(s^{(\mathrm{het})}_\infty\bigr)=R^{(\rm het)}(\infty)$.
Due to the monotonicity of $R^{(\rm het)}(s)$ in the second stage, 
the value $s^{(\mathrm{het})}_\infty$ can only correspond to the third stage, and it is equal to
\begin{equation*}
s^{(\mathrm{het})}_\infty=\arcosh\left[F_\infty\phi_0-\sqrt{1+4\eta F_\infty}\right],
\end{equation*}
where
\begin{equation*}
F_\infty=N+\frac{1+\eta}2-\sqrt{\eta N+\left(\frac{1+\eta}2\right)^2}.
\end{equation*}
In particular, we have the limits
$$\lim_{\eta\to\overline\eta^{(\mathrm{het})}+0}s^{(\mathrm{het})}_\infty=0$$
and
$$
\lim_{\eta\to1}s^{(\mathrm{het})}_\infty=\infty.
$$
Also, if
$\eta>\widetilde\eta^{(\mathrm{het})}$ there is a minimum of $R^{(\rm het)}$ in the third stage 
corresponding to the value
\begin{multline*}
s^{(\mathrm{het})}_\mathrm{R}=\arcosh\Bigl[\Bigl\{N+\frac{1+\eta}2\\
-\sqrt{\eta\,(1+N)-\EuScript N_\mathrm{env}(1+\EuScript N_\mathrm{env})(1-\eta)^2}\Bigr\}\phi_0\\
-(2\EuScript N_\mathrm{env}+1)(1-\eta)\Bigr],
\end{multline*}
which has its limits
$$
\lim_{\eta\to\widetilde\eta^{(\mathrm{het})}+0}s^{(\mathrm{het})}_\mathrm{R}=0
$$
and
$$\lim_{\eta\to1}s^{(\mathrm{het})}_\mathrm{R}=\infty.$$
Taking into account the above considerations we have for optimal squeezing in environment
$s^{(\mathrm{het})}_\star$ (providing the highest heterodyne rate for a given transmissivity) the equality
\begin{equation*}
s^{(\mathrm{het})}_\star=
\begin{cases}
\infty,&\mbox{if}\quad0<\eta\leqslant\overline\eta^{(\mathrm{het})},\\
0,&\mbox{if}\quad\overline\eta^{(\mathrm{het})}\leqslant\eta<1,
\end{cases}
\end{equation*}
which is similar to the analogous relation for capacity~\eqref{sstar}.

\section{Multiple channel uses}\label{MemoryChannelsSec}

Let us now move to the case of multiple uses (multi-mode) of the lossy
bosonic channel.  We will consider those types of memory channel
environments which give rise to spectral problems (in general, symplectic
eigenvalues are not functions of matrix spectrum). One of the simplest models of
this class is
\begin{equation}
V_\mathrm{env}=\bigoplus_{k=1}^nV_{\mathrm{env},k},
\label{envmmode}
\end{equation}
where each
$V_{\mathrm{env},k}=V(\EuScript{N}_{\mathrm{env},k},s_{\mathrm{env},k})$ (see
Eq.~\eqref{ind1mode}) is the single-mode environment corresponding to $k$th
channel use.  
It follows from~\cite{HiroshimaPRA}
that optimal matrices $V_\mathrm{in}$ and
$V_\mathrm{mod}$ have the same form as~\eqref{envmmode}, \emph{i.e.} they are
direct sums of some single-mode matrices.  Then, the average amount of photons
per mode in $V_\mathrm{in}$ is related with the amount taken for each mode (see
Eq.~\eqref{indNsm}) as
\begin{equation}
N=\frac1n\sum_{k=1}^nN_{k}.
\label{indNmulti-mode}
\end{equation}
In the following it will be useful to work with \emph{total amount of input
photons}
\begin{equation*}
\mathcal{N}:=nN,
\end{equation*}
which will always be written in calligraphic font.  Note, that $\mathcal{N}=N$
for the single channel use. Similarly, we will search the maximum for
\emph{total capacity}
\begin{equation*}
\mathcal{C}:=n\,\underline C_n,
\end{equation*}
where
\begin{equation}
\underline C_n=\frac1n\sum_{k=1}^nC_{k}
\label{DefCk}
\end{equation}
with $C_k$ the capacity of the single ($k$th) channel use (mode) as studied in
Sec.~\ref{MLChannel}. Below we use the system of notations introduced in
Subsec.~\ref{SystemOfNotations} for the case of single channel use by adding
extra index (usually, $k$) to all quantities in order to indicate which channel
use the quantities are referred to.

Notice, that apart from the model~\eqref{envmmode}, also environment model of the
form~\eqref{ind00} (with commuting blocks $V_{\mathrm{env},qq}$ and
$V_{\mathrm{env},pp}$) gives rise to spectral problem.  It particular, in this case 
it also follows from~\cite{HiroshimaPRA} that the maximum of $\chi$-quantity~\eqref{HolDef} is
achieved with matrices $V_\mathrm{in}$ and $V_\mathrm{mod}$ of the same form
as~\eqref{ind00}, \emph{i.e.} with null off-diagonal blocks.
Furthermore, all diagonal blocks of all matrices will be mutually commuting. 
Such form of covariance matrices makes symplectic eigenvalues functions of the usual
eigenvalues, specifically
\begin{equation}
\nu_k=
\sqrt{o_{qk}o_{pk}},\qquad\overline\nu_k=\sqrt{\overline o_{qk}\overline o_{pk}},
\label{NusesnunuovMMC}
\end{equation}
where
\begin{equation*}
\begin{split}
&o_{uk}=\eta\,i_{uk} + (1-\eta)\,e_{uk},\\
&\overline o_{uk} = \eta\,(i_{uk}+m_{uk})+ (1-\eta)\,e_{uk}.
\end{split}
\end{equation*}
Both energy constraint~\eqref{enrestr} and symplectic
spectrum~\eqref{NusesnunuovMMC} are preserved under orthogonal transformations.
Thus, without affecting the final result, below we can consider all the involved
matrices to be diagonal (see also the discussion in the appendix of~\cite{NJP}).
Notice, that if all matrices are diagonal, then the optimal input state is pure (it
straightforwardly follows from the theorem~\ref{InpPurTheor2} applied to each
channel use).

More generally, according to the Williamson decomposition thereom, any
covariance matrix can be put in a diagonal form by acting with a symplectic
transformation~\cite{Williamson1936}. However, such a symplectic transformation
may not preserve the energy contraint.  One can hence restrict the consideration
to the class of models for which the symplectic transformation preserves the
energy constraint (these are jointly symplectic and orthogonal). In particular,
the models~\eqref{ind00} belong to this class. The general form of such
$V_\mathrm{env}$ matrices is presented in the Appendix of~\cite{NJP} (see
also~\cite{MemUnraveling}).

\subsection{Convex separable programming}\label{ConvSepProgrSubSec}

The optimization problem for multiple channel uses is formulated as follows.
One needs to find the maximum over the variables $i_{uk}$, $m_{uk}$, and
$m_{u_\star k}$ for the following functions\footnote{Here the homodyne rate
corresponds to the measurement of (generally) different quadratures for
different channel uses, where less noisy quadratures are used for information
transmission. Such definition of homodyne rate is different from those given by
the relation~\eqref{HomR}, where the same quadrature is measured in all modes.}:
\begin{align*}
&\underline{C}_n=
\frac1n\sum_{k=1}^n\left[g\left(\overline\nu_k-\frac12\right)-
g\left(\nu_k-\frac12\right)\right],
\\
&R^{(\mathrm{het})}_n=
\frac1n\sum_{k=1}^n\left[\log_2\overline\nu_k^{(\rm het)}-
\log_2\nu_k^{(\rm het)}\right],
\\
&R^{(\mathrm{hom})}_n=
\frac1{2n}\sum_{k=1}^n\left[\log_2\overline o_{u_\star k}-\log_2o_{u_\star k}\right]
\end{align*}
with the constraints
\begin{eqnarray*}
&&i_{uk}>0,\nonumber\\
&&m_{uk},m_{u_{\star} k}\geqslant0,\nonumber\\
&&\frac{1}{n}\sum_{k=1}^n\left[i_{uk}+\frac{1}{4i_{uk}}+m_{uk}+m_{u_{\star} k}\right]=2N+1.\qquad
\end{eqnarray*}
Then, the problem of finding the capacity\footnote{The case of rates is completely
analogous to that of capacity, therefore here it is omitted.} can be reformulated as
finding the maximum for sum of concave\footnote{The concavity of single-use
capacity $C_k$ over its energy constraint $N_k$ was proved in
Subsec.~\ref{ConcavityOfSolution}.} functions (each of them depending on one
variable)
\begin{equation}
\mathcal C(N)=\sum_{k=1}^nC_k
\label{ExtProb1}
\end{equation}
over the distribution $P(N_k)$ of positive numbers $N_k$ satisfying the constraint
\begin{align}
&\mathcal N=\sum_{k=1}^nN_k, 
\label{ExtProb2}\\
&N_k=\frac1{2}\left[i_{uk}+\frac1{4i_{uk}}+m_{uk}+m_{u_\star k}-1\right]\geqslant0, \nonumber
\end{align}
where $N_k$ is the amount of energy granted for $k$th mode (see
Eq.~\eqref{indNmulti-mode}), and $C_k=C_k(i_{uk},m_{uk},m_{u_\star k})$ (see the
definition~\eqref{DefCk}) is parametrized by fixed parameters $e_{uk}$,
$e_{u_\star k}$ and $\eta$, \emph{i.e.} $C_k$ only depends on the eigenvalues
belonging to $k$th mode.  Thus, the total optimization problem is splitted in
two tasks: the first task is the ``internal optimization'' solved in
Sec.~\ref{MLChannel}, \emph{i.e.} optimization inside each mode (see
``box''~\eqref{blackbox1use}) and the second task is the ``external
optimization'', \emph{i.e.} finding the optimal distribution $P(N_k)$ of the
total energy $\mathcal N$ over ``boxes'' to get maximal output sum
$\sum_{k=1}^nC_k$: 
\begin{align*}
&N_1\longrightarrow\boxed{C_1=C_1(N_1)}\longrightarrow C_1\\
&\;\;\;\quad\dots\dots\dots\dots\dots\dots\dots\\
&N_n\longrightarrow\boxed{C_n=C_n(N_n)}\longrightarrow C_n
\end{align*}

This ``external optimization'' problem is known in mathematics as \emph{convex
separable programming} which was solved in~\cite{StefStef},~\cite{StefBook}.  In
particular, the following theorem based on concavity of target function was
proved~\cite{StefStef}:

\bigskip

\begin{theorem}\label{StefTheor}

A feasible solution $\{N_k\}$ is an optimal solution to the
problem~\eqref{ExtProb1},~\eqref{ExtProb2} if and only if there exists a
$\lambda\in\mathbb{R}$ such that
\begin{equation}
N_k=0,\quad\qquad\qquad\qquad
\mbox{if}\quad\lambda\geqslant\frac{\partial C_k}{\partial N_{k}}(N_k=0),
\label{Theor1}
\end{equation}
\begin{equation}
N_k\,|\,\lambda=\frac{\partial C_k}{\partial N_{k}}(N_k),\:
\qquad\mbox{if}\quad\lambda<\frac{\partial C_k}{\partial N_{k}}(N_k=0).
\label{Theor2}
\end{equation}
\end{theorem}

\bigskip

\noindent Thus, the theorem states that any solution of ``external
optimization'' problem satisfying its Lagrange equations is optimal because it
is unique.  Also, it follows from the theorem that the dependence $\lambda(N)$ is
monotonic. Indeed, if $\lambda$ is increasing, then some modes can change
their ``case'' from~\eqref{Theor2} to~~\eqref{Theor1}, which results to zeroing
their contribution to $N=\sum_{k=1}^nN_k$.  Even if some modes remain in the
case~\eqref{Theor2}, their contribution $N_k$ is decreasing because of the concavity
and the monotonically increasing behavior of functions $C_k(N_k)$. Analogously,
lower $\lambda$ corresponds to higher $N$.

Below it will be convenient to use the \emph{threshold functions} (see also~\cite{NewNoiseChannel})
\begin{equation}
\begin{split}
&\lambda_{1\to2,k}\equiv\frac{dC_k}{dN_k}(N_k=0)={\frac\eta{o_{u_\star k}}g_1(\nu_k)},\\
&\lambda_{2\to3,k}\equiv\frac{dC_k}{dN_k}(N_{2\to3,k})=
\frac\eta{\overline\nu_k}g_1(\overline\nu_k)
\end{split}
\label{ThrFunsMultmode}
\end{equation}
defined analogously to single-mode relations~\eqref{Thr12Gen}
and~\eqref{Thr23Gen}, where quantity $N_{2\to3,k}$ is given by
Eq.~\eqref{Nthr23} applied to $k$th mode. Thus, the threshold functions are
generalizations of the single-use threshold values written in
$\lambda$-representation (see Subsec.~\ref{AlternativeRepr}).  
Taking into account~\eqref{DerThrRestr} one can see that
$\lambda\in(0,\lambda_\mathrm{max})$ for $N>0$, where
\begin{equation*}
\lambda_\mathrm{max}=\max_k\frac{\partial C_k}{\partial N_k}(N_k=0)
=\max_k\,{\lambda_{1\to2,k}}.
\end{equation*}

In the following the notion of \emph{stage} will be referred to each mode (in
complete analogy with the single use case presented in Sec.~\ref{MLChannel}).
It~allows the optimization problem to be interpreted as the search for the
optimal \emph{distribution of modes across stages}.  In particular, the case
$N_k=0$ holds if and only if $k$th mode belongs to the first stage, and the case
$\lambda=\lambda_\mathrm{max}$ corresponds to zero capacity, where all modes are
in the first stage.  Analogously, it follows from theorem~\ref{StefTheor}, that
if it is
\begin{equation*}
\lambda<\min_k\frac{\partial C_k}{\partial N_k}(N_k=0)=\min_k\,{\lambda_{1\to2,k}},
\end{equation*}
only the second and third stages exist (by comparing $N_k$ granted for $k$th
mode with its threshold value $N_{2\to3,k}$ one can obtain its actual stage). 

The proposition~\ref{NoiseEffectPropos} (see Subsec.~\ref{AlternativeRepr})
applied to multiple uses threshold functions~\eqref{ThrFunsMultmode} shows the
relationship between the level of noise in particular quadratures and their
participation to information transmission. For example, for fixed value of total
energy $N$, the $k$th mode can change its stage from first to second if the noise in
quadrature $e_{uk}$ or $e_{u_\star k}$ is sufficiently decreased. More
generally, one can say that it is the most optimal case when less noisy modes
get more input energy and thus transfer more information, which is similar to
the case of classical channels.

The ``external optimization'' problem is reducible to
single transcendental equation on $\lambda$
\begin{equation}
\mathcal N=\sum_{k=1}^nN_k(\lambda), 
\label{MMTrEq}
\end{equation}
which has single root because of theorem~\ref{StefTheor}. It can be solved
by using, e.g., method of bisection. Remember, that $N_k(\lambda)=0$ if
$\lambda>\lambda_{1\to2,k}$. Then, $N_k(\lambda)$ is given by
Eq.~\eqref{N2stLamRepr} if $\lambda_{2\to3,k}<\lambda<\lambda_{1\to2,k}$.
Finally, $N_k(\lambda)$ is given by Eq.~\eqref{Planck3st} if
$\lambda<\lambda_{2\to3,k}$.  Thus, equation~\eqref{MMTrEq} can be
considered as giving feasible solution for \emph{any} $\lambda$, the only
difference is that such a solution corresponds to another value of $\mathcal N$.

As far as the solution is unique it is sufficient to prove the convergence of the
bisection method applied to Eq.~\eqref{MMTrEq}, which can be done as follows.
Notice, that $\lambda$ and $\mathcal N$ are related each other by
one-to-one correspondence, and the dependence $\lambda(\mathcal N)$ is
monotonic. In particular, the limit $\lambda\to0$ corresponds to the limit
$\mathcal N\to\infty$, and the value $\lambda=\lambda_\mathrm{max}$ corresponds
to $\mathcal N=0$. Thus, as far as a unique $\lambda$ corresponds to a given
$\mathcal N$, the method of bisection applied to the transcendental
equation~\eqref{MMTrEq} for the variable $\lambda\in\bigl(0,\lambda_\mathrm{max}\bigr]$
always converges to the solution.

Apart from the considered ``blackbox'' approach, the given optimization problem
can be also interpreted in the following way.  There are two effective unknown
``variables" for the systems of Lagrange equations\footnote{Note, that each
distribution of modes across stages results to its own system of Lagrange
equations, where unknown variables are the eigenvalues of $V_\mathrm{in}$ and
$V_\mathrm{mod}$. As far as the system of Lagrange equations itself does not
provide effective method to find distribution of modes across stages, some
\emph{a proiri} properties are necessary to write a fast algorithm. In
particular, concavity and monotonic behavior of capacity are such properties
for the given problem.}: distribution of modes across stages and~$\lambda$. In the
simplest case, one of these variables can be set as internal and the another one
as external during optimization process. The algorithm proposed
in~\cite{StefStef} uses $\lambda$ as internal variable, while the above
algorithm uses distribution of modes across stages for that.  Since the latter
algorithm is usually faster, below we will make use of it.

\subsection{Classical capacity and rates}\label{ClCapAndRatesSubSec}

Remember, that explicit analytical solution of the optimization problem is not
possible and depends on the form of the threshold functions $\lambda_{1\to2,k}$,
$\lambda_{2\to3,k}$ defined by environment matrix $V_\mathrm{env}$ and
transmissivity $\eta$. However, if we are interested in finding approximate
values of capacity, e.g. $\underline C_n^{(0)}$ or $\underline C_n^{(1)}$
relying on quantity $\lambda^{(0)}(N)$ (see Eq.~\eqref{lambda0}), some
simplification of general method is possible. Below we show this using
$\underline C_n^{(0)}$ and $\underline C_n^{(1)}$ as examples, but
the generalization to the case of rates is straightforward.

Notice, that mode transcendental equation~\eqref{ModeTrEq1use} can be formally
written as the dependence $\overline o_{u_\star h}=f_h(i_{uh})$ for the $h$th channel use.  
Then, remember, that $\lambda^{(0)}$ (which is the amount of input photons
granted for each channel use in the $\lambda$-representation) is the same for all
modes in the third and the second stages. As far as the variable $\overline
o_{u_\star k}$ for any mode $k$ can be used as an equivalent replacement of
$\lambda^{(0)}$ (see Eq.~\eqref{lambda0}), we can introduce a new variable 
\begin{align}
x:=\,&\overline o_{qm}=\overline o_{pm}=\overline o_{ql}=\overline o_{pl}=...
\nonumber\\
=\,&\overline o_{qh}=f_h(i_{ph})=\overline o_{pt}=f_t(i_{qt}),
\label{chain}
\end{align}
getting a chain of equalities linking \emph{all} modes of the second and third
stages.  Here modes $m$ and $l$ belong to the third stage, while modes $h$ and
$t$ to the second stage ($m_{ph}=m_{qt}=0$).  Modes of the first stage are not
included in~\eqref{chain} and all give $V_\mathrm{in}$-eigenvalues equal
to~$\frac12$. If~some mode belongs to the third stage, its $V_{\rm in}$-eigenvalues
can be found from the relations~\eqref{inputEigs}. If some mode belongs to the
second stage, its input eigenvalues are given in Subsec.~\ref{TheSecondStage}
(see Eqs.~\eqref{iuapproxIUO} and~\eqref{FirstOrdNuses}). 

Taking into account stages discrimination, equation~\eqref{VoutAv} can be
rewritten as
\begin{equation}
\sum_{\{2,3|m_{uk}\neq0\}}\bigl[\eta\,(i_{uk}+m_{uk})+(1-\eta)\,e_{uk}\bigr]
=\bigl[2n_3+n_2\bigr]x,
\label{ModesRes}
\end{equation}
where $n_j$~is the number of modes belonging to $j$-th stage ($j=1,2,3;\;
n=n_1+n_2+n_3$) and $\sum_{\{2,3|m_{uk}\neq0\}}$ stands for the summation over
all eigenvalues of second and third stages, except for the $uk$-th ones
corresponding to $m_{uk}=0$.  Also, the energy constraint~\eqref{enrestr} can be
rewritten as
\begin{equation}
\sum_{\{2,3|m_{uk}\neq0\}}\bigl[i_{uk}+m_{uk}\bigr]=
2n\left[N+\frac12\right]-n_1-
{\sum_{k}}^{\prime\prime} i_{uk},
\label{enconstr}
\end{equation}
where $i_{uk}=f_k^{-1}(x)$ and the double prime sum extends over $uk$-th
eigenvalues of the second stage, such that $m_{uk}=0$.  Substituting
Eq.~\eqref{enconstr} into Eq.~\eqref{ModesRes} we get a transcendental equation
for the single variable $x$. Since all unknown eigenvalues can be expressed
through $x$ (see Eqs.~\eqref{chain}) we can formally arrive at
$\underline{C}_n^{(0)}$ and $\underline C_n^{(1)}$.

Notice, that as far as the relation $i_{uk}=f_k^{-1}(x)$ is explicit in the
zeroth-order and the first-order approximations (see Eqs.~\eqref{iuapproxIUO},
and~\eqref{FirstOrdNuses}), one can express the quantities
$\underline{C}_n^{(0)}$ and $\underline{C}_n^{(1)}$ as functions of solution of
only one algebraic equation (see Eqs.~\eqref{ModesRes} and~\eqref{enconstr}) for
one variable~$x$. 

When all modes are in the third stage we have the explicit analytical
solution and the equalities $\underline C_n=\underline{C}_n^{(0)}=\underline{C}_n^{(1)}$. 
In particular, it is
\begin{align}
\underline{C}_n=g\bigl[\eta N+(1-\eta)N_\mathrm{env}\bigr]
-\frac1n\sum_{k=1}^ng\bigl[(1-\eta)\EuScript N_\mathrm{env,k}\bigr],
\label{capacita}
\end{align}
where
\begin{equation}
N_\mathrm{env}=\frac1n\sum_{k=1}^nN_{\mathrm{env},k}
\label{Menv}
\end{equation}
is the average number of photons in the multiple uses environment.  The
analytical lower bound given by Eq.\eqref{capacita} generalizes the expression
presented in~\cite{NJP}. Analogously, in the case of all modes belonging to the
third stage the heterodyne rate reads
\begin{align*}
&R_n^{(\mathrm{het})}=
\log_2\left[\eta N+(1-\eta)N_\mathrm{env}^\mathrm{(het)}+\frac12\right]
\nonumber\\
&\qquad\qquad\qquad\qquad
-\frac1n\sum_{k=1}^n\log_2\left[
(1-\eta)\EuScript{N}_{\mathrm{env},k}^\mathrm{(het)}+\frac12\right],
\end{align*}
where $N_\mathrm{env}^\mathrm{(het)}$ is defined similarly to
Eq.~\eqref{Menv}.

If all modes are in the second stage, the homodyne rate reads (see
Eq.~\eqref{SndStFou})
\begin{equation*}
R_n^\mathrm{(hom)}=
\frac1n\sum_{k=1}^n\log_2\left[
\sqrt{(\phi_k/4)^2+\phi_k\,T_1^\mathrm{(hom)}}-\phi_k/4\,\right],
\end{equation*}
where $T_1^\mathrm{(hom)}$  (see Eq.~\eqref{Tdefhomhet}) is given by the root of
equation (see the relations~\eqref{lambdaToN2st} and~\eqref{iuapproxIUOlambda})
\begin{equation*}
\frac1n\sum_{k=1}^n\bigl[i_{uk}^{\phantom{1}}-\phi_{k}^{-1}\bigr]=2N+1-T_1^\mathrm{(hom)}
\end{equation*}
with
\begin{equation*}
i_{uk}=\frac12\left[\sqrt{(\phi_k/4)^2+\phi_k\,T_1^\mathrm{(hom)}}-\phi_k/4\right].
\end{equation*}

If the number of channel uses tends to infinity the discussed procedure can be
properly generalized by changing the transcendental equations (e.g.
Eqs.~\eqref{ModesRes} and~\eqref{enconstr}) to equations on functions (spectral
densities). However, if the considered model has some symmetry over stages, the
general solution can be further simplified by considering some parameters which
mark the boundaries between regions of modes belonging to different stages.  In
Subsec.~\ref{OmegaModelSec} we will show an example along this line.

\subsection{Application to a particular memory channel}\label{OmegaModelSec}

In this subsection we look for the capacity of channels whose environment is
described by a covariance matrix of the form
\begin{equation}
V_\mathrm{env}=\left[\EuScript{N}_{\rm env}+\frac12\right]\left(
\begin{array}{cc}
e^{s\Omega} & 0\\
0 & e^{-s\Omega}
\end{array}
\right),
\label{Venv}
\end{equation}
where $\Omega$ is a real symmetric $n\times n$ matrix and $s\in\mathbb{R}$ is a
parameter describing the environment properties. In particular, we will consider
the case of environment model~\eqref{Venv} with
\begin{equation*}
\Omega_{ij}=\delta_{i,j+1}+\delta_{i,j-1};\quad i,j=1,\ldots,n
\end{equation*}
describing a specific lossy bosonic channel with memory~\cite{PRA}, which will
be referred to as \emph{$\Omega$-model of the environment}.  Notice, that by
taking $\Omega_{ij}=\delta_{ij}$ we recover the case of the memoryless channel.

The parameter $s$ in Eq.~\eqref{Venv} represents the degree of correlation among
environment modes. We are interested in the asymptotic behavior of this channel.
That implies to take the limit $n\to\infty$ in the equations of
Subsec.~\ref{ClCapAndRatesSubSec}.  It can be treated for some relations as the
limit of Riemann sums resulting to the integral expressions. Thus, instead of a
set of equations on eigenvalues we get a set of equations on functions which are
spectral densities for the involved (infinite-dimensional) matrices. Below we
denote the spectral densities by the same symbols as proper eigenvalues, but
written in calligraphic and replacing the mode number $h$ by a continuous
parameter $\xi$, \emph{i.e.}, $i_{uh}\to\mathcal I_{u\xi}$, $o_{qh}\to\mathcal
O_{q\xi}$, etc.

\begin{figure}[t]
\begin{center}
\includegraphics[scale=1]{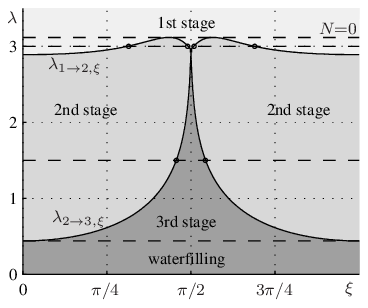}
\caption{Threshold functions $\lambda_{1\to2,\xi}$ and $\lambda_{2\to3,\xi}$ vs
$\xi$ for $\Omega$-model. The value of other parameters are $\eta=0.4$, $s=0.5$,
$\EuScript{N}_\mathrm{env}=0.01$. The horizontal dashed lines correspond to
different values of $N$, where the top line is the case of $N=0$ and the bottom
line is the border-case when all modes are in the third stage (below it we have
``waterfilling solution''). The points at which horizontal lines (corresponding
to some values of $\lambda$ or, equivalently, $N$) cross threshold functions
$\lambda_{1\to2,\xi}$ and $\lambda_{2\to3,\xi}$ mark borders between modes with
different stages. The regions corresponding to different stages are filled by
different grey colors.}
\label{thr_funcs_art}
\end{center}
\end{figure}

\begin{figure}[t]
\includegraphics{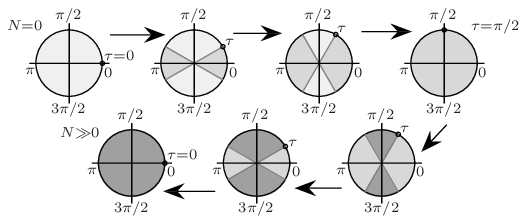}
\caption{Schematic representation of the ``quantum water filling" for the
capacities $\underline C^{(0)}$ and $\underline C^{(1)}$ and the environment
model $\Omega_{ij}=\delta_{i,j+1}+\delta_{i,j-1}$ (it follows from the threshold
functions $\lambda_{1\to2,\xi}^{(0)}$ and $\lambda_{2\to3,\xi}^{(0)}$ of
variable $\xi$). The angle $\xi$ parametrizing the spectral density corresponds
to polar angle.  White, grey and black sectors correspond to the first, second
and third stages, respectively. Arrows show change of stages with increasing of
$N$. The parameter $\tau$ marks the points of stage change.}
\label{tau_scheme}
\end{figure}

\begin{figure}[t]
\begin{center}
\includegraphics[scale=1]{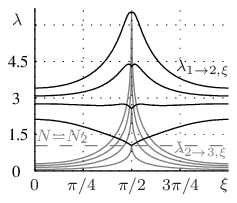}\quad\includegraphics[scale=1]{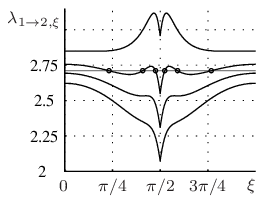}
\caption{Threshold functions $\lambda_{1\to2,\xi}$ (black) and $\lambda_{2\to3,\xi}$ (grey) 
are plotted vs $\xi$ for capacity of the channel with the
$\Omega$-model of environment. The values of parameters at the left are
$\EuScript N_\mathrm{env}=0.01$, $s=1$, $\eta$ (from bottom to top): $0.15$, $0.35$, $0.55$, $0.75$.
The values of parameters at the right are
$\EuScript N_\mathrm{env}=0.01$, $s=1$, $\eta$ (from bottom to top): $0.29$, $0.32$, $0.35$, $0.4$.
Horizontal lines correspond to particular chosen values of input energy.}
\label{thrfunvseta}
\end{center}
\end{figure}

\begin{figure}[t]
\begin{center}
\includegraphics[scale=1]{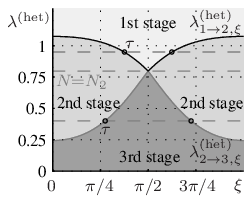}\quad\includegraphics[scale=1]{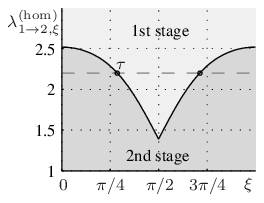}
\caption{Threshold functions for heterodyne (left) and homodyne (right) rates
are plotted for the parameters $\EuScript N_\mathrm{env}=0.5$, $s=1$, $\eta=0.65$
(channel environment is given by $\Omega$-model).
Horizontal lines correspond to particular chosen values of input energy.
These lines cross threshold functions at the points (marked by $\tau$ if
$\xi\in\left[0,\frac\pi2\right]$) corresponding to stage change.}
\label{thrfunrates}
\end{center}
\end{figure}

\begin{figure}[t]
\begin{center}
\includegraphics{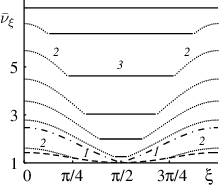}\quad\quad\includegraphics{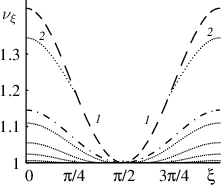}
\end{center}
\includegraphics{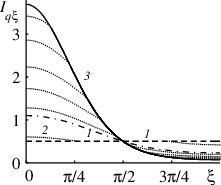}\quad\quad\includegraphics{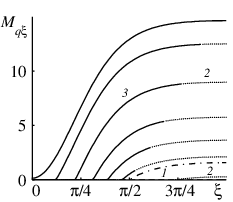}
\caption{Going from top-left clockwise the spectral densities
$\bar\nu_\xi,\nu_\xi,\mathcal M_{q\xi}, \mathcal I_{q\xi}$ (for
$\Omega_{ij}=\delta_{i,j+1}+\delta_{i,j-1}$) are plotted vs the parameter $\xi$
for $N=0,0.05,0.67,1,2,3.5,6,9,11$ (from bottom to top curve for quantities
$\bar\nu_\xi,\mathcal M_{q\xi}, \mathcal I_{q\xi}$, and from top to bottom curve
for quantity $\nu_\xi$).  Solid, dotted and dashed parts of curves correspond to
third, second and first stages, respectively.  Dash-dotted curve corresponds to
the case of all modes belonging to the second stage.  The values of other
parameters used are $\EuScript N_\mathrm{env}=s=1$, $\eta=0.5$. Numbers
\emph{1}, \emph{2} and \emph{3} are used to indicate the regions with
corresponding stages.}
\label{OmegaWaterFil}
\end{figure}

It is convenient to use the parameter $\xi$ as arising from the spectrum of
$V_\mathrm{env}$-matrix~\cite{PRA}
\begin{equation}
\mathcal E_{u\xi}=\left(\EuScript N_\mathrm{env}+\frac12\right)e^{\pm2s\cos\xi},
\label{Euxi}
\end{equation}
labeling both modes (if $\xi\in[0,\pi]$) and eigenvalues (if $\xi\in[0,2\pi]$).
Plus and minus in Eq.~\eqref{Euxi} stand for $u=q$ and $u=p$, respectively. Due
to the mirror symmetry of eigenvalues~\eqref{Euxi} over quadratures, the
symplectic spectrum and the distribution of modes across stages have to be
symmetric with respect to the point $\frac\pi2$, therefore we restrict ourselves to
consider spectral densities only defined in the interval $\bigl[0,\frac\pi2\bigr]$.

Threshold functions $\lambda_{1\to2,\xi}$ and
$\lambda_{2\to3,\xi}$ (and also their analogs for rates) for
$\Omega$-model are shown in Figs.~\ref{thr_funcs_art},
\ref{thrfunvseta} and~\ref{thrfunrates}. 
In general, the equation $\lambda_{1\to2,\xi}=\lambda$ (for the variable $\xi$) can have up to
three different roots in the interval $\left[0,\frac\pi2\right]$. Below we will calculate the
capacities $\underline C^{(0)}$ and $\underline C^{(1)}$ which are essentially
simpler as the equation $\lambda_{1\to2,\xi}^{(0)}=\lambda$ has at maximum a
single root $\tau$ which marks the boundary between the modes belonging to the
first and second stages (the equation $\lambda_{2\to3,\xi}=\lambda$ has at most one root).

Suppose that all modes belong to the third stage, which holds true if (it can be
obtained, e.g., from Eq.~\eqref{cueigs} or~\eqref{Nthr23} by combining it with
Eq.~\eqref{Euxi}, see also Appendix in Ref.~\cite{PRA})
\begin{equation*}
w:=\frac1{2|s|}\ln\frac{\eta\,(2N+1)+(1-\eta)(2N_{\rm env}+1)\,I_0(2s)}{\eta+(1-\eta)(2N_{\rm env}+1)}
\geqslant1,
\end{equation*}
where $I_0$ is the modified Bessel function of the first kind and zero-order.
The capacity $\underline{C}$ in this case is given by
Eq.~\eqref{CC1useSimple}, where the amount of environment photons $N_\mathrm{env}$
is given by Eq.~\eqref{NindDef} after a formal replacement $\cosh s\to I_0(2s)$.
This example explicitly shows the possibility of an enhancement of the capacity
with increasing degree of memory $s$ (however, at the cost of increasing the
amount of environment photons $N_\mathrm{env}$).

\begin{figure}[t]
\includegraphics{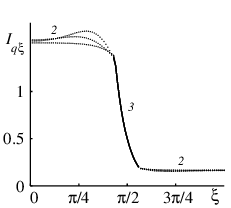}\quad\includegraphics{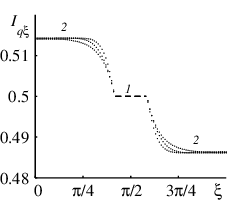}
\caption{Exact solution, first-order and zeroth-order approximations for
spectral density $\mathcal I_{q\xi}$ vs $\xi$ for $N=1$ (left) and $N=0.01$
(right). The values of other parameters are \mbox{$\EuScript
N_\mathrm{env}=0.5$}, $s=2.5$, $\eta=0.95$. Solid, dotted and dashed parts of
curves correspond to third, second and first stages, respectively. Functions
with maximum and minimum variations correspond to exact solution and
zeroth-order approximation, respectively.}
\label{IqApps}
\end{figure}

If $w<1$ we can have one of the following distributions of modes across stages
according to the properties of the threshold functions
$\lambda_{1\to2,\xi}^{(0)}$ and $\lambda_{2\to3,\xi}^{(0)}$ (see
Fig.~\ref{tau_scheme}):
\begin{itemize}
\item[\emph{i)}]
a mixture of the second and the third stages
\emph{(2,3,2)};
\item[\emph{ii)}]
a mixture of the second and the first stages \emph{(2,1,2)};
\item[\emph{iii)}]
all modes belonging to the second stage \emph{(2,2,2)} which happens for a
single value\footnote{Do not confuse this definition of $N_2$ with that used in
Subsec.~\ref{ConvSepProgrSubSec}.} $N_2$ of the parameter $N$, given $s$, $\eta$
and $\EuScript N_\mathrm{env}$.
\end{itemize}
If $N>N_2$ or $N<N_2$ we have the \emph{(2,3,2)} or \emph{(2,1,2)} case with the
center of the interval $[0,\pi]$ filled by the third or the first stage,
respectively.  We label by $\tau\in\left[0,\frac\pi2\right]$ the point corresponding to the
boundary between the regions of modes corresponding to different stages.  The
possible distributions of modes across stages and the dependence of $\tau$ from
$N$ are sketched in Fig.\ref{tau_scheme}. Notice, that at the point $\tau$ we
must have $\mathcal{\overline O}_{u\tau}=\mathcal O_{u\tau}$ which can be
rewritten as
\begin{equation}
x=x(\tau)=\eta\,\mathcal I_{u\tau}+(1-\eta)\,\mathcal E_{u\tau}.
\label{continuation}
\end{equation}
Here $u=q$ gives $\mathcal I_{q\tau}=e^{2s\cos\tau}/2$ (see
Eq.~\eqref{inputEigs}) for \emph{(2,3,2)} case and $u=p$ gives $\mathcal
I_{p\tau}=\frac12$ for \emph{(2,1,2)} case (we use different quadratures in these
cases because of either $q$ or $p$ quadrature changes its stage in the interval
which contains $\tau$).

Then, the transcendental equation for $x$ (see Eqs.~\eqref{ModesRes}
and~\eqref{enconstr}) can be rewritten as an equation for $\tau$
\begin{equation}
\eta\left[N+\frac{\tau_1}\pi-\frac1\pi\int_0^\tau\mathcal{I}_{q\xi}\,d\xi\right]
+\frac{1-\eta}\pi\int_0^{\tau_2}\mathcal E_{p\xi}d\xi=\frac{\tau_2}{\pi}x,
\label{OmegaTrEq}
\end{equation}
where $(\tau_1,\tau_2)$ is equal to $(\tau,\tau)$ for \emph{(2,1,2)} and to
\mbox{$\left(\frac\pi2,\pi-\tau\right)$} for \emph{(2,3,2)}. Moreover, $x$ is given by
Eq.~\eqref{continuation} and $\mathcal{I}_{q\xi}$ is the spectral density for
the second stage which can be found as solution of functional equation obtained
from Eq.~\eqref{iuapproxIUO} (or Eq.~\eqref{FirstOrdNuses} in the case of
$\underline C^{(1)}$) after the replacements discussed at the begining of this
subsection. By substituting $\tau=\frac\pi2$ in Eq.~\eqref{OmegaTrEq} we find $N_2$.
Comparing it with the actual energy restriction $N$ we get the correct value of
$\lambda^{(0)}$ and the distribution of modes across stages.  Then, solving
Eq.~\eqref{OmegaTrEq} with the found distribution of modes across stages we
arrive at $\tau$ and $x$. Finally, $\underline{C}^{(0)}$ is expressed through
these parameters as follows (see Eqs.~\eqref{HGaussian}, \eqref{underC}
and~\eqref{HolDef}):
\begin{align*}
\underline{C}^{(0)}&=\left(1-\frac2\pi\tau_3^{\phantom{1}}\right)\left[g\left(x-\frac12\right)
-g\bigl((1-\eta)\,\EuScript N_\mathrm{env}\bigr)\right]\\
&+\frac2\pi\int_0^\tau\left[g\left(\sqrt{x\mathcal{O}_{q\xi}}-\frac12\right)-
g\left(\sqrt{\mathcal{O}_{q\xi}\mathcal{O}_{p\xi}}-\frac12\right)\right]d\xi,
\end{align*}
where
\begin{align*}
&\mathcal{O}_{q\xi}=\eta\,\mathcal{I}_{q\xi}+(1-\eta)\,\mathcal E_{q\xi},\\
&\mathcal{O}_{p\xi}=\frac\eta4\,\mathcal{I}_{q\xi}^{-1}+(1-\eta)\,\mathcal E_{p\xi},
\end{align*}
$\tau_3^{\phantom{1}}$ is equal to $\pi/2$ for \emph{(2,1,2)} and to $\tau$ for \emph{(2,3,2)}.

The solution of the optimization problem for multiple channel uses can be
interpreted as ``quantum waterfilling" in analogy with usual (classical)
``waterfilling" introduced for classical Gaussian channels with memory (see
e.g.~\cite{NewNoiseChannel}, \cite{HSH} and~\cite{CoverThomas}).  The dependence
of the found spectral densities (also symplectic ones) from $N$ is similar to
filling a vessel with water. The form of the vessel is defined by the model
$V_\mathrm{env}$ and transmissivity $\eta$.  The symplectic spectral density
$\overline\nu_\xi$ goes always up by increasing $N$ (with respect to
$\nu_\xi(N=0)$), while $\nu_\xi$ goes always down (or does not change). For
environment models showing correlation (memory) among modes, the presence of the
second stage gives rise to capillary effects on the edges of the vessel
resulting to a ``water level" with meniscus form.  This ``quantum water filling"
effect for the considered model is shown in Fig.\ref{OmegaWaterFil} for
symplectic spectral densities $\overline\nu_\xi$, $\nu_\xi$ and spectral
densities $\mathcal M_{q\xi}$, $\mathcal I_{q\xi}$.  Graphs of $\mathcal
I_{q\xi}$ calculated through exact mode transcendental equation, zeroth-order
and first-order approximations are shown in Fig.\ref{IqApps}. Despite some
visible difference between exact and approximate spectral densities the
corresponding symplectic spectral densities are almost equal, thus resulting to
the difference less than 0.05\% between the capacities. The small value of this
difference comes from the fact that the Holevo-$\chi$ has zero derivative with respect to
the eigenvalues of $V_\mathrm{in}$ and $V_\mathrm{mod}$ in the neighborhood of
the solutions of Lagrange equations (as they are equations for optimization problem).

In Fig.\ref{MenvFun}-left the capacity $\underline C^{(1)}$ for $\Omega$-model
is plotted versus $s$ for different values of $\eta$. The universal
limit~\eqref{Climit} for $s\to\infty$ is still valid.

\subsection{Optimal channel memory and superadditivity}

\begin{figure}[t]
\begin{center}
\includegraphics[scale=1]{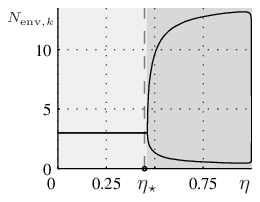}\quad\includegraphics[scale=1]{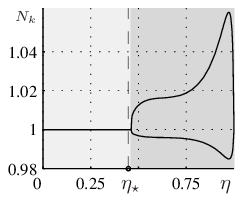}
\includegraphics[scale=1]{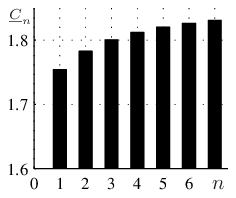}
\caption{Nontrivial behavior of optimal memory for capacity. 
Amounts of photons $N_{\mathrm{env},k}$ (top-left) and
$N_k$ (top-right) 
corresponding to optimal memory for capacity are plotted as functions of transmissivity $\eta$.
The values of other parameters are $N_\mathrm{env}=3$, $N=1$ and $n=5$.
The lighter and the darker backgrounds indicate the
additive and superadditive regions of transmissivity, correspondingly.
Vertical dashed lines at
$\eta=\eta_\star$ 
mark the analytically estimated boundary between additive and superadditive regions.
Bottom: capacity for optimal memory model is plotted vs amount of channel uses $n$.
The values of other parameters are $N_\mathrm{env}=3$, $N=1$ and $\eta=0.8$.} 
\label{optmemcap}
\end{center}
\end{figure}

\begin{figure}[t]
\begin{center}
\includegraphics[scale=1]{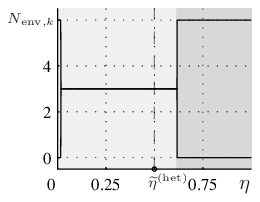}\quad\includegraphics[scale=1]{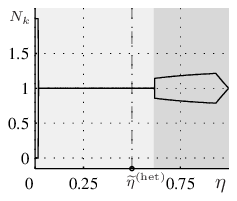}
\includegraphics[scale=1]{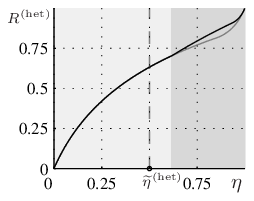}
\caption{Nontrivial behavior of optimal memory for heterodyne rate. 
Amounts of photons $N_{\mathrm{env},k}$ (top-left) and $N_k$ (top-right) 
corresponding to optimal memory for heterodyne rate
are plotted as functions of transmissivity $\eta$.
Bottom: heterodyne rate for memoryless model
(grey curve)
and for optimal memory model (black curve).
The values of other parameters for all three graphs are $N_\mathrm{env}=3$, $N=1$ and $n=2$.
The lighter and the darker backgrounds indicate the
additive and superadditive regions of transmissivity, correspondingly.
Vertical dashed lines at
$\eta=\widetilde\eta^\mathrm{(het)}$ 
mark the analytically estimated boundary between additive and superadditive regions.}
\label{optmemhet}
\end{center}
\end{figure}

Finally, let us discuss the role of squeezing and memory in lossy bosonic
channel.  Considering the capacity~\eqref{CC1useSimple} as a function on the
set of environment models with fixed $N_\mathrm{env}$, one can see that it shows
\emph{violation of quadrature symmetry}. In fact despite the symmetry of all
equations over quadratures, the maximum of $\underline{C}$ is achieved when
$e_q\ne e_p$ (see also~\cite{HSH}).  This also follows from the environment purity theorem proved for
the single channel use (see Appendix~\ref{EnvPurTh}). By applying this theorem
to each channel use for the case of memory channel one can see that optimal
environment can always be chosen pure.

\begin{figure}[t]
\includegraphics{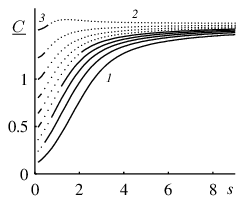}\quad\includegraphics{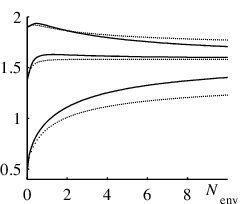}
\caption{On the left, the quantity  $\underline{C}^{(1)}$ is plotted vs $s$ for
values of $\eta$ going from $0.1$ (bottom curve) to $0.9$ (top curve) with step
$0.1$. The values of the other parameters are $N=\EuScript N_\mathrm{env}=1$.
Solid parts of curves correspond to the third and first stages, respectively.
Dotted part of curves correspond to the second stage. Numbers \emph{1}, \emph{2}
and \emph{3} are used to indicate the regions corresponding to the cases
\emph{(2,1,2)}, \emph{(2,3,2)} and \emph{(3,3,3)}, respectively.  On the right,
the maximum of $\underline{C}^{(1)}$ over $V_\mathrm{env}$ (\emph{i.e.} over
parameters $s$ and $\EuScript N_\mathrm{env}$) is plotted vs $N_\mathrm{env}$
for values of $\eta=0.1,0.5,0.9$ going from bottom to top curve. Solid and
dotted curves corresponds to \mbox{$\Omega=\delta_{ij}$} and
$\Omega_{ij}=\delta_{i,j+1}+\delta_{i,j-1}$, respectively. The value of  the
other parameter is $N=1$.}
\label{MenvFun}
\end{figure}

Now let us analyze the symmetry of the capacity over modes.  Suppose, that the
average (per mode) amount of photons in the environment $N_\mathrm{env}$ is
fixed and the capacities for the single use of memoryless channel and multiple
uses of memory channel (e.g. for $\Omega$-model) are compared.  As far as the
Holevo-$\chi$ quantity~\eqref{HolDef} is symmetric over modes, one can expect
that the capacity for the single channel use will always be higher. However,
this is not true as results from the \emph{violation of mode symmetry}.
Indeed, this can be seen in Fig.\ref{MenvFun}-right where the capacity
$\underline{C}^{(1)}$ maximized over parameters $s$ and $\EuScript
N_\mathrm{env}$ (thus, we have always $\EuScript N_\mathrm{env}=0$) for memory
and for memoryless cases is plotted versus~$N_\mathrm{env}$.
We can see that the $\Omega$-model for some parameters values provides higher
capacity than memoryless model. Unfortunately the form of the \emph{optimal} (in terms of
capacity) memory for the channel is still unknown.
We consider the finding of the optimal channel memory to be important and challenging
problem.

As far as the optimal environment $V_\mathrm{env}$ can be always chosen in pure
state, each its $k$th mode can be completely characterized by its squeezing
$s_k$. Hence, the problem of finding optimal channel memory 
can be reformulated as finding the form of the function $s(k)$ (or
$s(\xi)$ for the case of $n\to\infty$).
This function is not a constant, but numerical study of this problem
in simplest situations shows that only two different values of $s(k)$ are
possible for all $k$ and given values of $\eta$, $N$ and $N_\mathrm{env}$.

The above properties can be also treated from the superadditivity viewpoint.
First, let us discuss the memoryless channel capacity.  It was proved in
Subsec.~\ref{ConcavityOfSolution} that the one-shot capacity is monotonically
increasing and concave function of $N$. In this case convex separable
programming method (see Subsec.~\ref{ConvSepProgrSubSec}) guarantees that
optimal input state is the direct sum of identical single-use matrices.  It
automatically implies additivity of memoryless capacity.  As far as concavity
was also proved for rates, the conclusions valid for capacity are also
applicable for the rates.

However, the problem of additivity can be posed in another way. Quite generally
one can compare different multi-mode environments containing (in average) the
same amount of photons and having the same purity. In particular, the case of
pure states is the most optimal as it is supported by environment purity
theorem. In this case it straightforwardly follows from the dependence
$\underline C(s)$ studied in Sec.~\ref{CritParsSec} that the dependence of
$\underline C(N_\mathrm{env})$ (e.g. for pure environment state, see
Eq.~\eqref{NindDef}) is in general non-monotonic, which guarantees optimality of
non-homogenuous distribution of photons $N_\mathrm{env}$ over environment modes
for some channel parameters. 

In particular, one can expect that if $\eta<\eta_\infty$ and
$N\geqslant\widetilde N(0)\approx0.3578$, then capacity is additive. In fact, in
this case the dependence $\underline C(N_\mathrm{env})$ corresponds to the
concave and monotonically growing functions.  Numerical calculations shows that
in this region of parameters capacity is indeed additive. Similarly, if
$\eta>\eta_\infty$, then $\underline C(N_\mathrm{env})$ has local maximum in the
interval $0<N_\mathrm{env}<\infty$ and numerical calculations shows that
capacity is superadditive (non-homogenuous distribution of environment energy of
modes is optimal) for some values of input energy. 
This allows us to conjecture, that the transitions between
superadditive and additive cases happen at critical and supercritical parameters
of single channel use. Notice, that heterodyne rate is in general also
non-monotonic function of $N_\mathrm{env}$, therefore it is also subjected to
superadditivity property (see Fig.\ref{optmemhet}).

The value of transmissivity $\eta$ corresponding to transition from
additive to superadditive region for given parameters $N$, $\EuScript N_\mathrm{env}$
can be qualitatively estimated in the following way. In the case of capacity,
this transition may happen close to the point $\eta=\eta_\star$ corresponding
to the maximum of $\underline C(N_\mathrm{env})$ (or $\underline C(s)$ because of
purity) for fixed value of $N$. Similarly, in the case of heterodyne rate,
one can roughly use $\eta=\widetilde\eta^\mathrm{(het)}$ (see Eq.~\eqref{HetToldeEta})
to estimate the transition point.

\section{Conclusion}\label{ConclusionSec}

\IEEEPARstart{I}{n} this paper, we have developed powerful and versatile
optimization methods for the estimation of Gaussian quantum channels' capacities
and rates.  We have applied them to the lossy bosonic channel in both memoryless
and memory setting by restricting to Gaussian states.

First, we have thoroughly characterized the memoryless channel, thus
generalizing the results of \cite{HW,GG2}.  To do that we have exploited the
single-mode channel whose environment's covariance matrix $V_\mathrm{env}$ can
be described by two parameters: squeezing $s$ and average amount of thermal
photons $\EuScript{N}_\mathrm{env}$. Then, to completely specify the channel
usage we have fixed the values of transmissivity $\eta$ and input energy $N$.
It is the latter value that defines the kind of solution for the capacity $C$.
For $N$ increasing from $0$ to $+\infty$ we have found three different
\emph{stages}, each characterized by a solution of a given form.

We have proved that the one-shot capacity is a concave and monotonically
increasing function of $N$.  Thus, as byproduct we have gotten the additivity of
the memoryless capacity
assuming covariance matrices for modulation, channel
environment and input states to be mutually commuting.
Moreover,  due to this
property the derivative $dC(N)/dN$ can be used as the equivalent replacement for
the amount $N$ of photons granted for channel input, thus providing another
channel's representation.  Within this representation (called
$\lambda$-representation) is easily visualizable the geometry of the stages
transitions. 

The one-shot capacity turns out to be a monotonic function of all
parameters, except of environment squeezing. This makes the latter a special
parameter.  In particular taking the limit $s\to+\infty$  we have defined
different \emph{regimes} depending on how the capacity tends this limit. This is
determined by the value of transmissivity and amount of environment thermal
photons.  \emph{Critical} values for these parameters can be defined at boundary
of different regimes. Similarly, other regimes and critical parameters can be
considered analyzing the other properties of $C(s)$ function.  Totally we have
defined five different regimes and four triads of critical parameters, which
characterize the existence and values of specific points of $C(s)$.

Already from that we can draw some general conclusions about the channel's
properties.  For instance, if
\begin{equation}
N\geqslant
\bigl[{\scriptstyle{\displaystyle\sqrt{3/2+5/({\scriptstyle2\sqrt{3}})}-1}}\bigr]/2
\approx0.3578,
\label{FC1}
\end{equation}
then $C(s)$ is always monotonic
over $0<s<+\infty$ if 
\begin{equation}
\eta\leqslant1-1/\sqrt{3}
\label{FCa}
\end{equation}
and has no more than one maximum in this interval otherwise. Also, $C(s)$ has no
more than one maximum if 
\begin{equation}
\EuScript{N}_\mathrm{env}\geqslant
\bigl[\bigl(\sqrt{3}-2/\sqrt{5}\bigr)^{-1}-1\bigr]/2\approx0.0969.
\label{FC2}
\end{equation}
Another example is the case of $C(0)=C(\infty)$ for $N\to\infty$, 
which is possible only if 
\begin{equation}
\label{FCb}
\eta\geqslant2/e,
\end{equation}
where inequality is saturated by pure environment state.

As far as the critical parameters in general depend on $N$, $\EuScript
N_\mathrm{env}$ (or $N,\eta$ --- depending on the parameter varied in analyzing
the behavior of $C(s)$) and not all of them exist in all the regimes, we have
defined three \emph{domains}.  Each domain is characterized by existence and/or
relations among critical parameters.  In turn, \emph{supercritical} values for
$N$ and $\EuScript N_\mathrm{env}$ can then be defined at boundary of different
domains.  The nontrivial global maximal or minimal values of critical and
supercritical parameters must be intended as fundamental constants
characterizing the channel.  Few of such constants which can be expressed in
radicals are the above numbers~\eqref{FC1}, \eqref{FC2} for supercritical
and~\eqref{FCa}, \eqref{FCb} for critical parameters.

Summarizing, in the space of parameters $N$, $\EuScript N_\mathrm{env}$ we have
defined two functions and by equating them to zero  we have divided the space
into three parts (domains).  The boundaries of domains define the supercritical
parameters. In turn domains define the possible regimes (five at maximum).
Critical parameters come out at the boundaries of regimes (this time in the
space of parameters $\eta$, $N$, $\EuScript N_\mathrm{env}$).  Then, the
towering achievement is the following route to determine the channel's
``state'':

\begin{itemize}
\item
find the channel \emph{domain} by comparing the actual
$N$, $\EuScript N_\mathrm{env}$ with their 
supercritical values (it gives the set of possible
\emph{regimes});
\item
find the channel \emph{regime} by comparing the actual $\eta$, $\EuScript
N_\mathrm{env}$ with their critical values;
\item
find the relevant values of squeezing parameter for the given channel
\emph{regime} and compare them with the actual $s$;
\item
find the channel \emph{stage}.
\end{itemize}

The above steps tell us the type of the curve $C(s)$, how many extremal and
specific points it has, in which interval we are in this curve and what is the
type of solution (stage).  This is particularly relevant to characterize
channels and might be useful in practical situations to determine the optimal
`work point' of a channel by having some freedom in its parameters values.

\medskip

Then, we have presented the solution for the memory channel, thus generalizing
the results of~\cite{GM,PRA,NJP,CosmoRates}.  Here, the problem of finding the
capacity has been reformulated in a multi-mode setting as a total optimization
problem split in two tasks: the first task is the ``internal optimization'',
\emph{i.e.} optimization inside each mode and the second task is the ``external
optimization'', \emph{i.e.} finding the optimal distribution of the total input
energy $\sum_{k=1}^nN_k$ over ``boxes'' (modes) to get maximal output sum
$\sum_{k=1}^nC_k(N_k)$.  Then, the first task has been addressed using the
techniques developed for the single-mode channel, while the second one using
convex separable programming techniques \cite{StefStef,StefBook}. For the latter
we have also given formal proofs of both the uniqueness of the solution and the
convergence of the proposed algorithm.

The above splitting has become possible because we were confined to the class
of memory models which make the optimization problem spectral. 

In the case of single-mode channel we have derived theorems about the
optimality of pure states showing that for any given $V_\mathrm{env}$ the
optimal input channel state is pure, and for any fixed $\Tr(V_\mathrm{env})$ the
optimal $V_\mathrm{env}$ is pure.  In particular, purity of $V_\mathrm{env}$
once $\Tr(V_\mathrm{env})$ is fixed, results in a violation of quadrature
symmetry.  When this result is extended to the memory channel (\emph{i.e.}
non identical multiple modes environment), with optimization over distribution
of input energies, we have discovered violation of mode symmetry too.  That is
to say, optimization inside each box gives us ``violation of quadrature
symmetry'' with ``input and environment purity theorems''; then, maximization
over our blackboxes gives us ``violation of mode symmetry'' and ``optimal
channel memory''.

In this context the enhancement (superadditivity) of classical capacity is
possible (for only some values of the memory channel parameters), if energy is
redistributed between environment modes to become (in general) different in
different modes. This possible violation of mode symmetry points out the
existence of nontrivial optimal channel's environment (memory).  Such
environment can always be chosen pure.  One can also say that capacity is
superadditive if mode symmetry is violated and additive otherwise, where
transition between additive and superadditive cases happens at critical and
supercritical parameters found for the single-use of the single-mode channel.

Notice that the main feature of the considered memory model is to be symbol independent,
\emph{i.e.} the action of the channel at a given use does not depend on the
previous inputs, and without a causal structure.  That made its characterization
a daunting task, which nevertheless has been accomplished.

Transmission rates for heterodyne and homodyne measurements have been treated
parallelly to the capacity because they can be considered as its logarithmic
approximations. In the case of heterodyne it has done by introducing heterodyne
variables. Thus, most of the capacity properties can be also found analyzing the
rates. In particular, it was shown that homodyne measurement for the single-use
of the single-mode channel gives a rate which is always monotonically growing
function of environment squeezing. However, this is not the case for heterodyne
measurement which is monotonically growing function of squeezing only in
neighborhood of $s\to\infty$, therefore its critical parameters were also
calculated and its regimes were studied to provide complete characterization. 

Finally, besides a thorough characterization of the lossy channel, we have
provided mathematical techniques for the solution of optimization problems in
information transmission with Gaussian channels.  The machinery developed herein
seems applicable  to other capacities and other Gaussian channels as witnessed
by the similarities with a recent study on additive Gaussian noise channel
\cite{NewNoiseChannel}, which can be characterized as well by critical
parameters~\cite{SchaferKarpovSPIE}.  Above all extension to the amplification
channel seems within reach and is planned as a future work.

\appendices
\section{Proof of the input purity theorem for capacity}\label{InPurTh}

Let us prove the theorem~\ref{InpPurTheor}. Since the dimension of matrices is
$2\times2$, there is a symplectic transformation $S$ which is orthogonal and
diagonalizes $V_\mathrm{env}$. Let us apply $S$ to matrices $V_\mathrm{out}$ and
$\overline{V}_\mathrm{out}$ (see Eqs.~\eqref{VoutInit} and~\eqref{VoutAv}).  The
transformation~$S$ preserves energy constraint~\eqref{enrestr}, symplectic
eigenvalues\footnote{As far as only the single-mode case is discussed, index $k$
(see Eq.~\eqref{HolDef}) is omitted for symplectic
eigenvalues.}~$\nu$,~$\overline\nu$ and does not change the Holevo function. If
$V_\mathrm{in}$, $V_\mathrm{mod}$ and $V_\mathrm{env}$ are taken in the form
\begin{align*}
V_\mathrm{in}=
\left(\begin{array}{cc}
i_q&i_{qp}\\
i_{qp}&i_p
\end{array}\right)&,\qquad
V_\mathrm{mod}=
\left(\begin{array}{cc}
m_q&m_{qp}\\
m_{qp}&m_p
\end{array}\right),\\
V_\mathrm{env}=
&\left(\begin{array}{cc}
e_q&e_{qp}\\
e_{qp}&e_p
\end{array}\right)
\end{align*}
we get for symplectic eigenvalues the relations
\begin{multline*}
\overline\nu^2=\bigl[\eta\,(i_q+m_q)+(1-\eta)\,e_q\bigr]\times\\
\bigl[\eta\,(i_p+m_p)+(1-\eta)\,e_p\bigr]-
\eta^2(i_{qp}+m_{qp})^2,
\end{multline*}
\begin{equation*}
\nu^2=\bigl[\eta\,i_q+(1-\eta)\,e_q\bigr]
\bigl[\eta\,i_p+(1-\eta)\,e_p\bigr]-\eta^2i_{qp}^2.
\end{equation*}
By setting $m_{qp}$ to zero we cannot violate positivity of $V_\mathrm{mod}$ or
change energy constraint~\eqref{enrestr}, which is equivalent to write
\begin{equation}
i_q+i_p+m_q+m_p=2N+1,
\label{PrConstr}
\end{equation}
but we always increase maximum in the Holevo function.  Thus, the optimal
$V_\mathrm{mod}$ must have $m_{qp}=0$. 

Also, it is evident that the case of mixed input
(\mbox{$i_q^{\phantom{1}}i_p^{\phantom{1}}-i_{qp}^2>\frac14$}) is not optimal.  Indeed,
in this case there is a value $i'_{qp}>i_{qp}^{\phantom{1}}$ which gives
$i_q^{\phantom{1}}i_p^{\phantom{1}}-i'^2_{qp}=\frac14$ and does not change the
constraint~\eqref{PrConstr}.  Because of the monotonic behavior and the
concavity over~$y$ of function $g\left(\sqrt{y}-\frac12\right)$, the matrix $V_\mathrm{in}$
with $i_{qp}$ replaced by $i'_{qp}$ gives higher maximum for capacity. Thus,
optimal input state must be pure. 

Following~\cite{NewNoiseChannel,SchaferKarpov} one can consider the Lagrange
equations for the variables $i_q$, $i_p$, $i_{qp}$, $m_q$, $m_p$, $m_{qp}$ with
constraints~\eqref{PrConstr} and
$i_q^{\phantom{1}}i_p^{\phantom{1}}-i_{qp}^2=\frac14$.  This is resonable because
the case of $N$ higher than some threshold value always gives the solution with
positive values of $i_q$, $i_p$, $m_q$, $m_p$ and positive matrix
$V_\mathrm{mod}$, therefore the corresponding constraints (requiring positivity)
can be omitted.  In particular, the derivative of the Lagrange function with
respect to $m_{qp}$ gives $i_{qp}+m_{qp}=0$. Taking into account that the
optimal $m_{qp}$ is $0$, we get that  $i_{qp}=0$ is also optimal.

Analogously, if $N$ is below that threshold, either the value of $m_q$ or $m_p$
found according to the above approach is negative. It means that for a given
$N$, the solution with positive values of both $m_q$ and $m_p$ does not exist
and single zero-equal $m$-value is the only possibility allowed by the
restriction $m_q,m_p\geqslant0$ (the trivial case $m_q=m_p=0$ gives zero
capacity and therefore is excluded from consideration).  

Let us consider the case of
$m_q=0$ (the case $m_p=0$ 
can be proved analogously). Notice, that $m_q=0$ implies that the
covariance $m_{qp}$ is not defined, \emph{i.e.} matrix $V_\mathrm{mod}\equiv
m_p$ is a scalar. Then, one can consider the Lagrange equations for the
variables $i_q$, $i_{p}$, $i_{qp}$, $m_{p}$ with constraints~\eqref{PrConstr}
and $i_q^{\phantom{1}}i_p^{\phantom{1}}-i_{qp}^2=\frac14$ (one can show that the
solution always gives $i_q,i_{p}>0$). Taking the Lagrange function in the form
\begin{multline*}
L=g\left(\overline\nu-\frac12\right)-g\left(\nu-\frac12\right)\\
-\lambda\,(i_q+i_{p}+m_{p}-2N-1)
-\gamma\left(i_q^{\phantom{1}}i_p^{\phantom{1}}-i_{qp}^2-\frac14\right),
\end{multline*}
where $\lambda$ and $\gamma$ are the Lagrange multipliers,
one can get, from $\partial L/\partial i_{qp}=0$, that
\begin{equation}
i_{qp}\left(\frac{g_1(\overline\nu)}{\overline\nu^2}-
\frac{g_1(\nu)}{\nu^2}-2\gamma\right)=0.
\label{iqp0}
\end{equation}
By expressing~$\gamma$ from Eq.~\eqref{iqp0} and substituting it in the relation 
\begin{equation*}
\frac{\partial L}{\partial\,i_q}-\frac{\partial L}{\partial\,i_{p}}=0
\end{equation*}
we arrive at
\begin{equation}
\frac{g_1(\overline\nu)}{\overline\nu^2}\left(\beta-\eta^2m_{p}\right)=
\frac{g_1(\nu)}{\nu^2}\beta,
\label{inconsist}
\end{equation}
where
\begin{equation*}
\beta=\eta\,(1-\eta)(e_q-e_{p})-(1-\eta^2)(i_q-i_{p}).
\end{equation*}
As far as the function $y^{-2}g_1(y)$ is monotonically decreasing and $m_p>0$,
Eq.~\eqref{inconsist} results not consistent. Thus, $i_{qp}=0$ is the only
possibility in Eq.~\eqref{iqp0}.

\section{Proof of the input purity theorem for rates}\label{InPurTh2}

Let us prove the theorem~\ref{InpPurTheor2}.  If quadrature $p$ is measured for
homodyne rate~\eqref{HomR} (the case of $q$-quadrature is analogous),
then one needs to maximize the quantity
\begin{equation}
R^\mathrm{(hom)}=\frac12\log_2\overline o_p-\frac12\log_2 o_p,
\label{HomAppx}
\end{equation}
where $\overline o_p$ and $o_p$ are diagonal elements of matrices $\overline
V_\mathrm{out}= \diag(\overline o_q,\overline o_p)$ and
$V_\mathrm{out}=\diag(o_q,o_p)$ (see Eqs.~\eqref{VoutInit} and~\eqref{VoutAv}).
Similarly, we shall denote input and modulation matrices as
$V_\mathrm{in}=\diag(i_q,i_p)$ and $V_\mathrm{mod}=\diag(m_q,m_p)$.
Analogously, to find the heterodyne rate~\eqref{HetR} one needs to maximize the
quantity
\begin{multline}
R^\mathrm{(het)}=
\log_2\sqrt{\bigl(\overline o_q+1/2\bigr)\bigl(\overline o_p+1/2\bigr)}\\
-\log_2\sqrt{\bigl(o_q+1/2\bigr)\bigl(o_p+1/2\bigr)}.
\label{HetAppx}
\end{multline}
The maximum for both functions~\eqref{HomAppx} and~\eqref{HetAppx} is taken over
the variables $i_q$, $i_p$, $m_q$ and $m_p$.

Suppose, that the maximum is achieved with a non pure state having $i_qi_p>\frac14$.
This means that some real number $\varepsilon>0$ exist, such that
$i_p^{\phantom{a}}=i'_p+\varepsilon$, where $i'_p=(4i_q^{\phantom{a}})^{-1}$.  New
variables denoted with primes and defined by transformations
\begin{align*}
&i'_p=i_p^{\phantom{a}}-\varepsilon,&&i'_q=i_q^{\phantom{a}},\\
&m'_p=m_p^{\phantom{a}}+\varepsilon,&&m'_q=m_q^{\phantom{a}},
\end{align*}
make $V_\mathrm{in}$ pure and preserve the energy constraint~\eqref{enrestr}.
They also preserve the values of the first terms and decrease the values of the
second terms in Eqs.~\eqref{HomAppx} and~\eqref{HetAppx}, thus providing higher
maximum than initial variables.  Hence, the theorem is proved by contradiction.

\section{Proof of the proposition~\ref{EuNeqEust}}\label{EuNeqEustProof}

Let us prove the proposition~\ref{EuNeqEust}.  Suppose, that $m_u>0$ and $m_{u_\star}=0$
are optimal for $e_u>e_{u_\star}$ in second stage. We will consider three possible cases
\mbox{$\overline o_u>\overline o_{u_\star}$}, \mbox{$\overline o_u<\overline o_{u_\star}$}
and \mbox{$\overline o_u=\overline o_{u_\star}$} separately.  If $\overline o_u>\overline
o_{u_\star}$, then our assumption leads to contradiction due to
proposition~\ref{EuNeqEustGen}.  In what follows we will use the equivalence between $\overline
o_u\leqslant\overline o_{u_\star}$ and $i_{u_\star}\geqslant i_{u_\star\min}$,
where
\begin{equation*}
i_{u_\star\min}=N+\frac12+\frac{1-\eta}{2\eta}\left(e_u-e_{u_\star}\right). 
\label{iuneqvol}
\end{equation*}

Notice, that our condition $e_u>e_{u_\star}$ leads to $i_u<i_{u_\star}$, where the latter
is equivalent to $i_u<\frac12$ due to optimality of pure input state (see
theorem~\ref{InpPurTheor}). For the interval $i_u<\frac12$ one can show that $\nu^2$ is a
decreasing function of $i_u$. In addition, for $i_{u_\star}>i_{u_\star\min}$ one can see
that $\overline\nu^2$ is a decreasing function of $i_{u_\star}$.  Indeed, for these
intervals the derivatives of $\nu^2$ and $\overline\nu^2$ are negative:
\begin{equation}
\begin{split}
&\frac{d\nu^2}{di_{u_{\phantom\star}}}=
\eta\,(1-\eta)\left(e_{u_\star}-\frac{e_u}{4i_u^2}\right)<0,\\
&\frac{d\overline\nu^2}{di_{u_\star}}=
2\eta^2\left(i_{u_\star\min}-i_{u_\star}\right)<0.
\end{split}
\label{dnudiu}
\end{equation}

First, let us consider the strict inequality $\overline o_u<\overline o_{u_\star}$. If
the variables $i_u$, $i_{u_\star}$ and $m_u$ are changed according to transformations
\begin{align}
&i'_{u_{\phantom\star}}=\frac1{4\left(i_{u_\star}-\varepsilon\right)},
\label{transformIU}\\
&i'_{u_\star}=i^{\phantom{'}}_{u_\star}-\varepsilon,
\label{transformIUST}\\
&m'_u=2N+1-i_{u_\star}+\varepsilon-\frac1{4\left(i_{u_\star}-\varepsilon\right)},
\nonumber
\end{align}
where $0<\varepsilon<i_{u_\star}-i_{u_\star\min}$, then the energy
constraint~\eqref{1useConstr} is preserved (the variable $m_{u_\star}=0$ remains
unchanged).  Since $i'_u>i_u^{\phantom{'}}$ and $i'_{u_\star}<i_{u_\star}^{\phantom{'}}$
the new symplectic eigenvalues satisfy $\nu'^2<\nu^2$ and $\overline\nu'^2>\overline\nu^2$
(see Eqs.~\eqref{dnudiu}). As far as $g$ is increasing function, the new variables
increase the first term in~Eq.~\eqref{defcap} and decrease the second term thus providing
higher capacity.

Next, we consider the case  $\overline o_u=\overline o_{u_\star}$.  Now we change the
variables $i_u$, $i_{u_\star}$ according to transformations~\eqref{transformIU},
\eqref{transformIUST} and variables $m_u$, $m_{u_\star}$ as follows:
\begin{align*}
&m'_{u_{\phantom{\star}}}=2N+1-i_{u_\star}-\frac1{4\left(i_{u_\star}-\varepsilon\right)},\\
&m'_{u_\star}=\varepsilon,
\end{align*}
where we choose $\varepsilon$ (also for Eqs.~\eqref{transformIU}
and~\eqref{transformIUST}) from the interval $0<\varepsilon<\frac12\left(\overline
o_{u_\star}-\overline o_u\right)$.  Since  $i'_u>i_u^{\phantom{'}}$ and
$i'_{u_\star}<i_{u_\star}^{\phantom{'}}$ we have $\nu'^2<\nu^2$.  In addition, the
equalities $\overline o_u=\overline o_{u_\star}=\overline o'_u=\overline o'_{u_\star}$
lead to $\overline\nu'^2=\overline\nu^2$. Thus, the new variables preserve the first term
and decrease the second term in Eq.~\eqref{defcap} thus providing higher capacity.

Finally, we have shown that for all possible cases (\mbox{$\overline o_u>\overline
o_{u_\star}$}, \mbox{$\overline o_u<\overline o_{u_\star}$} and \mbox{$\overline
o_u=\overline o_{u_\star}$}) the capacity can be increased by a suitable change of variables.
Hence, the proposition is proved by contradiction.

\section{Proof of the environment purity theorem}\label{EnvPurTh}

Let us prove the theorem~\ref{EnvPurTheor}.
At first, notice that the following Lemma holds.

\bigskip

\begin{lemma}
Suppose one has real positive numbers $a,b,c,d$, where 
$c>a$, $b-a>d-c$ and $f(x)$ is a
monotonically growing concave function in the interval $x\in(0,\infty)$, then
$$f(b)-f(a)>f(d)-f(c).$$
\end{lemma}

\bigskip

In the case of the first stage $\underline C\equiv0$.  In the case of the third
stage
$$
\max_{\EuScript{N}_\mathrm{env}}\underline C=\underline C(\EuScript{N}_\mathrm{env}=0),
$$
\emph{i.e.} it is optimal to make the environment pure. Then, suppose that we
have the case of second stage and environment in mixed state is optimal.
Remember, that it was proved for \mbox{$e_{q}>e_{p}$} that $m_{q}=0$ and
$o_{q}>\overline o_{p}>o_{p}$ (see proposition~\ref{EuNeqEust} and Eq.~\eqref{ineqs2ndStage}).  
Let us now change the environment variables by preserving $N_\mathrm{env}$ and making the new
environment state pure
($\EuScript{N}'_\mathrm{env}=0$). It corresponds to the change of variables
$e_{q}\to e'_{q}$, $e_{p}\to e'_{p}$ (the eigenvalues $i_{u}$ and $m_{u}$ remain the
same), where the new value of 
squeezing $s'$ is given by the relation
$$\cosh s'=\left(2\EuScript N_\mathrm{env}+1\right)\cosh s.$$ 
This results to $o'_{q}>o_{q}$ and
$o'_{p}<o_{p}$, {\it i.e.} $o'_{q}-o'_{p}>o_{q}-o_{p}$, while
$o'_{q}+o'_{p}=o_{q}+o_{p}$. It means that {$\nu'<\nu$} (see analogous proofs in
Subsec.~\ref{TheSolStages}).  One can then write down:
$$
o'_{q}\,(o'_{p}+\eta\,m_{p}^{\phantom{1}})-o'_{q}o'_{p}>o_{q}\,(o_{p}+\eta\,m_{p})-o_{q}o_{p},
$$
which is equivalent to $o'_{q}>o_{q}^{\phantom{1}}$. Taking into account the above inequality
and applying the Lemma for $f(x)={\sqrt{x}}$ one gets
\begin{multline*}
\sqrt{o'_{q}\,(o'_{p}+\eta\,m_{p}^{\phantom{1}})}-\sqrt{o'_{q}o'_{p}}
>\sqrt{o_{q}\,(o_{p}+\eta\,m_{p})}-\sqrt{o_{q}o_{p}\phantom{'}},
\end{multline*}
\emph{i.e.} $\overline\nu'-\nu'>\overline\nu-\nu$.  Finally, applying again the
Lemma for the function $f(x)=g\left(x-\frac12\right)$ one gets $\underline C'>\underline C$.
Hence, the theorem is proved by contradiction.

\section{The second derivative of solution over input energy}\label{App2ndDer}

Let us show that $d^2\underline C/d{}^{}N^2<0$. In the second stage it is
\begin{equation}
\frac{d^2\underline C}{d{}^{}N^2}=\frac{\partial^2\underline C}{\partial N^2}+
2\frac{\partial^2\underline C}{\partial N\partial{}^{}i_u}
\frac{\partial{}^{}i_u}{\partial N}
+\frac{\partial^2\underline C}{\partial{}^{}i_u^2}
\left(\frac{\partial{}^{}i_u}{\partial N}\right)^2+
\frac{\partial{}^{}\underline C}{\partial{}^{}i_u}\frac{\partial^2i_u}{\partial N^2}
\label{FullSndDir}
\end{equation}
Taking into account that $\partial\underline{}^{}C/\partial{}^{}i_u=0$ for any
values of $N$, we get an equality
\begin{equation*}
\frac{d}{d{}^{}N}\left(\frac{\partial{}^{}\underline C}{\partial{}^{}i_u}\right)=
\frac{\partial^2\underline C}{\partial N\partial{}^{}i_u}+
\frac{\partial^2\underline C}{\partial{}^{}i_u^2}\frac{\partial{}^{}i_u}{\partial N},
\end{equation*}
which allows us to rewrite the derivative~\eqref{FullSndDir} as
\begin{equation}
\frac{d^2\underline C}{d{}^{}N^2}=\frac{\partial^2\underline C}{\partial N^2}+
\frac{\partial^2\underline C}{\partial N\partial{}^{}i_u}
\frac{\partial{}^{}i_u}{\partial N}=
\frac{\partial^2\underline C}{\partial N^2}-
\frac{\partial^2\underline C}{\partial{}^{}i_u^2}
\left(\frac{\partial{}^{}i_u}{\partial N}\right)^2,
\label{SndDerReduced}
\end{equation}
where
\begin{equation*}
\frac{\partial{}^{}i_u}{\partial N}=
-\frac{\partial\mathcal F}{\partial N}
\left(\frac{\partial\mathcal F}{\partial{}^{}i_u}\right)^{-1}
\end{equation*}
and
\begin{equation*}
\frac{\partial^2\underline{C}}{\partial N^2}=
\bigl[g_2(\overline\nu)-g_1(\overline\nu)\bigr]
\left(\frac{\eta}{\overline o_{u_\star}}\right)^2.
\end{equation*}

One can show that 
\begin{align}
&\frac{\partial^2\underline C}{\partial N\partial{}^{}i_u}=
\frac{\eta^2}2\mathcal{L},
&&\frac{\partial\mathcal F}{\partial N}=\eta{}^{}\mathcal{L},
\label{DersExtra}
\end{align}
where
\begin{equation*}
\mathcal L=
g_1(\overline\nu)\left(\frac1{\overline\nu^2}+\frac1{\overline o_{u_\star}^2}\right)
+g_2(\overline\nu)\left(\frac1{\overline\nu^2}-\frac1{\overline o_{u_\star}^2}\right).
\end{equation*}
Since it always is $\overline\nu^2>\overline o_{u_\star}^2$, $g_1>0$ and $g_2<0$
(see Eqs.~\eqref{gExplicit}),
the quantity $\mathcal L$ and the derivatives~\eqref{DersExtra} are positive.
Also it can be found that
\begin{align*}
\frac{\partial\mathcal F}{\partial{}^{}i_u}=
-&\frac\eta2\Biggl[
g_1(\overline\nu)\left(\frac1{o_u}+\frac1{\overline o_{u_\star}}\right)^2-
g_1(\nu)\left(\frac1{o_u}+\frac1{4i_u^2o_{u_\star}^{\phantom{1}}}\right)^2\\
&\,-g_2(\overline\nu)\left(\frac1{o_u}-\frac1{\overline o_{u_\star}}\right)^2+
g_2(\nu)\left(\frac1{o_u}-\frac1{4i_u^2o_{u_\star}^{\phantom{1}}}\right)^2\\
&\,+\frac{g_1(\nu)}{\eta{}^{}o_{u_\star}^{\phantom{1}}i_u^3}
\Biggr].
\end{align*}

It was shown in~\cite{NewNoiseChannel} for additive noise channel that
\begin{align*}
&\frac{\partial{}^{}i_u}{\partial N}>0,&&
\frac{d^2\underline C}{d{}^{}N^2}<0
\end{align*}
in the second stage, which can be similarly proved
also for lossy channel.
In addition, it is evident from Eq.~\eqref{iuapproxiu} that
$\partial{}^{}i_u/\partial N>0$ in the zeroth-order approximation.
Then, in the third stage we have
\begin{equation}
\frac{d^2\underline C}{d{}^{}N^2}=\frac{\partial^2\underline C}{\partial N^2}=
\frac{\eta^2}{\overline\nu^2}g_2(\overline\nu)<0.
\label{2ndDer3st}
\end{equation}
Thus, we have shown that the second derivative of capacity is negative in the
case of both the second and the third stages.

The derivative~\eqref{2ndDer3st} also holds for rates if the replacement~\eqref{Replg} 
is applied. Besides it, for the heterodyne rate the replacement~\eqref{Replnu} must be applied.

\section*{Acknowledgment}
The research leading to these results has received funding from the European
Commission's seventh Framework Programme (FP7/2007-2013) under grant agreement
no. 213681.  O.~P.~thanks Zborovskii~V.~G., Karpov~E.~A. and Sch\"afer J. for
fruitful discussions.
\ifCLASSOPTIONcaptionsoff
  \newpage
\fi



%


\begin{IEEEbiographynophoto}{Oleg Pilyavets}
was born in Frunze, USSR in 1983. He received his B.Sc. and M.Sc. degree in Applied
Physics and Mathematics from Moscow Institute of Physics and Technology, the deaprtment of
Problems of Physics and Power Engineering. In 2009 he earned the Ph.D. in
Physics from the P.~N.~Lebedev Physical Institute, Moscow. In 2010 he received the Ph.D.
in Physics also from the University of Camerino, Italy.  Then he spent one year at
University of Camerino as a PostDoctoral fellow.

Now he has a postdoctoral position at Centre for Quantum Information and Communication
of the Universit\'e Libre de Bruxelles, Belgium. During the last years his research interest
was mainly on information transmission through Gaussian quantum channels.
\end{IEEEbiographynophoto}

\begin{IEEEbiographynophoto}{Cosmo Lupo}
received the Ph.D. degree in Fundamental and Applied Physics from the
University of Napoli ``Federico II'', Italy, in 2007. He was Marie Curie Fellow at the
Research Center for Quantum Information (RCQI) in 2008. From 2008 he is a PostDoctoral
fellow at the School of Science and Technology, University of Camerino, Italy.
He has been engaged in geometric quantum computation, then in entanglement
characterization and more recently in quantum channels capacities.
\end{IEEEbiographynophoto}

\begin{IEEEbiographynophoto}{Stefano Mancini}
received the Ph.D. degree in physics from the University of Perugia,
Italy, in 1998. He was Postdoctoral Fellow at the University of Milan for three years.
Subsequently, with temporary lecturer positions held at University of Milan and at
University of Camerino, Italy, he have contributed to establish the first Italian academic
courses on quantum information and computation.

From 2004 to 2010 he has been researcher
of theoretical physics and mathematical methods at Faculty of Science, University of
Camerino. Since September 2010 he is professor of theoretical physics and mathematical
methods at School of Science and Technology, University of Camerino. He has been involved
in the fields of theoretical quantum optics, quantum control theory and quantum information
theory. He has given significant contributions to the chgaracterization of entanglement, to
the formalism of quantum feedback, to the models of quantum memory channels and to the development of
quantum cryptographic protocols.

He has authored or coauthored more than 150 papers
published in leading international journals. He has been Editor of four journals for
special issues devoted to quantum information topics. He is currently a member of the
Editorial Board of the International Journal of Quantum Information. Dr. Mancini was
awarded two times by the Italian Ministry of Research under Young Researchers Program.
\end{IEEEbiographynophoto}




\vfill


\end{document}